# Mathematical Foundations of the Relativistic Theory of Quantum Gravity

**Fran De Aquino**
Maranhao State University, Physics Department, S.Luis/MA, Brazil.

**Abstract:** Starting from the *action function*, we have derived a theoretical background that leads to the *quantization of gravity* and the deduction of a correlation between the gravitational and the inertial masses, which depends on the *kinetic momentum* of the particle. We show that the strong equivalence principle is reaffirmed and, consequently, Einstein's equations are preserved. In fact, such equations are deduced here directly from this new approach to Gravitation. Moreover, we have obtained a generalized equation for the inertial forces, which incorporates the Mach's principle into Gravitation. Also, we have deduced the equation of Entropy; the Hamiltonian for a particle in an electromagnetic field and the reciprocal fine structure constant directly from this new approach. It was also possible to deduce the expression of the *Casimir force* and to explain the *Inflation Period* and the *Missing Matter,* without assuming existence of *vacuum fluctuations*. This new approach to Gravitation will allow us to understand some crucial matters in Cosmology.

**Contents**

## 1. INTRODUCTION

Quantum Gravity was originally studied, by Dirac and others, as the problem of quantizing General Relativity. This approach presents many difficulties, detailed by Isham [1]. In the 1970's, physicists tried an even more conventional approach: simplifying Einstein's equations by assuming that they are *almost linear*, and then applying the standard methods of quantum field theory to the thus oversimplified equations. But this method, too, failed. In the 1980's a very different approach, known as string theory, became popular. Thus far, there are many enthusiasts of string theory. But the mathematical difficulties in string theory are formidable, and it is far from clear that they will be resolved any time soon. At the end of 1997, Isham [2] pointed out several "Structural Problems Facing Quantum Gravity Theory". At the beginning of this new century, the problem of quantizing the gravitational field was still open.

In this work, we propose a new approach to Quantum Gravity. Starting from the generalization of the *action function* we have derived a theoretical background that leads to the quantization of gravity. Einstein's General Relativity equations are deduced directly from this theory of Quantum Gravity. Also, this theory leads to a complete description of the Electromagnetic Field, providing a consistent unification of gravity with electromagnetism.

## 2. THEORY

We start with the *action* for a free-particle that, as we know, is given by

$$S = -\alpha \int_a^b ds$$

where $\alpha$ is a quantity which characterizes the particle.

In Relativistic Mechanics, the action can be written in the following form [3]:

$$S = \int_{t_1}^{t_2} L dt = -\int_{t_1}^{t_2} \alpha c \sqrt{1 - V^2/c^2}\, dt$$

where

$$L = -\alpha c \sqrt{1 - V^2/c^2}$$

is the Lagrange's function.

In Classical Mechanics, the Lagrange's function for a free-particle is, as we know, given by: $L = aV^2$ where $V$ is the speed of the particle and $a$ is a quantity *hypothetically* [4] given by:

$$a = m/2$$

where $m$ is the mass of the particle. However, there is no distinction about the kind of mass (if *gravitational mass*, $m_g$, or *inertial mass* $m_i$) neither about its sign $(\pm)$.

The correlation between $a$ and $\alpha$ can be established based on the fact that, on the limit $c \to \infty$, the relativistic expression for $L$ must be reduced to the classic expression $L = aV^2$. The result [5] is: $L = \alpha V^2/2c$. Therefore, if $\alpha = 2ac = mc$, we obtain $L = aV^2$. Now, we must decide if $m = m_g$ or $m = m_i$. We will see in this work that the definition of $m_g$ includes $m_i$. Thus, the right option is $m_g$, i.e.,

$$a = m_g/2.$$

Consequently, $\alpha = m_g c$ and the generalized expression for the action of a free-particle will have the following form:

$$S = -m_g c \int_a^b ds \qquad (1)$$

or

$$S = -\int_{t_1}^{t_2} m_g c^2 \sqrt{1 - V^2/c^2}\, dt \qquad (2)$$

where the Lagrange's function is

$$L = -m_g c^2 \sqrt{1 - V^2/c^2}. \qquad (3)$$

The integral $S = \int_{t_1}^{t_2} m_g c^2 \sqrt{1 - V^2/c^2}\, dt$, preceded by the *plus* sign, cannot have a *minimum*. Thus, the integrand of Eq.(2) must be *always positive*. Therefore, if $m_g > 0$, then necessarily $t > 0$; if $m_g < 0$, then $t < 0$. The possibility of $t < 0$ is based on the well-known equation $t = \pm t_0 \big/ \sqrt{1 - V^2/c^2}$ of Einstein's Theory.

Thus if the *gravitational mass* of a particle is *positive,* then $t$ is also *positive* and, therefore, given by $t = +t_0 \big/ \sqrt{1 - V^2/c^2}$. This leads to the well-known relativistic prediction that the particle goes to the *future,* if $V \to c$. However, if the *gravitational mass* of the particle is *negative,* then $t$ is *negative* and given by $t = -t_0 \big/ \sqrt{1 - V^2/c^2}$. In this case, the prediction is that the particle goes to the *past,* if $V \to c$. Consequently, $m_g < 0$ is the necessary condition for the particle to go to the *past*. Further on, a correlation between the *gravitational* and the *inertial* masses will be derived, which contains the possibility of $m_g < 0$.

The Lorentz's transforms follow the same rule for $m_g > 0$ and $m_g < 0$, i.e., the sign before $\sqrt{1 - V^2/c^2}$ will be $(+)$ when $m_g > 0$ and $(-)$ if $m_g < 0$.

The *momentum*, as we know, is the vector $\vec{p} = \partial L / \partial \vec{V}$.Thus, from Eq.(3) we obtain

$$\vec{p} = \frac{m_g \vec{V}}{\pm\sqrt{1 - V^2/c^2}} = M_g \vec{V}$$

The $(+)$ sign in the equation above will be used when $m_g > 0$ and the $(-)$ sign if $m_g < 0$. Consequently, we will express the momentum $\vec{p}$ in the following form

$$\vec{p} = \frac{m_g \vec{V}}{\sqrt{1 - V^2/c^2}} = M_g \vec{V} \qquad (4)$$

The derivative $d\vec{p}/dt$ is the *inertial* force $F_i$ which acts on the particle. If the force is perpendicular to the speed, we have

$$\vec{F}_i = \frac{m_g}{\sqrt{1 - V^2/c^2}} \frac{d\vec{V}}{dt} \qquad (5)$$

However, if the force and the speed have the same direction, we find that

$$\vec{F}_i = \frac{m_g}{\left(1 - V^2/c^2\right)^{3/2}} \frac{d\vec{V}}{dt} \qquad (6)$$

From Mechanics [6], we know that $\vec{p} \cdot \vec{V} - L$ denotes the *energy* of the particle. Thus, we can write

$$E_g = \vec{p} \cdot \vec{V} - L = \frac{m_g c^2}{\sqrt{1 - V^2/c^2}} = M_g c^2 \qquad (7)$$

Note that $E_g$ is not null for $V = 0$, but that it has the finite value

$$E_{g0} = m_{g0} c^2 \qquad (8)$$

Equation (7) can be rewritten in the following form:

$$E_g = m_g c^2 - \frac{m_g c^2}{\sqrt{1 - V^2/c^2}} - m_g c^2 =$$

$$= \frac{m_g}{m_i} \left[ m_i c^2 + \underbrace{\left( \frac{m_i c^2}{\sqrt{1 - V^2/c^2}} - m_i c^2 \right)}_{E_{Ki}} \right] =$$

$$= \frac{m_g}{m_i} \left( E_{i0} + E_{Ki} \right) = \frac{m_g}{m_i} E_i \qquad (9)$$

By analogy to Eq. (8), $E_{i0} = m_{i0} c^2$ into the equation above, is the inertial energy *at rest*. Thus, $E_i = E_{i0} + E_{Ki}$ is the *total* inertial energy, where $E_{Ki}$ is the *kinetic* *inertial energy*. From Eqs. (7) and (9) we thus obtain

$$E_i = \frac{m_{i0}c^2}{\sqrt{1-V^2/c^2}} = M_i c^2. \qquad (10)$$

For small velocities $(V << c)$, we obtain

$$E_i \approx m_{i0}c^2 + \tfrac{1}{2} m_i V^2 \qquad (11)$$

where we recognize the classical expression for the *inertial kinetic energy* of the particle.

The expression for the *gravitational kinetic energy*, $E_{Kg}$, is easily deduced by comparing Eq.(7) with Eq.(9). The result is

$$E_{Kg} = \frac{m_g}{m_i} E_{Ki}. \qquad (12)$$

In the presented picture, we can say that the *gravity*, $\vec{g}$, in a gravitational field produced by a particle of gravitational mass $M_g$, depends on the particle's gravitational energy, $E_g$ (given by Eq.(7)), because we can write

$$g = -G\frac{E_g}{r^2c^2} = -G\frac{M_g c^2}{r^2c^2} \qquad (13)$$

Due to $g = \partial\Phi/\partial r$, the expression of the *relativistic gravitational potential*, $\Phi$, is given by

$$\Phi = -\frac{GM_g}{r} = -\frac{Gm_g}{r\sqrt{1-V^2/c^2}}$$

Then, it follows that

$$\Phi = -\frac{GM_g}{r} = -\frac{Gm_g}{r\sqrt{1-V^2/c^2}} = \frac{\phi}{\sqrt{1-V^2/c^2}}$$

where $\phi = -Gm_g/r$.

Then we get

$$\frac{\partial\Phi}{\partial r} = \frac{\partial\phi}{\partial r\sqrt{1-V^2/c^2}} = \frac{Gm_g}{r^2\sqrt{1-V^2/c^2}}$$

whence we conclude that

$$\frac{\partial\Phi}{\partial r} = \frac{Gm_g}{r^2\sqrt{1-V^2/c^2}}$$

By definition, *the gravitational potential energy per unit of gravitational mass* of a particle inside a gravitational field is equal to the *gravitational potential* $\Phi$ *of the field*. Thus, we can write that

$$\Phi = \frac{U(r)}{m'_g}$$

Then, it follows that

$$F_g = -\frac{\partial U(r)}{\partial r} = -m'_g\frac{\partial\Phi}{\partial r} = -G\frac{m_g m'_g}{r^2\sqrt{1-V^2/c^2}}$$

If $m_g > 0$ and $m'_g < 0$, or $m_g < 0$ and $m'_g > 0$ the force will be *repulsive*; *the force will never be null* due to the existence of a *minimum value* for $m_g$ (see Eq. (24)). However, if $m_g < 0$ and $m'_g < 0$, or $m_g > 0$ and $m'_g > 0$ the force will be *attractive*. Just for $m_g = m_i$ and $m'_g = m'_i$ we obtain the *Newton's attraction law*.

On the other hand, as we know, the gravitational force is *conservative*. Thus, gravitational energy, in agreement with the energy conservation law, can be expressed by the *decrease* of the inertial energy, i.e.,

$$\Delta E_g = -\Delta E_i \qquad (14)$$

This equation expresses the fact that a decrease of gravitational energy corresponds to an increase of the inertial energy.

Therefore, a variation $\Delta E_i$ in $E_i$ yields a variation $\Delta E_g = -\Delta E_i$ in $E_g$. Thus $E_i = E_{i0} + \Delta E_i$; $E_g = E_{g0} + \Delta E_g = E_{g0} - \Delta E_i$ and

$$E_g + E_i = E_{g0} + E_{i0} \qquad (15)$$

Comparison between (7) and (10) shows that $E_{g0} = E_{i0}$, i.e., $m_{g0} = m_{i0}$. Consequently, we have

$$E_g + E_i = E_{g0} + E_{i0} = 2E_{i0} \qquad (16)$$

However $E_i = E_{i0} + E_{Ki}$. Thus, (16) becomes

$$E_g = E_{i0} - E_{Ki}. \qquad (17)$$

Note the *symmetry* in the equations of $E_i$ and $E_g$. Substitution of $E_{i0} = E_i - E_{Ki}$ into (17) yields

$$E_i - E_g = 2E_{Ki} \qquad (18)$$

Squaring the Eqs.(4) and (7) and comparing the result, we find the following correlation between gravitational energy and *momentum* :

$$\frac{E_g^2}{c^2} = p^2 + m_g{}^2 c^2. \qquad (19)$$

The energy expressed as a function of the *momentum* is, as we know, called *Hamiltonian* or Hamilton's function:

$$H_g = c\sqrt{p^2 + m_g{}^2 c^2}. \qquad (20)$$

Let us now consider the problem of quantization of gravity. Clearly there is something unsatisfactory about the whole notion of quantization. It is important to bear in mind that the quantization process is a series of rules-of-thumb rather than a well-defined algorithm, and contains many ambiguities. In fact, for electromagnetism we find that there are (at least) two different approaches to quantization and that while they appear to give the same theory they may lead us to very different quantum theories of gravity. Here we will follow a new theoretical strategy: It is known that starting from the Schrödinger equation we may obtain the well-known expression for the energy of a particle in periodic motion inside a cubical box of edge length $L$ [ 7 ]. The result now is

$$E_n = \frac{n^2 h^2}{8 m_g L^2} \qquad n = 1,2,3,... \qquad (21)$$

Note that the term $h^2 / 8m_g L^2$ (energy) will be minimum for $L = L_{max}$ where $L_{max}$ is the maximum edge length of a cubical box whose maximum diameter

$$d_{max} = L_{max}\sqrt{3} \qquad (22)$$

is equal to *the maximum* length scale *of the Universe*.

The minimum energy of a particle is obviously its inertial energy at rest $m_g c^2 = m_i c^2$. Therefore we can write

$$\frac{n^2 h^2}{8 m_g L_{max}^2} = m_g c^2$$

Then from the equation above it follows that

$$m_g = \pm \frac{nh}{c L_{max} \sqrt{8}} \qquad (23)$$

whence we see that there is a *minimum value* for $m_g$ given by

$$m_{g(min)} = \pm \frac{h}{c L_{max} \sqrt{8}} \qquad (24)$$

The *relativistic* gravitational mass $M_g = m_g \left(1 - V^2/c^2\right)^{-1/2}$, defined in the Eqs.(4), shows that

$$M_{g(\min)} = m_{g(\min)} \qquad (25)$$

The *box normalization* leads to the conclusion that the *propagation number* $k = \left|\vec{k}\right| = 2\pi/\lambda$ is restricted to the values $k = 2\pi n / L$. This is deduced assuming an *arbitrarily large but finite* cubical box of volume $L^3$ [8]. Thus, we have

$$L = n\lambda$$

From this equation, we conclude that

$$n_{max} = \frac{L_{max}}{\lambda_{min}}$$

and

$$L_{min} = n_{min}\lambda_{min} = \lambda_{min}$$

Since $n_{min} = 1$. Therefore, we can write that

$$L_{max} = n_{max} L_{min} \qquad (26)$$

From this equation, we thus conclude that

$$L = n L_{min} \qquad (27)$$

or

$$L = \frac{L_{max}}{n} \qquad (28)$$

Multiplying (27) and (28) by $\sqrt{3}$ and reminding that $d = L\sqrt{3}$, we obtain

$$d = nd_{min} \qquad or \qquad d = \frac{d_{max}}{n} \qquad (29)$$

Equations above show that the length (and therefore the *space*) is *quantized*.

By analogy to (23) we can also conclude that

$$M_{g(max)} = \frac{n_{max} h}{cL_{min}\sqrt{8}} \qquad (30)$$

since the relativistic gravitational mass, $M_g = m_g\left(1 - V^2/c^2\right)^{-1/2}$, is just a multiple of $m_g$.

Equation (26) tells us that $L_{min} = L_{max}/n_{max}$. Thus, Eq.(30) can be rewritten as follows

$$M_{g(max)} = \frac{n_{max}^2 h}{cL_{max}\sqrt{8}} \qquad (31)$$

Comparison of (31) with (24) shows that

$$M_{g(\max)} = n_{\max}^2 m_{g(\min)} \qquad (32)$$

which leads to following conclusion that

$$M_g = n^2 m_{g(\min)} \qquad (33)$$

This equation shows that *the gravitational mass is quantized*.

Substitution of (33) into (13) leads to *quantization of gravity*, i.e.,

$$g = \frac{GM_g}{r^2} = n^2\left(\frac{Gm_{g(\min)}}{\left(r_{\max}/n\right)^2}\right) = $$
$$= n^4 g_{\min} \qquad (34)$$

From the Hubble's law, it follows that

$$V_{max} = \tilde{H} l_{max} = \tilde{H}\left(d_{max}/2\right)$$
$$V_{min} = \tilde{H} l_{min} = \tilde{H}\left(d_{min}/2\right)$$

whence

$$\frac{V_{max}}{V_{min}} = \frac{d_{max}}{d_{min}}$$

Equations (29) tell us that $d_{max}/d_{min} = n_{max}$. Thus the equation above gives

$$V_{min} = \frac{V_{max}}{n_{max}} \qquad (35)$$

which leads to following conclusion

$$V = \frac{V_{max}}{n} \qquad (36)$$

this equation shows that *velocity* is also quantized.

From this equation one concludes that we can have $V = V_{max}$ or $V = V_{max}/2$, but there is nothing in between. This shows clearly that $V_{max}$ cannot be equal to $c$ (speed of light in vacuum). Thus, it follows that

$n = 1 \qquad V = V_{max}$

$n = 2 \qquad V = V_{max}/2$

$n = 3 \qquad V = V_{max}/3 \qquad$ *Tachyons*

........ .................

$n = n_x - 1 \qquad V = V_{max}/(n_x - 1)$

---------------------------

$n = n_x \qquad V = V_{max}/n_x = c \quad \leftarrow$

$n = n_x + 1 \qquad V = V_{max}/(n_x + 1) \qquad$ *Tardyons*

$n = n_x + 2 \qquad V = V_{max}/(n_x + 2)$

............. .........................

*where* $n_x$ *is a big number.*

Then $c$ is the speed *upper limit* of the *Tardyons* and also the speed *lower limit* of the *Tachyons*. Obviously, this limit is *always the same in all inertial frames*. Therefore $c$ can be used as a *reference speed*, to which we may compare any speed $V$, as occurs for the relativistic factor $\sqrt{1 - V^2/c^2}$. Thus, in this factor, $c$ does not refer to maximum propagation speed of the interactions such as some authors suggest; $c$ is just a speed limit which remains the same in any inertial frame.

The temporal coordinate $x^0$ of space-time is now $x^0 = V_{max} t$ ( $x^0 = ct$ is then obtained when $V_{max} \rightarrow c$ ). Substitution of $V_{max} = nV = n\left(\tilde{H} l\right)$ into this equation yields $t = x^0/V_{max} = \left(1/n\tilde{H}\right)\left(x^0/l\right)$. On the other hand, since $V = \tilde{H} l$ and $V = V_{max}/n$ we can write that $l = V_{max}\tilde{H}^{-1}/n$. Thus $\left(x^0/l\right) = \tilde{H}(nt) = \tilde{H} t_{max}$. Therefore, we can finally write

$$t = \left(1/n\tilde{H}\right)\left(x^0/l\right) = t_{max}/n \qquad (37)$$

which shows the quantization of *time*.

From Eqs. (27) and (37) we can easily conclude that the *spacetime is not continuous* it is *quantized*.

Now, let us go back to Eq. (20) which will be called the *gravitational* Hamiltonian to distinguish it from the *inertial* Hamiltonian $H_i$ :

$$H_i = c\sqrt{p^2 + m_{i0}{}^2c^2} \qquad (38)$$

Consequently, Eq. (18) can be rewritten in the following form:

$$H_i - H_g = 2\Delta H_i \qquad (39)$$

where $\Delta H_i$ is the *variation on the inertial Hamiltonian* or *inertial kinetic energy*. A *momentum* variation $\Delta p$ yields a variation $\Delta H_i$ given by:

$$\Delta H_i = \sqrt{(p+\Delta p)^2c^2 + m_{i0}{}^2c^4} - \sqrt{p^2c^2 + m_{i0}{}^2c^4} \qquad (40)$$

By considering that the particle is *initially at rest* $(p=0)$. Then, Eqs. (20), (38) and (39) give respectively: $H_g = m_g c^2$, $H_i = m_{i0}c^2$ and

$$\Delta H_i = \left[\sqrt{1+\left(\frac{\Delta p}{m_{i0}c}\right)^2} - 1\right] m_{i0}c^2$$

By substituting $H_g$, $H_i$ and $\Delta H_i$ into Eq.(39), we get

$$m_g = m_{i0} - 2\left[\sqrt{1+\left(\frac{\Delta p}{m_{i0}c}\right)^2} - 1\right] m_{i0}. \qquad (41)$$

This is the *general expression of the correlation between the gravitational and inertial mass*. Note that for $\Delta p > m_{i0}c\left(\sqrt{5}/2\right)$, the value of $m_g$ becomes *negative*.

Equation (41) shows that $m_g$ decreases of $\Delta m_g$ for an increase of $\Delta p$. Thus, starting from (4) we obtain

$$p + \Delta p = \frac{\left(m_g - \Delta m_g\right)V}{\sqrt{1-(V/c)^2}}$$

By considering that the particle is *initially at rest* $(p=0)$, the equation above gives

$$\Delta p = \frac{\left(m_g - \Delta m_g\right)V}{\sqrt{1-(V/c)^2}}$$

From the Eq.(16) we obtain: $E_g = 2E_{i0} - E_i = 2E_{i0} - \left(E_{i0} + \Delta E_i\right) = E_{i0} - \Delta E_i$

However, Eq.(14) tells us that $-\Delta E_i = \Delta E_g$; what leads to $E_g = E_{i0} + \Delta E_g$ or $m_g = m_{i0} + \Delta m_g$. Thus, in the expression of $\Delta p$ we can replace $\left(m_g - \Delta m_g\right)$ for $m_{i0}$, i.e,

$$\Delta p = \frac{m_{i0}V}{\sqrt{1-(V/c)^2}}$$

We can therefore write

$$\frac{\Delta p}{m_{i0}c} = \frac{V/c}{\sqrt{1-(V/c)^2}} \qquad (42)$$

By substitution of the expression above into Eq. (41), we thus obtain:

$$m_g = m_{i0} - 2\left[\left(1 - V^2/c^2\right)^{-\frac{1}{2}} - 1\right] m_{i0} \qquad (43)$$

For $V=0$ we obtain $m_g = m_{i0}$. Then,

$$m_{g(min)} = m_{i0(min)}$$

Substitution of $m_{g(min)}$ into the *quantized* expression of $M_g$ (Eq. (33)) gives

$$M_g = n^2 m_{i0(min)}$$

where $m_{i0(min)}$ is the *elementary quantum of inertial mass* to be determined.

For $V=0$, the *relativistic* expression $M_g = m_g\Big/\sqrt{1-V^2/c^2}$ becomes $M_g = M_{g0} = m_{g0}$. However, Eq. (43) shows that $m_{g0} = m_{i0}$. Thus, the *quantized* expression of $M_g$ reduces to

$$m_{i0} = n^2 m_{i0(min)}$$

In order to define the *inertial quantum number*, we will change $n$ in the expression above for $n_i$. Thus we have

$$m_{i0} = n_i^2 m_{i0(\min)} \qquad (44)$$

which shows the quantization of *inertial mass*; $n_i$ is the *inertial quantum number*.

We will change $n$ in the quantized expression of $M_g$ for $n_g$ in order to define the *gravitational quantum number*. Thus, we have

$$M_g = n_g^2 m_{i0(min)} \qquad (44a)$$

Finally, by substituting $m_g$ given by Eq. (43) into the relativistic expression of $M_g$, we readily obtain

$$M_g = \frac{m_g}{\sqrt{1-V^2/c^2}} = \\ = M_i - 2\left[\left(1-V^2/c^2\right)^{-\frac{1}{2}} - 1\right]M_i \qquad (45)$$

By expanding in power series and neglecting infinitesimals, we arrive at:

$$M_g = \left(1 - \frac{V^2}{c^2}\right)M_i \qquad (46)$$

Thus, the well-known expression for the simple pendulum period, $T = 2\pi\sqrt{(M_i/M_g)(l/g)}$, can be rewritten in the following form:

$$T = 2\pi\sqrt{\frac{l}{g}}\left(1 + \frac{V^2}{2c^2}\right) \qquad for \quad V \ll c$$

Now, it is possible to learn why Newton's experiments using simple penduli do not find any difference between $M_g$ and $M_i$. The reason is due to the fact that, in the case of penduli, the ratio $V^2/2c^2$ is less than $10^{-17}$, which is much smaller than the accuracy of the mentioned experiments.

Newton's experiments have been improved upon (one part in 60,000) by Friedrich Wilhelm Bessel (1784–1846). In 1890, Eötvos confirmed Newton's results with accuracy of one part in $10^7$. Posteriorly, Eötvos experiment has been repeated with accuracy of one part in $10^9$. In 1963, the experiment was repeated with an even greater accuracy, one part in $10^{11}$. The result was the same previously obtained.

In all these experiments, the ratio $V^2/2c^2$ is less than $10^{-17}$, which is much smaller than the accuracy of $10^{-11}$ obtained in the previous more precise experiment.

Then, we arrive at the conclusion that all these experiments say nothing in regard to the relativistic behavior of masses in relative motion.

Let us now consider a planet in the Sun's gravitational field to which, in the absence of external forces, we apply Lagrange's equations. We arrive at the well-known equation:

$$\left(\frac{dr}{dt}\right)^2 + r^2\left(\frac{d\varphi}{dt}\right)^2 - \frac{2GM_i}{r} = \mathrm{E}$$

$$r^2\frac{d\varphi}{dt} = \mathrm{h}$$

where $M_i$ is the inertial mass of the Sun. The term $\mathrm{E} = -GM_i/a$, as we know, is called the *energy constant*; $a$ is the semiaxis major of the Kepler-ellipse described by the planet around the Sun.

By replacing $M_i$ into the differential equation above for the expression given by Eq. (46), and expanding in power series, neglecting infinitesimals, we arrive, at:

$$\left(\frac{dr}{dt}\right)^2 + r^2\left(\frac{d\varphi}{dt}\right)^2 - \frac{2GM_g}{r} = \mathrm{E} + \frac{2GM_g}{r}\left(\frac{V^2}{c^2}\right)$$

Since $V = \omega r = r(d\varphi/dt)$, we get

$$\left(\frac{dr}{dt}\right)^2 + r^2\left(\frac{d\varphi}{dt}\right)^2 - \frac{2GM_g}{r} = \mathrm{E} + \frac{2GM_g r}{c^2}\left(\frac{d\varphi}{dt}\right)^2$$

which is the *Einsteinian equation of the planetary motion*.

Multiplying this equation by $(dt/d\varphi)^2$ and remembering that $(dt/d\varphi)^2 = r^4/\mathrm{h}^2$, we obtain

$$\left(\frac{dr}{d\varphi}\right)^2 + r^2 = \mathrm{E}\left(\frac{r^4}{\mathrm{h}^2}\right) + \frac{2GM_g r^3}{\mathrm{h}^2} + \frac{2GM_g r}{c^2}$$

Making $r = 1/u$ and multiplying both members of the equation by $u^4$, we get

$$\left(\frac{du}{d\varphi}\right)^2 + u^2 = \frac{\mathrm{E}}{\mathrm{h}^2} + \frac{2GM_g u}{\mathrm{h}^2} + \frac{2GM_g u^3}{c^2}$$

This leads to the following expression

$$\frac{d^2u}{d\varphi^2} + u = \frac{GM_g}{\mathrm{h}^2}\left(1 + \frac{3u^2\mathrm{h}^2}{c^2}\right)$$

In the absence of term $3\mathrm{h}^2u^2/c^2$, the integration of the equation should be immediate, leading to $2\pi$ period. In order to obtain the value of the perturbation we can use any of the well-known methods, which lead to an angle $\varphi$, for two successive perihelions, given by

$$2\pi + \frac{6G^2M_g^2}{c^2\mathrm{h}^2}$$

Calculating per century, in the case of Mercury, we arrive at an angle of 43" for the perihelion advance.

This result is the best theoretical proof of the accuracy of Eq. (45).

Now consider a relativistic particle inside a gravitational field. The condition for it to escape from the gravitational field is that its *inertial kinetic energy* becomes equal to the absolute value of the *gravitational energy of the field*, which is given

$$U(r) = -\frac{GM_gM'_g}{r} =$$

$$= -\frac{Gm_gM'_g}{r\sqrt{1-V^2/c^2}}$$

Since $\Phi = U(r)/M'_g$ and $g = \partial\Phi/\partial r$ then, we get

$$g = -\frac{Gm_g}{r^2\sqrt{1-V^2/c^2}}$$

where $V$ is the velocity of the mass $m_g$, in respect to the observer. $V$ is also the velocity with which the observer moves away from $m_g$. If the observer is inside the gravitational field produced by $m_g$, then, $V$ is the velocity with which the observer *escapes* from $m_g$ (or the escape velocity from the gravitational field of $m_g$). Since the gravitational field is created by a particle with *non-null* gravitational mass, then obviously, $V < c$. If $V << c$ the escape velocity is given by

$$\tfrac{1}{2}M'_gV^2 = G\frac{m_gM'_g}{r}$$

whence we obtain

$$V^2 = \frac{2Gm_g}{r}$$

By substituting this expression into the equation of $g$, above obtained, the result is

$$g = \frac{\partial\Phi}{\partial r} = \frac{Gm_g}{r^2\sqrt{1-2Gm_g/rc^2}}$$

whence we recognize the *Schwzarzschilds' equation*. Note in this equation the presence of $m_g$, whose value, according to Eq. (41) can be reduced or made *negative*. In this case*, the singularity $g \to \infty$, produced by *Schwzarzschilds' radius* $r = 2Gm_g/c^2$, $(m_g = m_i)$, obviously does not occur. Consequently, *Black Hole does not exist*.

For $V << c$ we get $\sqrt{1-V^2/c^2} \cong 1 + V^2/2c^2$. Since $V^2 = 2Gm_g/r$, then we can write that

$$\sqrt{1-V^2/c^2} \cong 1 + \frac{Gm_g}{rc^2} = 1 + \frac{\phi}{c^2}$$

Substitution of $\sqrt{1-V^2/c^2} = 1 + \phi/c^2$ into the well-known expression below

$$T = t\sqrt{1-V^2/c^2}$$

which expresses the relativistic correlation between *own time* (*T*) and *universal time* (*t*), gives

$$T = t\left(1 + \frac{\phi}{c^2}\right)$$

It is known from the Optics that the frequency of a wave, measured in units of *universal time,* remains constant during its propagation, and that it can be expressed by

$$\omega_0 = \frac{\partial \psi}{\partial t}$$

where $d\psi/dt$ is the derivative of the *eikonal* $\psi$ with respect to the time.

On the other hand, the frequency of the wave measured in units of *own time* is given by

$$\omega = \frac{\partial \psi}{\partial T}$$

Thus, we conclude that

$$\frac{\omega}{\omega_0} = \frac{\partial t}{\partial T}$$

whence we obtain

$$\frac{\omega}{\omega_0} = \frac{t}{T} = \frac{1}{\left(1 + \frac{\phi}{c^2}\right)}$$

By expanding in power series, neglecting infinitesimals, we arrive at:

$$\omega = \omega_0\left(1 - \frac{\phi}{c^2}\right)$$

In this way, if a light ray with a frequency $\omega_0$ is emitted from a point where the gravitational potential is $\phi_1$, it will have a frequency $\omega_1$. Upon reaching a point where the gravitational potential is $\phi_2$ its frequency will be $\omega_2$. Then, according to equation above, it follows that

$$\omega_1 = \omega_0\left(1 - \frac{\phi_1}{c^2}\right) \quad \textit{and} \quad \omega_2 = \omega_0\left(1 - \frac{\phi_2}{c^2}\right)$$

Thus, from point 1 to point 2 the frequency will be shifted in the interval $\Delta\omega = \omega_1 - \omega_2$, given by

$$\Delta\omega = \omega_0\left(\frac{\phi_2 - \phi_1}{c^2}\right)$$

If $\Delta\omega < 0$, $(\phi_1 > \phi_2)$, the shift occurs in the direction of the decreasing frequencies (*red-shift*). If $\Delta\omega > 0$, $(\phi_1 < \phi_2)$ the *blue-shift* occurs.

Let us now consider another consequence of the existence of correlation between $M_g$ and $M_i$.

*Lorentz's force* is usually written in the following form:

---

* This can occur, for example, in a stage of gravitational contraction of a neutron star (mass > 2.4M$_\odot$), when the gravitational masses of the neutrons, in the core of star, are progressively turned *negative*, as a consequence of the increase of the density of magnetic energy inside the neutrons, $W_n = \frac{1}{2}\mu_0 H_n^2$, reciprocally produced by the *spin* magnetic fields of the own neutrons, $\vec{H}_n = \frac{1}{2}\mu_0\left(\vec{M}_n/2\pi r_n^3\right) = \gamma_n\left(e\vec{S}_n/4\pi m_n r_n^3\right)$ due to the decrease of the neutrons radii, $r_n$, along the very strong compression at which they are subjected. Since $W_n \propto r_n^{-6}$, and $\rho_n \propto r_n^{-3}$, then $W_n$ increases much more rapidly – with the decrease of $r_n$ – than $\rho_n$. Consequently, the ratio $W_n/\rho_n$ increases progressively with the compression of the neutrons star. According to Eq. (41), the gravitational masses of the neutrons can be turned *negative* at given stage of the compression. Thus, due to the difference of pressure, the value of $W_n/\rho_n$ in the crust is smaller than the value in the core. This means that, the gravitational mass of the core becomes negative *before* of the gravitational mass of the crust. This makes the gravitational contraction culminates with an explosion, due to the *repulsive* gravitational forces between the core and the crust. Therefore, the contraction has a limit and, consequently, the singularity does not occur.

$$d\,\vec{p}/dt = q\vec{E} + q\vec{V}\times\vec{B}$$

where $\vec{p} = m_{i0}\vec{V}\big/\sqrt{1-V^2/c^2}$. However, Eq.(4) tells us that $\vec{p} = m_g V\big/\sqrt{1-V^2/c^2}$. Therefore, the expressions above must be corrected by multiplying its members by $m_g/m_{i0}$,i.e.,

$$\vec{p}\frac{m_g}{m_{i0}} = \frac{m_g}{m_{i0}}\frac{m_{i0}\vec{V}}{\sqrt{1-V^2/c^2}} = \frac{m_g\vec{V}}{\sqrt{1-V^2/c^2}} = \vec{p}$$

and

$$\frac{d\vec{p}}{dt} = \frac{d}{dt}\left(\vec{\mathrm{p}}\frac{m_g}{m_{i0}}\right) = \left(q\vec{E} + q\vec{V}\times\vec{B}\right)\frac{m_g}{m_{i0}}$$

That is now the *general expression* for Lorentz's force. Note that it depends on $m_g$.

When the force is perpendicular to the speed, Eq. (5) gives $d\vec{p}/dt = m_g\left(d\vec{V}/dt\right)\big/\sqrt{1-V^2/c^2}$.By comparing with Eq.(46), we thus obtain

$$\left(m_{i0}\big/\sqrt{1-V^2/c^2}\right)\left(d\vec{V}/dt\right) = q\vec{E} + q\vec{V}\times\vec{B}$$

Note that this equation is the expression of an *inertial* force.

Starting from this equation, well-known experiments have been carried out in order to verify the relativistic expression: $m_i\big/\sqrt{1-V^2/c^2}$. In general, the *momentum* variation $\Delta p$ is expressed by $\Delta p = F\Delta t$ where $F$ is the applied force during a time interval $\Delta t$. Note that there is no restriction concerning the *nature* of the force $F$, i.e., it can be mechanical, electromagnetic, etc.

For example, we can look on the *momentum* variation $\Delta p$ as due to absorption or emission of *electromagnetic energy* by the particle (by means of *radiation* and/or by means of *Lorentz's force* upon the *charge* of the particle).

In the case of radiation (any type), $\Delta p$ can be obtained as follows. It is known that the *radiation pressure* , $dP$, upon an area $dA = dxdy$ of a volume $dV = dxdydz$ of a particle( the incident radiation normal to the surface $dA$ )is equal to the energy $dU$ absorbed per unit volume $(dU/dV)$.i.e.,

$$dP = \frac{dU}{dV} = \frac{dU}{dxdydz} = \frac{dU}{dAdz} \qquad (47)$$

Substitution of $dz = vdt$ ( $v$ is the speed of radiation) into the equation above gives

$$dP = \frac{dU}{dV} = \frac{(dU/dAdt)}{v} = \frac{dD}{v} \qquad (48)$$

Since $dPdA = dF$ we can write:

$$dFdt = \frac{dU}{v} \qquad (49)$$

However we know that $dF = dp/dt$, then

$$dp = \frac{dU}{v} \qquad (50)$$

From Eq. (48), it follows that

$$dU = dPdV = \frac{dVdD}{v} \qquad (51)$$

Substitution into (50) yields

$$dp = \frac{dVdD}{v^2} \qquad (52)$$

or

$$\int_0^{\Delta p} dp = \frac{1}{v^2}\int_0^D\int_0^V dVdD$$

whence

$$\Delta p = \frac{VD}{v^2} \qquad (53)$$

This expression is general for all types of waves including *non-electromagnetic waves* such as *sound waves.* In this case, $v$ in Eq.(53), will be the speed of sound in the medium and $D$ the *intensity* of the sound radiation.

In the case of *electromagnetic waves*, the Electrodynamics tells us that $v$ will be given by

$$v = \frac{dz}{dt} = \frac{\omega}{\kappa_r} = \frac{c}{\sqrt{\frac{\varepsilon_r\mu_r}{2}\left(\sqrt{1+(\sigma/\omega\varepsilon)^2}+1\right)}}$$

where $k_r$ is the real part of the *propagation vector* $\vec{k}$ ; $k = \left|\vec{k}\right| = k_r + ik_i$ ; $\varepsilon$ , $\mu$ and $\sigma$, are the electromagnetic characteristics of the medium in which the incident (or emitted) radiation is propagating ( $\varepsilon = \varepsilon_r\varepsilon_0$ where $\varepsilon_r$ is the *relative dielectric permittivity* and $\varepsilon_0 = 8.854\times10^{-12}F/m$ ; $\mu = \mu_r\mu_0$ where $\mu_r$ is the *relative magnetic permeability* and $\mu_0 = 4\pi \times 10^{-7} H/m$; $\sigma$ is the *electrical conductivity*). For an *atom* inside a body, the incident (or emitted) radiation on this atom will be propagating inside the body, and consequently, $\sigma = \sigma_{body}$, $\varepsilon = \varepsilon_{body}$, $\mu = \mu_{body}$.

It is then evident that the *index of refraction* $n_r = c/v$ will be given by

$$n_r = \frac{c}{v} = \sqrt{\frac{\varepsilon_r \mu_r}{2}\left(\sqrt{1+(\sigma/\omega\varepsilon)^2}+1\right)} \qquad (54)$$

On the other hand, from Eq. (50) follows that

$$\Delta p = \frac{U}{v}\left(\frac{c}{c}\right) = \frac{U}{c} n_r$$

Substitution into Eq. (41) yields

$$m_g = \left\{1 - 2\left[\sqrt{1+\left(\frac{U}{m_{i0}c^2} n_r\right)^2} - 1\right]\right\} m_{i0} \qquad (55)$$

If the body is *also* rotating, with an angular speed $\omega$ around its central axis, then it acquires an additional energy equal to its rotational energy $\left(E_k = \frac{1}{2} I\omega^2\right)$. Since this is an increase in the internal energy of the body, and this energy is basically electromagnetic, we can assume that $E_k$, such as $U$, corresponds to an amount of electromagnetic energy absorbed by the body. Thus, we can consider $E_k$ as an increase $\Delta U = E_k$ in the electromagnetic energy $U$ absorbed by the body. Consequently, in this case, we must replace $U$ in Eq. (55) for $(U + \Delta U)$. If $U << \Delta U$, the Eq. (55) reduces to

$$m_g \cong \left\{1 - 2\left[\sqrt{1+\left(\frac{I\omega^2 n_r}{2m_{i0}c^2}\right)^2} - 1\right]\right\} m_{i0}$$

For $\sigma << \omega\varepsilon$, Eq. (54) shows that $n_r = c/v = \sqrt{\varepsilon_r \mu_r}$ and $n_r = \sqrt{\mu\sigma c^2/4\pi f}$ in the case of $\sigma >> \omega\varepsilon$. In this case, if the body is a *Mumetal* disk $\left(\mu_r = 105{,}000\, at\, 100 gauss; \sigma = 2.1\times 10^7 S.m^{-1}\right)$ with radius $R$, $\left(I = \frac{1}{2} m_{i0} R^2\right)$, the equation above shows that the *gravitational mass* of the disk is

$$m_{g(disk)} \cong \left\{1 - 2\left[\sqrt{1 + 1.12\times 10^{-13}\frac{R^4\omega^4}{f}} - 1\right]\right\} m_{i0(disk)}$$

Note that the effect of the electromagnetic radiation applied upon the disk is highly relevant, because in the absence of this radiation the index of refraction, present in equations above, becomes equal to 1. Under these circumstances, the possibility of strongly reducing the gravitational mass of the disk practically disappears. In addition, the equation above shows that, in practice, the frequency $f$ of the radiation cannot be high, and that *extremely-low frequencies* (ELF) are most appropriated. Thus, if the frequency of the electromagnetic radiation applied upon the disk is $f = 0.1 Hz$ (See Fig. I (a)) and the radius of the disk is $R = 0.15 m$, and its angular speed $\omega = 1.05\times 10^4 rad/s\,(\sim 100{,}000\ rpm)$, the result is

$$m_{g(disk)} \cong -2.6 m_{i0(disk)}$$

This shows that the gravitational mass of a body can also be controlled by means of its *angular velocity*.

In order to satisfy the condition $U << \Delta U$, we must have $dU/dt << d\Delta U/dt$, where $P_r = dU/dt$ is the radiation power. By integrating this expression, we get $\overline{U} = P_r/2f$. Thus we can conclude that, for $U << \Delta U$, we must have $P_r/2f << \frac{1}{2} I\omega^2$, i.e.,

$$P_r << I\omega^2 f$$

By dividing both members of the expression above by the area $S = 4\pi r^2$, we obtain

$$D_r << \frac{I\omega^2 f}{4\pi r^2}$$

Therefore, this is the necessary condition in order to obtain $U << \Delta U$. In the case of the Mumetal disk, we must have

$$D_r << 10^5/r^2 \qquad \left(watts/m^2\right)$$

From Electrodynamics, we know that a radiation with frequency $f$ propagating within a material with electromagnetic characteristics $\varepsilon$, $\mu$ and $\sigma$ has the amplitudes of its waves attenuated by $e^{-1}$=0.37 (37%) when it penetrates a distance $z$, given by †

$$z=\frac{1}{\omega\sqrt{\frac{1}{2}\varepsilon\mu\left(\sqrt{1+(\sigma/\omega\varepsilon)^2}-1\right)}}$$

For $\sigma >> \omega\varepsilon$, equation above reduces to

$$z=\frac{1}{\sqrt{\pi\mu f\sigma}}$$

In the case of the *Mumetal* subjected to an ELF radiation with frequency $f=0.1Hz$, the value is $z=1.07mm$. Obviously, the thickness of the Mumetal disk must be less than this value.

Equation (55) is general for all types of electromagnetic fields including *gravitoelectromagnetic* fields (See Fig. I (b)).

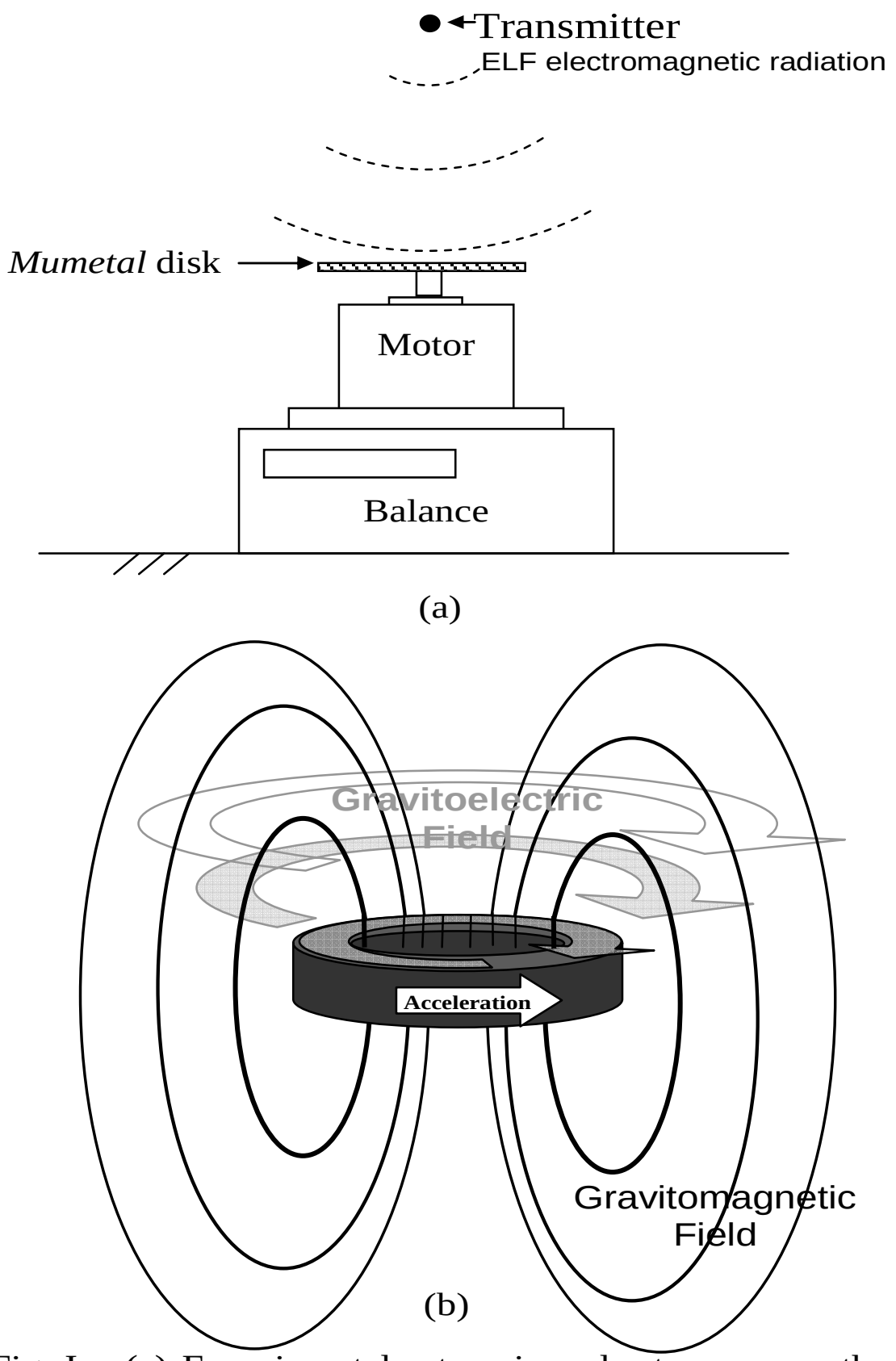

Fig. I – (a) Experimental set-up in order to measure the gravitational mass decreasing in the rotating Mumetal disk. A sample connected to a dynamometer can measure the decreasing of gravity above the disk. (b) Gravitoelectromagnetic Field.

The Maxwell-like equations for weak gravitational fields are [9]

$$\nabla . \boldsymbol{D_G} = -\rho$$

$$\nabla \times \boldsymbol{E_G} = -\frac{\partial \boldsymbol{B_G}}{\partial \boldsymbol{t}}$$

$$\nabla . \boldsymbol{B_G} = 0$$

$$\nabla \times \boldsymbol{H_G} = -\boldsymbol{j_G} + \frac{\partial \boldsymbol{D_G}}{\partial \boldsymbol{t}}$$

where $D_G = 4\varepsilon_{rG}\varepsilon_{0G}E_G$ is the *gravitodisplacement* field ( $\varepsilon_{rG}$ is the *gravitoelectric relative permittivity* of the medium; $\varepsilon_{0G}$ is the *gravitoelectric permittivity* for free space and $\boldsymbol{E_G} = \boldsymbol{g}$ is the *gravitoelectric* field intensity); $\rho$ is the density of local rest mass in the local rest frame of the matter; $B_G = \mu_{rG}\mu_{0G}H_G$ is the *gravitomagnetic* field ($\mu_{rG}$ is the *gravitomagnetic relative permeability*, $\mu_{0G}$ is the *gravitomagnetic permeability* for free space and $\boldsymbol{H_G}$ is the *gravitomagnetic* field intensity; $\boldsymbol{j_G} = -\boldsymbol{\sigma_G E_G}$ is the local rest-mass current density in this frame ($\sigma_G$ is the *gravitoelectric conductivity* of the medium).

Then, for *free space* we can write that

$$D_G = 4\varepsilon_{0G}E_G = 4\varepsilon_{0G}\,g = 4\varepsilon_{0G}\left(\frac{GM}{r^2}\right)$$

But from the electrodynamics we know that

$$D = \varepsilon E = \frac{q}{4\pi r^2}$$

By analogy we can write that

$$D_G = \frac{M_g}{4\pi r^2}$$

By comparing this expression with the previous expression of $D_G$ , we get

$$\varepsilon_{0G} = \frac{1}{16\pi G} = 2.98\times10^{8}\,kg^2.N^{-1}.m^{-2}$$

which is the expression of the *gravitoelectric permittivity* for free space.

The *gravitomagnetic permeability* for free space [10,11] is

$$\mu_{0G} = \frac{16\pi G}{c^2} = 3.73\times10^{-26}\,m/kg$$

We then convert Maxwell-like equations

---

† Quevedo, C. P. (1978) *Eletromagnetismo* McGraw-Hill, p.269-270.

for weak gravity into a wave equation for free space in the standard way. We conclude that *the speed* of *Gravitational Waves* in free space is

$$v=\frac{1}{\sqrt{\varepsilon_{0G}\mu_{0G}}}=c$$

This means that *both electromagnetic and gravitational plane waves propagate at the free space with the same speed*.

Thus, the impedance for free space is

$$Z_G=\frac{E_G}{H_G}=\sqrt{\mu_{0G}/\varepsilon_{0G}}=\mu_{0G}c=\frac{16\pi G}{c}$$

and the Poynting-like vector is

$$\vec{S}=\vec{E}_G\times\vec{H}_G$$

For a plane wave propagating in the vacuum, we have $|E_G|=Z_G|H_G|$. Then, it follows that

$$\left|\vec{S}\right|=\frac{1}{2Z_G}\left|\vec{E}_G\right|^2=\frac{\omega^2}{2Z_G}\left|\vec{h}\right|^2=\frac{c^2\omega^2}{32\pi G}\left|h_{0i}\right|^2$$

which is *the power per unit area* of a harmonic plane wave of angular frequency $\omega$.

In classical electrodynamics the density of energy in an *electromagnetic* field, $W_e$, has the following expression

$$W_e=\tfrac{1}{2}\varepsilon_r\varepsilon_0E^2+\tfrac{1}{2}\mu_r\mu_0H^2$$

In analogy with this expression we define the energy density in a *gravitoelectromagnetic* field, $W_G$, as follows

$$W_G=\tfrac{1}{2}\varepsilon_{rG}\varepsilon_{0G}E_G^2+\tfrac{1}{2}\mu_{rG}\mu_{0G}H_G^2$$

For free space we obtain

$$\mu_{rG}=\varepsilon_{rG}=1$$

$$\varepsilon_{0G}=1/\mu_{0G}\,c^2$$

$$E_G/H_G=\mu_{0G}c$$

and

$$\boldsymbol{B}_G=\mu_{0G}\,\boldsymbol{H}_G$$

Thus, we can rewrite the equation of $W_G$ as follows

$$W_G=\tfrac{1}{2}\left(\frac{1}{\mu_{0G}c^2}\right)c^2B_G^2+\tfrac{1}{2}\mu_{0G}\left(\frac{B_G}{\mu_{0G}}\right)^2=\frac{B_G^2}{\mu_{0G}}$$

Since $U_G=W_GV$, ($V$ is the *volume* of the particle) and $n_r=1$ for free space we can write (55) in the following form

$$m_g=\left\{1-2\left[\sqrt{1+\left(\frac{W_G}{\rho\,c^2}\right)^2}-1\right]\right\}m_{i0}$$

$$=\left\{1-2\left[\sqrt{1+\left(\frac{B_G^2}{\mu_{0G}\rho\,c^2}\right)^2}-1\right]\right\}m_{i0}\quad(55a)$$

where $\rho=m_{i0}/\mathrm{V}$.

This equation shows how the gravitational mass of a particle is altered by a *gravitomagnetic* field.

A gravitomagnetic field, according to Einstein's theory of general relativity, arises from moving matter (matter current) just as an ordinary magnetic field arises from moving charges. The Earth rotation is the source of a very weak gravitomagnetic field given by

$$B_{G,Earth}=-\frac{\mu_{0G}}{16\pi}\left(\frac{M\omega}{r}\right)_{Earth}\approx10^{-14}\,rad.s^{-1}$$

Perhaps ultra-fast rotating stars can generate very strong gravitomagnetic fields, which can make the gravitational mass of particles inside and near the star *negative*. According to (55a) this will occur if $B_G>1.06c\sqrt{\mu_{0G}\rho}$. Usually, however, gravitomagnetic fields produced by *normal* matter are very weak.

Recently Tajmar, M. et al., [12] have proposed that in addition to the *London moment*, $B_L$, ($B_L=-\left(2m^*/e^*\right)\omega\cong1.1\times10^{-11}\omega$ ; $m^*$ and $e^*$ are the Cooper-pair mass and charge respectively), a rotating superconductor should exhibit also a large *gravitomagnetic* field, $B_G$, to explain an apparent mass increase of Niobium Cooper-pairs discovered by Tate et al[13,14]. According to Tajmar and Matos [15], in the case of *coherent* matter, $B_G$ is given by: $B_G=-2\omega\rho_c\mu_{0G}\lambda_{gr}^2$ where $\rho_c$ is the mass density of *coherent* matter and $\lambda_{gr}$ is the *graviphoton* wavelength. By choosing $\lambda_{gr}$ proportional to the local density of *coherent* matter, $\rho_c$. i.e.,

$$\frac{1}{\lambda_{gr}^2}=\left(\frac{m_{gr}c}{\hbar}\right)=\mu_{0G}\rho_c$$

we obtain

$$B_G=-2\omega\rho_c\mu_{0G}\lambda_{gr}^2=-2\omega\rho_c\mu_{0G}\left(\frac{1}{\mu_{0G}\rho_c}\right)=$$
$$=-2\omega$$

and the graviphoton mass, $m_{gr}$ , is

$$m_{gr}=\mu_{0G}\rho_c\hbar/c$$

Note that if we take the case of *no* local sources of *coherent* matter $(\rho_c=0)$, the graviphoton mass will be *zero*. However, graviphoton will have non-zero mass inside coherent matter $(\rho_c\neq 0)$. This can be interpreted as a consequence of the graviphoton gaining mass inside the superconductor via the Higgs mechanism due to the breaking of gauge symmetry.

It is important to note that the *minus* sign in the expression for $B_G$ can be understood as due to the change from the normal to the coherent state of matter, i.e., a switch between real and *imaginary* values for the particles inside the material when going from the normal to the coherent state of matter. Consequently, in this case the variable $U$ in (55) must be replaced by $iU_G$ and not by $U_G$ only. Thus we obtain

$$m_g=\left\{1-2\left[\sqrt{1-\left(\frac{U_G}{m_{i0}c^2}n_r\right)^2}-1\right]\right\}m_{i0} \quad (55b)$$

Since $U_G=W_GV$, we can write (55b) for $n_r=1$, in the following form

$$m_g=\left\{1-2\left[\sqrt{1-\left(\frac{W_G}{\rho_c c^2}\right)^2}-1\right]\right\}m_{i0}$$
$$=\left\{1-2\left[\sqrt{1-\left(\frac{B_G^2}{\mu_{0G}\rho_c c^2}\right)^2}-1\right]\right\}m_{i0} \quad (55c)$$

where $\rho_c=m_{i0}/\mathrm{V}$ is the local density of *coherent* matter.

Note the different sign (inside the square root) with respect to (55a).

By means of (55c) it is possible to check the changes in the gravitational mass of the *coherent part* of a given material (e.g. the Cooper-pair fluid). Thus for the *electrons* of the Cooper-pairs we have

$$m_{ge}=m_{ie}+2\left[1-\sqrt{1-\left(\frac{B_G^2}{\mu_{0G}\rho_e c^2}\right)^2}\right]m_{ie}=$$
$$=m_{ie}+2\left[1-\sqrt{1-\left(\frac{4\omega^2}{\mu_{0G}\rho_e c^2}\right)^2}\right]m_{ie}=$$
$$=m_{ie}+\chi_e m_{ie}$$

where $\rho_e$ is the mass density of the electrons.

In order to check the changes in the gravitational mass of *neutrons* and *protons* (non-coherent part) inside the superconductor, we must use Eq. (55a) and $B_G=-2\omega\rho\mu_{0G}\lambda_{gr}^2$ [Tajmar and Matos, op.cit.]. Due to $\mu_{0G}\rho_c\lambda_{gr}^2=1$, that expression of $B_G$ can be rewritten in the following form

$$B_G=-2\omega\rho\mu_{0G}\lambda_{gr}^2=-2\omega(\rho/\rho_c)$$

Thus we have

$$m_{gn}=m_{in}-2\left[\sqrt{1+\left(\frac{B_G^2}{\mu_{0G}\rho_n c^2}\right)^2}-1\right]m_{in}=$$
$$=m_{in}-2\left[\sqrt{1+\left(\frac{4\omega^2(\rho_n/\rho_c)^2}{\mu_{0G}\rho_n c^2}\right)^2}-1\right]m_{in}=$$
$$=m_{in}-\chi_n m_{in}$$

$$m_{gp}=m_{ip}-2\left[\sqrt{1+\left(\frac{B_G^2}{\mu_{0G}\rho_p c^2}\right)^2}-1\right]m_{ip}=$$
$$=m_{ip}-2\left[\sqrt{1+\left(\frac{4\omega^2(\rho_p/\rho_c)^2}{\mu_{0G}\rho_p c^2}\right)^2}-1\right]m_{ip}=$$
$$=m_{ip}-\chi_p m_{ip}$$

where $\rho_n$ and $\rho_p$ are the mass density of *neutrons* and *protons* respectively.

In Tajmar's experiment, induced accelerations fields outside the superconductor in the order of $100\mu g$, at angular velocities of about $500 rad.s^{-1}$ were observed.

Starting from $g = Gm_{g(initial)}/r$ we can write that $g + \Delta g = G(m_{g(initial)} + \Delta m_g)/r$. Then we get $\Delta g = G\Delta m_g/r$. For $\Delta g = \eta g = \eta Gm_{g(initial)}/r$ it follows that $\Delta m_g = \eta m_{g(initial)} = \eta m_i$. Therefore a variation of $\Delta g = \eta g$ corresponds to a gravitational mass variation $\Delta m_g = \eta m_{i0}$. Thus $\Delta g \approx 100\mu g = 1\times10^{-4} g$ corresponds to

$$\Delta m_g \approx 1\times10^{-4} m_{i0}$$

On the other hand, the total gravitational mass of a particle can be expressed by

$$m_g = N_n m_{gn} + N_p m_{gp} + N_e m_{ge} + N_p \Delta E/c^2 =$$
$$N_n(m_{in} - \chi_n m_{in}) + N_p(m_{ip} - \chi_p m_{ip}) +$$
$$+ N_e(m_{ie} - \chi_e m_{ie}) + N_p \Delta E/c^2 =$$
$$= (N_n m_{in} + N_p m_{ip} + N_e m_{ie}) + N_p \Delta E/c^2 -$$
$$-(N_n \chi_n m_{in} + N_p \chi_p m_{ip} + N_e \chi_e m_{ie}) + N_p \Delta E/c^2 =$$
$$= m_i - (N_n \chi_n m_{in} + N_p \chi_p m_{ip} + N_e \chi_e m_{ie}) + N_p \Delta E/c^2$$

where $\Delta E$ is the interaction energy; $N_n$, $N_p$, $N_e$ are the number of neutrons, protons and electrons respectively. Since $m_{in} \cong m_{ip}$ and $\rho_n \cong \rho_p$ it follows that $\chi_n \cong \chi_p$ and consequently the expression of $m_g$ reduces to

$$m_g \cong m_{i0} - (2N_p \chi_p m_{ip} + N_e \chi_e m_{ie}) + N_p \Delta E/c^2 \quad (55d)$$

Assuming that $N_e \chi_e m_{ie} << 2N_p \chi_p m_{ip}$ and $N_p \Delta E/c^2 << 2N_p \chi_p m_{ip}$ Eq. (55d) reduces to

$$m_g \cong m_{i0} - 2N_p \chi_p m_{ip} = m_i - \chi_p m_i \quad (55e)$$

or

$$\Delta m_g = m_g - m_{i0} = -\chi_p m_{i0}$$

By comparing this expression with $\Delta m_g \approx 1\times10^{-4} m_i$ which has been obtained from Tajmar's experiment, we conclude that at angular velocities $\omega \approx 500 rad.s^{-1}$ we have

$$\chi_p \approx 1\times10^{-4}$$

From the expression of $m_{gp}$ we get

$$\chi_p = 2\left[\sqrt{1 + \left(\frac{B_G^2}{\mu_{0G}\rho_p c^2}\right)^2} - 1\right] =$$
$$= 2\left[\sqrt{1 + \left(\frac{4\omega^2(\rho_p/\rho_c)^2}{\mu_{0G}\rho_p c^2}\right)^2} - 1\right]$$

where $\rho_p = m_p/V_p$ is the mass density of the protons.

In order to calculate $V_p$ we need to know the type of space (metric) inside the proton. It is known that there are just 3 types of space: the space of *positive* curvature, the space of *negative* curvature and the space of *null* curvature. The negative type is obviously excluded since the volume of the proton is *finite*. On the other hand, the space of null curvature is also excluded since the space inside the proton is strongly curved by its enormous mass density. Thus we can conclude that inside the proton the space has *positive* curvature. Consequently, the volume of the proton, $V_p$, will be expressed by the 3-dimensional space that corresponds to a *hypersphere* in a 4-dimentional space, i.e., $V_p$ will be the space of positive curvature the volume of which is [16]

$$V_p = \int_0^{2\pi}\int_0^{\pi}\int_0^{\pi} r_p^3 \sin^2\chi \sin\theta d\chi d\theta d\phi = 2\pi^2 r_p^3$$

In the case of Earth, for example, $\rho_{Earth} << \rho_p$. Consequently the curvature of the space inside the Earth is approximately *null* (space approximately *flat)*. Then $V_{Earth} \cong \frac{4}{3}\pi r_{Earth}^3$.

For $r_p = 1.4\times10^{-15} m$ we then get

$$\rho_p = \frac{m_p}{V_p} \cong 3\times10^{16}\,kg/m^3$$

Starting from the London moment it is easy to see that by precisely measuring the magnetic field and the angular velocity of the superconductor, one can calculate the mass of the Cooper-pairs. This has been done for both classical and high-Tc superconductors [17-20]. In the experiment with the highest precision to date, Tate et al, op.cit., reported a disagreement between the theoretically predicted Cooper-pair mass in Niobium of $m^*/2m_e = 0.999992$ and its experimental value of $1.000084(21)$, where $m_e$ is the electron mass. This anomaly was actively discussed in the literature without any apparent solution [21-24].

If we consider that the apparent mass increase from Tate's measurements results from an *increase* in the gravitational mass $m_g^*$ of the Cooper-pairs due to $B_G$, then we can write

$$\frac{m_g^*}{2m_e} = \frac{m_g^*}{m_i^*} = 1.000084$$

$$\Delta m_g^* = m_g^* - m_{g(initial)}^* = m_g^* - m_i^* =$$

$$= 1.000084\,m_i^* - m_i^* =$$

$$= +0.84\times10^{-4}\,m_i^* = \chi^* m_i^*$$

where $\chi^* = 0.84\times10^{-4}$.

From (55c) we can write that

$$m_g^* = m_i^* + 2\left[1-\sqrt{1-\left(\frac{4\omega^2}{\mu_{0G}\rho^* c^2}\right)^2}\,\right]m_i^* =$$

$$= m_i^* + \chi^* m_i^*$$

where $\rho^*$ is the Cooper-pair mass density.

Consequently we can write

$$\chi^* = 2\left[1-\sqrt{1-\left(\frac{4\omega^2}{\mu_{0G}\rho^* c^2}\right)^2}\,\right] = 0.84\times10^{-4}$$

From this equation we then obtain

$$\rho^* \cong 3\times10^{16}\,kg/m^3$$

Note that $\rho_p \cong \rho^*$.

Now we can calculate the graviphoton mass, $m_{gr}$, inside the Cooper-pairs fluid (coherent part of the superconductor) as

$$m_{gr} = \mu_{0G}\rho^*\hbar/c \cong 4\times10^{-52}\,kg$$

Outside the coherent matter $(\rho_c = 0)$ the graviphoton mass will be *zero* $(m_{gr} = \mu_{0G}\rho_c\hbar/c = 0)$.

Substitution of $\rho_p, \rho_c = \rho^*$ and $\omega \approx 500\,rad.s^{-1}$ into the expression of $\chi_p$ gives

$$\chi_p \approx 1\times10^{-4}$$

Compare this value with that one obtained from the Tajmar experiment.

Therefore, the decrease in the gravitational mass of the superconductor, expressed by (55e), is

$$m_{g,SC} \cong m_{i,SC} - \chi_p m_{i,SC}$$

$$\cong m_{i,SC} - 10^{-4} m_{i,SC}$$

This corresponds to a decrease of the order of $10^{-2}\%$ in respect to the initial gravitational mass of the superconductor. However, we must also consider the *gravitational shielding effect,* produced by this decrease of $\approx 10^{-2}\%$ in the gravitational mass of the particles inside the superconductor (see Fig. II). Therefore, the *total* weight decrease in the superconductor will be much greater than $10^{-2}\%$. According to Podkletnov experiment [25] it can reach up to 1% of the total weight of the superconductor at $523.6\,rad.s^{-1}$ $(5000rpm)$. In this experiment a slight decrease (up to $\approx 1\%$) in the weight of samples hung above the disk (rotating at 5000rpm) was observed. A smaller effect on the order of $0.1\%$ has been observed when the disk is not rotating. The percentage of weight decrease is the same for samples of different masses and chemical compounds. The effect does not seem to diminish with increases in elevation above the disk. There appears to be a "shielding cylinder" over the disk that extends upwards for at least 3 meters. No weight reduction has been observed under the disk.

It is easy to see that the decrease in the weight of samples hung above the disk (inside the "shielding cylinder" over the disk) in the Podkletnov experiment, is also a consequence of the *Gravitational Shielding Effect* showed in Fig. II.

In order to explain the *Gravitational Shielding Effect*, we start with the gravitational field, $\vec{g} = -\frac{GM_g}{R^2}\hat{\mu}$, produced by a particle with gravitational mass, $M_g$. The gravitational flux, $\phi_g$, through a spherical surface, with area $S$ and radius $R$, concentric with the mass $M_g$, is given by

$$\phi_g = \oint_S \vec{g}d\vec{S} = g\oint_S dS = gS =$$

$$= \frac{GM_g}{R^2}\left(4\pi R^2\right) = 4\pi GM_g$$

Note that the flux $\phi_g$ does not depend on the radius $R$ of the surface $S$, i.e., it is the *same* through any surface concentric with the mass $M_g$.

Now consider a particle with gravitational mass, $m'_g$, placed into the gravitational field produced by $M_g$. According to Eq. (41), we can have $m'_g/m'_{i0} = -1$, $m'_g/m'_{i0} \cong 0$ [‡], $m'_g/m'_{i0} = 1$, etc. In the first case, the gravity acceleration, $g'$, upon the particle $m'_g$, is $\vec{g}' = -g = +\frac{GM_g}{R^2}\hat{\mu}$. This means that in this case, the gravitational flux, $\phi'_g$, through the particle $m'_g$ will be given by $\phi'_g = g'S = -gS = -\phi_g$, i.e., it will be *symmetric* in respect to the flux when $m'_g = m'_{i0}$ (third case). In the second case $\left(m'_g \cong 0\right)$, the intensity of the gravitational force between $m'_g$ and $M_g$ will be very close to zero. This is equivalent to say that the gravity acceleration upon the particle with mass $m'_g$ will be $g' \cong 0$. Consequently we can write that $\phi'_g = g'S \cong 0$. It is easy to see that there is a correlation between $m'_g/m'_{i0}$ and $\phi'_g/\phi_g$, i.e.,

_ If $m'_g/m'_{i0} = -1 \quad\Rightarrow\quad \phi'_g/\phi_g = -1$

_ If $m'_g/m'_{i0} = 1 \quad\Rightarrow\quad \phi'_g/\phi_g = 1$

_ If $m'_g/m'_{i0} \cong 0 \quad\Rightarrow\quad \phi'_g/\phi_g \cong 0$

Just a simple algebraic form contains the requisites mentioned above, the correlation

$$\frac{\phi'_g}{\phi_g} = \frac{m'_g}{m'_{i0}}$$

By making $m'_g/m'_{i0} = \chi$ we get

$$\phi'_g = \chi\ \phi_g$$

This is the expression of the gravitational flux through $m'_g$. It explains the *Gravitational Shielding Effect* presented in Fig. II.

As $\phi_g = gS$ and $\phi'_g = g'S$, we obtain

$$g' = \chi\ g$$

This is the gravity acceleration inside $m'_g$.

Figure II (b) shows the gravitational shielding effect produced by two particles at the same direction. In this case, the gravity acceleration inside and above the second particle will be $\chi^2 g$ if $m_{g2} = m_{i1}$.

‡ The quantization of the gravitational mass (Eq.(33)) shows that for $n = 1$ the gravitational mass is not *zero* but equal to $m_{g(min)}$. Although the gravitational mass of a particle is never null, Eq.(41) shows that it can be turned very close to zero.

These particles are representative of any material particles or material *substance* (solid, liquid, gas, plasma, electrons flux, etc.), whose gravitational mass have been reduced by the factor $\chi$. Thus, *above* the substance, the gravity acceleration $g'$ is reduced at the same proportion $\chi = m_g / m_{i0}$, and, consequently, $g' = \chi\, g$, where $g$ is the gravity acceleration *below* the substance.

Figure III shows an experimental set-up in order to check the factor $\chi$ above *a high-speed electrons flux*. As we have shown (Eq. 43), the *gravitational mass* of a particle decreases with the increase of the velocity $V$ of the particle.

Since the theory says that the factor $\chi$ is given by the correlation $m_g / m_{i0}$ then, in the case of an electrons flux, we will have that $\chi = m_{ge} / m_{ie}$ where $m_{ge}$ as function of the velocity $V$ is given by Eq. (43). Thus, we can write that

$$\chi = \frac{m_{ge}}{m_{ie}} = \left\{1 - 2\left[\frac{1}{\sqrt{1 - V^2/c^2}} - 1\right]\right\}$$

Therefore, if we know the velocity $V$ of the electrons we can calculate $\chi$. ($m_{ie}$ is the electron mass at rest).

When an electron penetrates the electric field $E_y$ (see Fig. III) an electric force, $\vec{F}_E = -e\vec{E}_y$, will act upon the electron. The direction of $\vec{F}_E$ will be contrary to the direction of $\vec{E}_y$. The magnetic force $\vec{F}_B$ which acts upon the electron, due to the magnetic field $\vec{B}$, is $\vec{F}_B = eVB\hat{\mu}$ and will be opposite to $\vec{F}_E$ because the electron charge is negative.

By adjusting conveniently $B$ we can make $\left|\vec{F}_B\right| = \left|\vec{F}_E\right|$. Under these circumstances in which the total force is zero, the spot produced by the electrons flux on the surface $\alpha$ returns from $O'$ to $O$ and is detected by the galvanometer $G$. That is, there is no deflection for the cathodic rays. Then it follows that $eVB = eE_y$ since $\left|\vec{F}_B\right| = \left|\vec{F}_E\right|$. Then, we get

$$V = \frac{E_y}{B}$$

This gives a measure of the velocity of the electrons.

Thus, by means of the experimental set-up, shown in Fig. III, we can easily obtain the velocity $V$ of the electrons below the body $\beta$, in order to calculate the *theoretical* value of $\chi$. The *experimental* value of $\chi$ can be obtained by dividing the weight, $P'_\beta = m_{g\beta} g'$ of the body $\beta$ for a voltage drop $\widetilde{V}$ across the anode and cathode, by its weight, $P_\beta = m_{g\beta} g$, when the voltage $\widetilde{V}$ is *zero*, i.e.,

$$\chi = \frac{P'_\beta}{P_\beta} = \frac{g'}{g}$$

According to Eq. (4), the gravitational mass, $M_g$, is defined by

$$M_g = \left|\frac{m_g}{\sqrt{1 - V^2/c^2}}\right|$$

While Eq. (43) defines $m_g$ by means of the following expression

$$m_g = \left\{1 - 2\left[\frac{1}{\sqrt{1 - V^2/c^2}} - 1\right]\right\} m_{i0}$$

In order to check the gravitational mass of the electrons it is necessary to know the pressure $P$ produced by the electrons flux. Thus, we have put a piezoelectric sensor in the bottom of the glass tube as shown in Fig. III. The electrons flux radiated from the cathode is accelerated by the anode1 and strikes on the piezoelectric sensor yielding a pressure $P$ which is measured by means of the sensor.

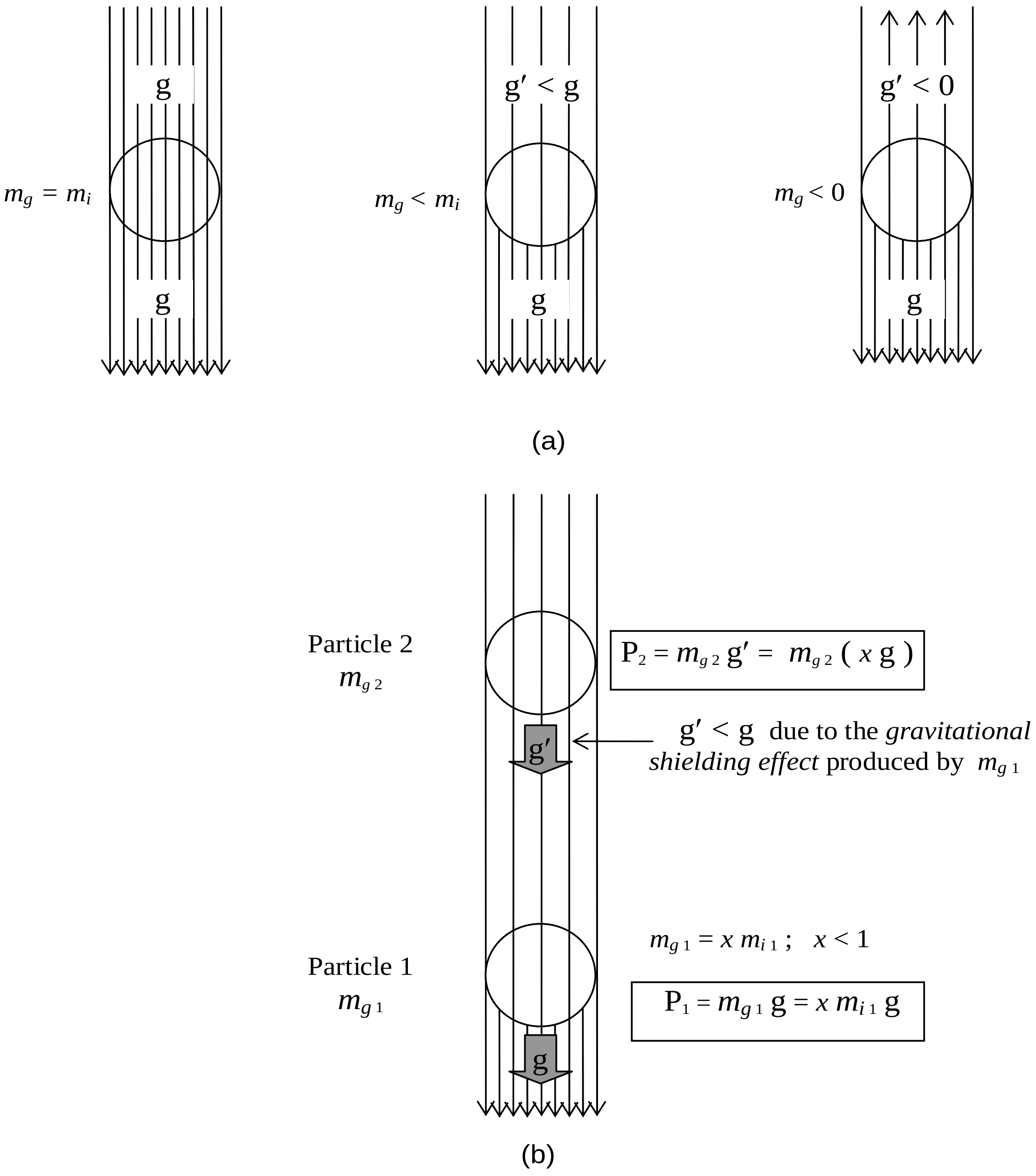

Fig. I I – The Gravitational Shielding effect.

Let us now deduce the correlation between $P$ and $M_{ge}$.

When the electrons flux strikes the sensor, the electrons transfer to it a *momentum* $Q = n_e q_e = n_e M_{ge} V$. Since $Q = F\Delta t = 2Fd/V$, we conclude that

$$M_{ge} = \frac{2d}{V^2}\left(\frac{F}{n_e}\right)$$

The amount of electrons, $n_e$, is given by $n_e = \rho S d$ where $\rho$ is the amount of electrons per unit of volume (electrons/m$^3$); $S$ is the cross-section of the electrons flux and $d$ the distance between cathode and anode.

In order to calculate $n_e$ we will start from the *Langmuir-Child law* and the *Ohm vectorial law*, respectively given by

$$J = \alpha\frac{\tilde{V}^{\frac{3}{2}}}{d} \text{ and } J = \rho_c V\,, \quad (\rho_c = \rho/e)$$

where $J$ is the thermoionic current density; $\alpha = 2.33\times10^{-6}\,A.m^{-1}.V^{-\frac{3}{2}}$ is the called *Child's constant*; $\tilde{V}$ is the voltage drop across the anode and cathode electrodes, and $V$ is the velocity of the electrons.

By comparing the Langmuir-Child law with the Ohm vectorial law we obtain

$$\rho = \frac{\alpha\tilde{V}^{\frac{3}{2}}}{ed^2V}$$

Thus, we can write that

$$n_e = \frac{\alpha\tilde{V}^{\frac{3}{2}}S}{edV}$$

and

$$M_{ge} = \left(\frac{2ed^2}{\alpha V\tilde{V}^{\frac{3}{2}}}\right)P$$

Where $P = F/S$, is the pressure to be measured by the piezoelectric sensor.

In the experimental set-up the total force $F$ acting on the piezoelectric sensor is the resultant of all the forces $F_\phi$ produced by each electrons flux that passes through each hole of area $S_\phi$ in the grid of the anode 1, and is given by

$$F = nF_\phi = n(PS_\phi) = \left(\frac{\alpha n S_\phi}{2ed^2}\right)M_{ge}V\tilde{V}^{\frac{3}{2}}$$

where $n$ is the number of holes in the grid. By means of the piezoelectric sensor we can measure $F$ and consequently obtain $M_{ge}$.

We can use the equation above to evaluate the magnitude of the force $F$ to be measured by the piezoelectric sensor. First, we will find the expression of $V$ as a function of $\tilde{V}$ since the electrons speed $V$ depends on the voltage $\tilde{V}$.

We will start from Eq. (46) which is the general expression for Lorentz's force, i.e.,

$$\frac{d\vec{p}}{dt} = \left(q\vec{E} + q\vec{V}\times\vec{B}\right)\frac{m_g}{m_{i0}}$$

When the force and the speed have the same direction Eq. (6) gives

$$\frac{d\vec{p}}{dt} = \frac{m_g}{\left(1 - V^2/c^2\right)^{\frac{3}{2}}}\frac{d\vec{V}}{dt}$$

By comparing these expressions we obtain

$$\frac{m_{i0}}{\left(1 - V^2/c^2\right)^{\frac{3}{2}}}\frac{d\vec{V}}{dt} = q\vec{E} + q\vec{V}\times\vec{B}$$

In the case of electrons accelerated by a sole electric field $(B = 0)$, the equation above gives

$$\vec{a} = \frac{d\vec{V}}{dt} = \frac{e\vec{E}}{m_{ie}}\left(1 - V^2/c^2\right)\sqrt{\frac{2e\tilde{V}}{m_{ie}}}$$

Therefore, the velocity $V$ of the electrons in the experimental set-up is

$$V = \sqrt{2ad} = \left(1 - V^2/c^2\right)^{\frac{3}{4}}\sqrt{\frac{2e\tilde{V}}{m_{ie}}}$$

From Eq. (43) we conclude that

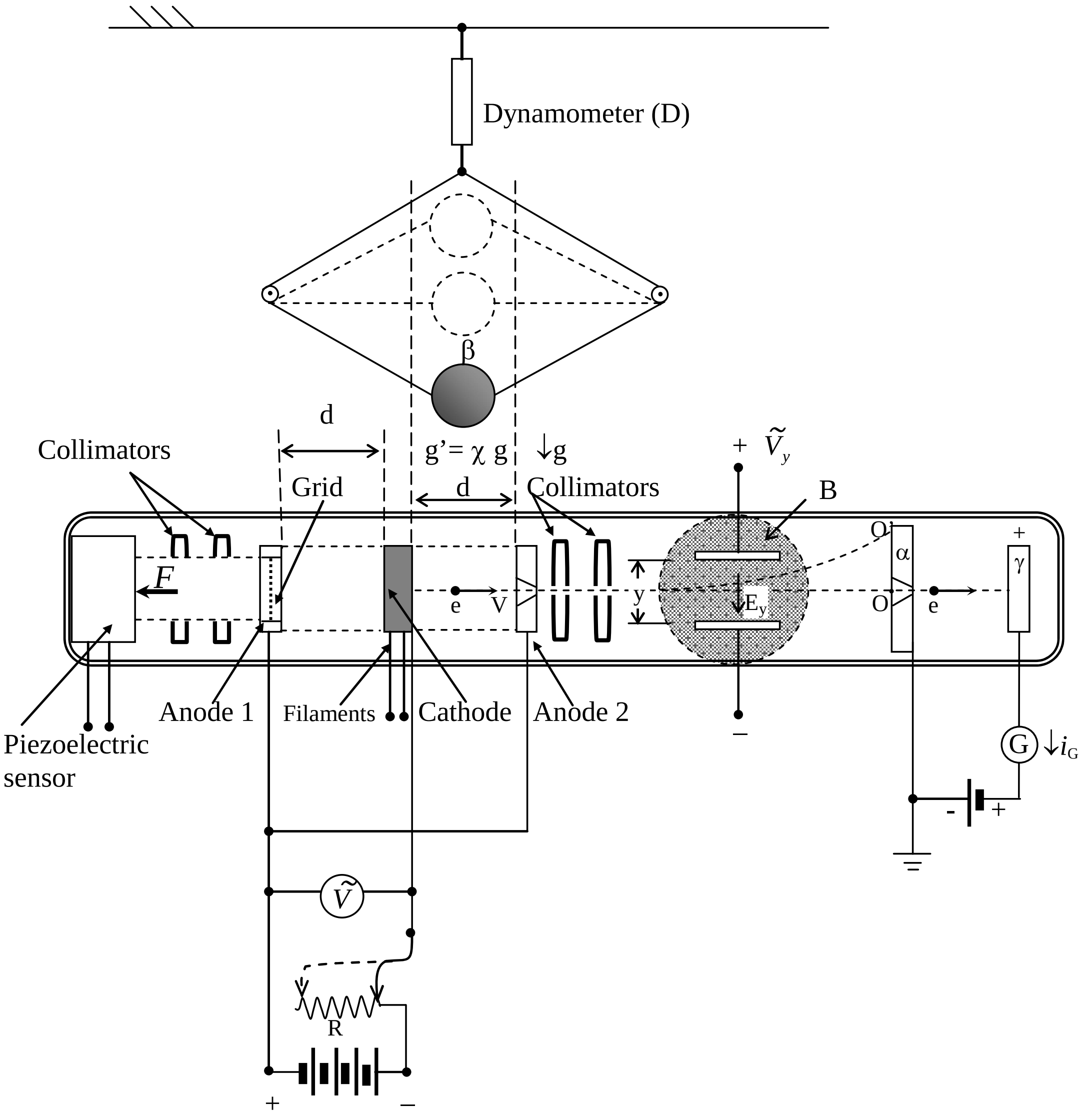

Fig. III – Experimental set-up in order to check the factor $\chi$ above a high-speed electrons flux. The set-up may also check the velocities and the gravitational masses of the electrons.

$m_{ge} \cong 0$ when $V \cong 0.745c$. Substitution of this value of $V$ into equation above gives $\widetilde{V} \cong 479.1KV$. This is the voltage drop necessary to be applied across the anode and cathode electrodes in order to obtain $m_{ge} \cong 0$.

Since the equation above can be used to evaluate the velocity $V$ of the electrons flux for a given $\widetilde{V}$, then we can use the obtained value of $V$ to evaluate the intensity of $\vec{B}$ in order to produce $eVB = eE_y$ in the experimental set-up. Then by adjusting $B$ we can check when the electrons flux is detected by the galvanometer $G$. In this case, as we have already seen, $eVB = eE_y$, and the velocity of the electrons flux is calculated by means of the expression $V = E_y / B$. Substitution of $V$ into the expressions of $m_{ge}$ and $M_{ge}$, respectively given by

$$m_{ge} = \left\{1 - 2\left[\frac{1}{\sqrt{1 - V^2/c^2}} - 1\right]\right\} m_{ie}$$

and

$$M_{ge} = \frac{m_{ge}}{\sqrt{1 - V^2/c^2}}$$

yields the corresponding values of $m_{ge}$ and $M_{ge}$ which can be compared with the values obtained in the experimental set-up:

$$m_{ge} = \chi m_{ie} = \left(P'_\beta / P_\beta\right) m_{ie}$$

$$M_{ge} = \frac{F}{V\widetilde{V}^{\frac{3}{2}}}\left(\frac{2ed^2}{\alpha n S_\phi}\right)$$

where $P'_\beta$ and $P_\beta$ are measured by the dynamometer $D$ and $F$ is measured by the piezoelectric sensor.

If we have $nS_\phi \cong 0.16m^2$ and $d = 0.08m$ in the experimental set-up then it follows that

$$F = 1.82 \times 10^{14} M_{ge} V\widetilde{V}^{\frac{3}{2}}$$

By varying $\widetilde{V}$ from 10KV up to 500KV we note that the maximum value for $F$ occurs when $\widetilde{V} \cong 344.7KV$. Under these circumstances, $V \cong 0.7c$ and $M_{ge} \cong 0.28 m_{ie}$. Thus the maximum value for $F$ is

$$F_{max} \cong 1.9N \cong 190gf$$

Consequently, for $\widetilde{V}_{max} = 500KV$, the piezoelectric sensor must satisfy the following characteristics:

– Capacity 200gf
– Readability 0.001gf

Let us now return to the explanation for the findings of Podkletnov's experiment. Next, we will explain the decrease of 0.1% in the weight of the superconductor when the disk is only levitating but not rotating.

Equation (55) shows how the gravitational mass is altered by *electromagnetic* fields.

The expression of $n_r$ for $\sigma >> \omega\varepsilon$ can be obtained from (54), in the form

$$n_r = \frac{c}{v} = \sqrt{\frac{\mu\sigma c^2}{4\pi f}} \qquad (56)$$

Substitution of (56) into (55) leads to

$$m_g = \left\{1 - 2\left[\sqrt{1 + \frac{\mu\sigma}{4\pi f}\left(\frac{U}{m_i c}\right)^2} - 1\right]\right\} m_{i0}$$

This equation shows that *atoms of ferromagnetic materials with very-high* $\mu$ can have gravitational masses strongly reduced by means of *Extremely Low Frequency* (ELF) electromagnetic radiation. It also shows that atoms of *superconducting* *materials* (due to *very-high* $\sigma$ ) can also have its gravitational masses strongly reduced by means of ELF electromagnetic radiation.

Alternatively, we may put Eq.(55) as a function of the *power density* ( or intensity ), $D$, of the radiation. The integration of (51) gives $U = VD/v$. Thus, we can write (55) in the following form:

$$m_g = \left\{1 - 2\left[\sqrt{1 + \left(\frac{n_r^2 D}{\rho c^3}\right)^2} - 1\right]\right\} m_{i0} \qquad (57)$$

where $\rho = m_{i0}/V$.

For $\sigma >> \omega\varepsilon$, $n_r$ will be given by (56) and consequently (57) becomes

$$m_g = \left\{1 - 2\left[\sqrt{1 + \left(\frac{\mu\sigma D}{4\pi f \rho c}\right)^2} - 1\right]\right\} m_{i0} \qquad (58)$$

In the case of *Thermal radiation*, it is common to relate the energy of photons to *temperature*, *T*, through the relation,

$$\langle hf \rangle \approx \kappa T$$

where $\kappa = 1.38 \times 10^{-23} J/°K$ is the *Boltzmann's constant*. On the other hand it is known that

$$D = \sigma_B T^4$$

where $\sigma_B = 5.67 \times 10^{-8} watts/m^2 °K^4$ is the *Stefan-Boltzmann's constant*. Thus we can rewrite (58) in the following form

$$m_g = \left\{1 - 2\left[\sqrt{1 + \left(\frac{\mu\sigma\sigma_B h T^3}{4\pi\kappa\rho c}\right)^2} - 1\right]\right\} m_{i0} \qquad (58a)$$

Starting from this equation, we can evaluate the effect of the *thermal radiation* upon the gravitational mass of the Copper-pair fluid, $m_{g,CPfluid}$. Below the *transition temperature*, $T_c$, $(T/T_c < 0.5)$ the conductivity of the superconducting materials is usually larger than $10^{22} S/m$ [26]. On the other hand the *transition temperature*, for high critical temperature (HTC) superconducting materials, is in the order of $10^2 K$. Thus (58a) gives

$$m_{g,CPfluid} = \left\{1 - 2\left[\sqrt{1 + \frac{\sim 10^{-9}}{\rho^2_{CPfluid}}} - 1\right]\right\} m_{i,CPfluid} \qquad (58b)$$

Assuming that the number of Copper-pairs per unit volume is $N \approx 10^{26} m^{-3}$ [27] we can write that

$$\rho_{CPfluid} = Nm^* \approx 10^{-4} kg/m^3$$

Substitution of this value into (58b) yields

$$m_{g,CPfluid} = m_{i,CPfluid} - 0.1\, m_{i,CPfluid}$$

This means that the gravitational masses of the electrons are decreased of ~10%. This corresponds to a decrease in the gravitational mass of the superconductor given by

$$\frac{m_{g,SC}}{m_{i,SC}} = \frac{N\left(m_{ge} + m_{gp} + m_{gn} + \Delta E/c^2\right)}{N\left(m_{ie} + m_{ip} + m_{in} + \Delta E/c^2\right)} =$$

$$= \left(\frac{m_{ge} + m_{gp} + m_{gn} + \Delta E/c^2}{m_{ie} + m_{ip} + m_{in} + \Delta E/c^2}\right) =$$

$$= \left(\frac{0.9 m_{ie} + m_{ip} + m_{in} + \Delta E/c^2}{m_{ie} + m_{ip} + m_{in} + \Delta E/c^2}\right) =$$

$$= 0.999976$$

Where $\Delta E$ is the interaction energy. Therefore, a decrease of $(1 - 0.999976) \approx 10^{-5}$, i.e., approximately $10^{-3}\%$ in respect to the initial gravitational mass of the superconductor, due to the local *thermal radiation* only. However, here we must also consider the *gravitational shielding effect* produced, in this case, by the decrease of $\approx 10^{-3}\%$ in the gravitational mass of the particles inside the superconductor (see Fig. II). Therefore the *total* weight decrease in the superconductor will be much greater than $\approx 10^{-3}\%$. This can explain the smaller effect on the order of $0.1\%$ observed in the Podkletnov measurements when the disk is not rotating.

Let us now consider an electric current $I$ through a conductor subjected to electromagnetic radiation with power density $D$ and frequency $f$.

Under these circumstances the *gravitational mass* $m_{ge}$ of the *electrons* of the conductor, according to Eq. (58), is given by

$$m_{ge} = \left\{1 - 2\left[\sqrt{1 + \left(\frac{\mu\sigma D}{4\pi f\rho c}\right)^2} - 1\right]\right\} m_e$$

where $m_e = 9.11\times10^{-31}kg$.

Note that if the radiation upon the conductor has extremely-low frequency (ELF radiation) then $m_{ge}$ can be strongly reduced. For example, if $f \approx 10^{-6}Hz$, $D \approx 10^5 W/m^2$ and the conductor is made of *copper* ($\mu \cong \mu_0; \sigma = 5.8\times10^7 S/m$ and $\rho = 8900 kg/m^3$) then

$$\left(\frac{\mu\sigma D}{4\pi f\rho c}\right) \approx 1$$

and consequently $m_{ge} \approx 0.1 m_e$.

According to Eq. (6) the force upon each *free electron* is given by

$$\vec{F}_e = \frac{m_{ge}}{\left(1 - V^2/c^2\right)^{3/2}}\frac{d\vec{V}}{dt} = e\vec{E}$$

where $E$ is the applied electric field. Therefore, the decrease of $m_{ge}$ produces an increase in the velocity $V$ of the free electrons and consequently the *drift velocity* $V_d$ is also increased. It is known that the density of electric current $J$ through a conductor [28] is given by

$$\vec{J} = \Delta_e \vec{V}_d$$

where $\Delta_e$ is the density of the free electric charges ( For cooper conductors $\Delta_e = 1.3\times10^{10} C/m^3$ ). Therefore increasing $V_d$ produces an increase in the electric current $I$. Thus if $m_{ge}$ is reduced 10 times $\left(m_{ge} \approx 0.1 m_e\right)$ the drift velocity $V_d$ is increased 10 times as well as the electric current. Thus we conclude that strong fluxes of ELF radiation upon electric/electronic circuits can suddenly increase the electric currents and consequently damage these circuits.

Since the *orbital electrons* moment of inertia is given by $I_i = \Sigma(m_i)_j r_j^2$, where $m_i$ refers to *inertial mass* and not to gravitational mass, then the *momentum* $L = I_i\omega$ of the conductor *orbital electrons* are not affected by the ELF radiation. Consequently, this radiation just affects the conductor's *free electrons* velocities. Similarly, in the case of superconducting materials, the *momentum*, $L = I_i\omega$, of the *orbital electrons* are not affected by the gravitomagnetic fields.

The vector $\vec{D} = (U/V)\vec{v}$, which we may define from (48), has the same direction of the propagation vector $\vec{k}$ and evidently corresponds to the *Poynting vector*. Then $\vec{D}$ can be replaced by $\vec{E}\times\vec{H}$. Thus we can write $D = \frac{1}{2}EH = \frac{1}{2}E(B/\mu) = \frac{1}{2}E[(E/v)/\mu] = \frac{1}{2}(1/v\mu)E^2$. For $\sigma >> \omega\varepsilon$ Eq. (54) tells us that $v = \sqrt{4\pi f/\mu\sigma}$. Consequently, we obtain

$$D = \tfrac{1}{2}E^2\sqrt{\frac{\sigma}{4\pi f\mu}}$$

This expression refers to the instantaneous values of $D$ and $E$. The average value for $E^2$ is equal to $\frac{1}{2}E_m^2$ because $E$ varies sinusoidaly

($E_m$ is the maximum value for $E$). Substitution of the expression of $D$ into (58) gives

$$m_g=\left\{1-2\left[\sqrt{1+\frac{\mu}{4c^2}\left(\frac{\sigma}{4\pi f}\right)^3\frac{E^2}{\rho^2}}-1\right]\right\}m_{i0} \qquad (59a)$$

Since $E_{rms}=E_m/\sqrt{2}$ and $E^2=\frac{1}{2}E_m^2$ we can write the equation above in the following form

$$m_g=\left\{1-2\left[\sqrt{1+\frac{\mu}{4c^2}\left(\frac{\sigma}{4\pi f}\right)^3\frac{E_{rms}^2}{\rho^2}}-1\right]\right\}m_{i0} \qquad (59a)$$

Note that for *extremely-low frequencies* the value of $f^{-3}$ in this equation becomes highly expressive.

Since $E=vB$ equation (59a) can also be put as a function of $B$, i.e.,

$$m_g=\left\{1-2\left[\sqrt{1+\left(\frac{\sigma}{4\pi f\mu c^2}\right)\frac{B^4}{\rho^2}}-1\right]\right\}m_{i0} \qquad (59b)$$

For *conducting* materials with $\sigma\approx10^7\,S/m$; $\mu_r=1$; $\rho\approx10^3\,kg/m^3$ the expression (59b) gives

$$m_g=\left\{1-2\left[\sqrt{1+\left(\frac{\approx10^{-12}}{f}\right)B^4}-1\right]\right\}m_{i0}$$

This equation shows that the decreasing in the *gravitational mass* of these conductors can become experimentally detectable for example, starting from 100Teslas at 10mHz.

One can then conclude that an interesting situation arises when a body penetrates a magnetic field in the direction of its center. The *gravitational mass* of the body decreases progressively. This is due to the intensity increase of the magnetic field upon the body while it penetrates the field. In order to understand this phenomenon we might, based on (43), think of the inertial mass as being formed by two parts: one *positive* and another *negative*. Thus, when the body penetrates the magnetic field, its negative inertial mass increases, but its total inertial mass decreases, i.e., although there is an increase of inertial mass, the total inertial mass (which is equivalent to *gravitational mass*) will be reduced.

On the other hand, Eq.(4) shows that the *velocity of the body must increase* as consequence of the gravitational mass decreasing since the *momentum* is conserved. Consider for example a spacecraft with velocity $V_s$ and gravitational mass $M_g$. If $M_g$ is reduced to $m_g$ then the velocity becomes

$$V_s'=\left(M_g/m_g\right)V_s$$

In addition, Eqs. 5 and 6 tell us that the *inertial forces* depend on $m_g$. Only in the particular case of $m_g=m_{i0}$ the expressions (5) and (6) reduce to the well-known Newtonian expression $F=m_{i0}a$. Consequently, one can conclude that the *inertial effects* on the spacecraft will also be reduced due to the *decreasing of its gravitational mass*. Obviously this leads to a new concept of aerospace flight.

Now consider an electric current $i=i_0 sin2\pi ft$ through a conductor. Since the current density, $\vec{J}$, is expressed by $\vec{J}=di/d\vec{S}=\sigma\vec{E}$, then we can write that $E=i/\sigma S=\left(i_0/\sigma S\right)sin2\pi ft$. Substitution of this equation into (59a) gives

$$m_g=\left\{1-2\left[\sqrt{1+\frac{i_0^4\mu}{64\pi^3c^2\rho^2S^4f^3\sigma}sin^4 2\pi ft}-1\right]\right\}m_{i0} \qquad (59c)$$

If the conductor is a *supermalloy* rod $(1\times1\times400mm)$ then $\mu_r=100{,}000$ (initial); $\rho=8770kg/m^3$; $\sigma=1.6\times10^6\,S/m$ and $S=1\times10^{-6}m^2$. Substitution of these values into the equation above yields the following expression for the *gravitational mass* of the supermalloy rod

$$m_{g(sm)} = \left\{1 - 2\left[\sqrt{1 + \left(5.71\times10^{-12}\, i_0^4 / f^3\right)\sin^4 2\pi ft} - 1\right]\right\} m_{i(sm)}$$

Some oscillators like the HP3325A (Op.002 High Voltage Output) can generate sinusoidal voltages with *extremely-low* frequencies down to $f = 1\times10^{-6}\,Hz$ and amplitude up to 20V (into $50\Omega$ load). The maximum output current is $0.08A_{pp}$ .

Thus, for $i_0 = 0.04A$ $\left(0.08A_{pp}\right)$ and $f < 2.25\times10^{-6}\,Hz$ the equation above shows that the *gravitational mass* of the rod becomes *negative* at $2\pi ft = \pi/2$; for $f \cong 1.7\times10^{-6}\,Hz$ at $t = 1/4f = 1.47\times10^{5}\,s \cong 40.8h$ it shows that $m_{g(sm)} \cong -m_{i(sm)}$.

This leads to the idea of the *Gravitational Motor*. See in Fig. IV a type of gravitational motor (Rotational Gravitational Motor) based on the possibility of gravity control on a ferromagnetic wire.

It is important to realize that this is not the unique way of decreasing the gravitational mass of a body. It was noted earlier that the expression (53) is general for all types of waves including non-electromagnetic waves like sound waves for example. In this case, the velocity $v$ in (53) will be the *speed of sound in the body* and $D$ the *intensity* of the sound radiation. Thus from (53) we can write that

$$\frac{\Delta p}{m_i c} = \frac{VD}{m_i c} = \frac{D}{\rho c v^2}$$

It can easily be shown that $D = 2\pi^2 \rho f^2 A^2 v$ where $A = \lambda P / 2\pi\rho v^2$ ; $A$ and $P$ are respectively the amplitude and maximum pressure variation of the sound wave. Therefore we readily obtain

$$\frac{\Delta p}{m_{i0}c} = \frac{P^2}{2\rho^2 c v^3}$$

Substitution of this expression into (41) gives

$$m_g = \left\{1 - 2\left[\sqrt{1 + \left(\frac{P^2}{2\rho^2 c v^3}\right)^2} - 1\right]\right\} m_{i0} \qquad (60)$$

This expression shows that in the case of sound waves the decreasing of gravitational mass is relevant for *very strong pressures* only.

It is known that in the nucleus of the Earth the pressure can reach values greater than $10^{13}\,N/m^2$. The equation above tells us that sound waves produced by pressure variations of this magnitude can cause strong decreasing of the *gravitational mass* at the surroundings of the point where the sound waves were generated. This obviously must cause an abrupt decreasing of the pressure at this place since pressure = weight /area = $m_g g$/area). Consequently a local instability will be produced due to the opposite internal pressure. The conclusion is that this effect may cause Earthquakes.

Consider a sphere of radius $r$ around the point where the sound waves were generated (at $\approx 1{,}000km$ depth; the Earth's radius is $6{,}378km$). If the *maximum* pressure, at the explosion place ( sphere of radius $r_0$), is $P_{max} \approx 10^{13}\,N/m^2$ and the pressure at the distance $r = 10km$ is $P_{min} = (r_0/r)^2 P_{max} \approx 10^9\,N/m^2$ then we can consider that in the sphere $P = \sqrt{P_{max}P_{min}} \approx 10^{11}\,N/m^2$ .Thus assuming $v \approx 10^3\,m/s$ and $\rho \approx 10^3\,kg/m^3$ we can calculate the variation of gravitational mass in the sphere by means of the equation of $m_g$ , i.e.,

$$\alpha = \left\{1 - 2\left[\sqrt{1 + \frac{i^4 \mu}{64\pi^3 c^2 \rho^2 S^4 f^3 \sigma}} - 1\right]\right\}$$

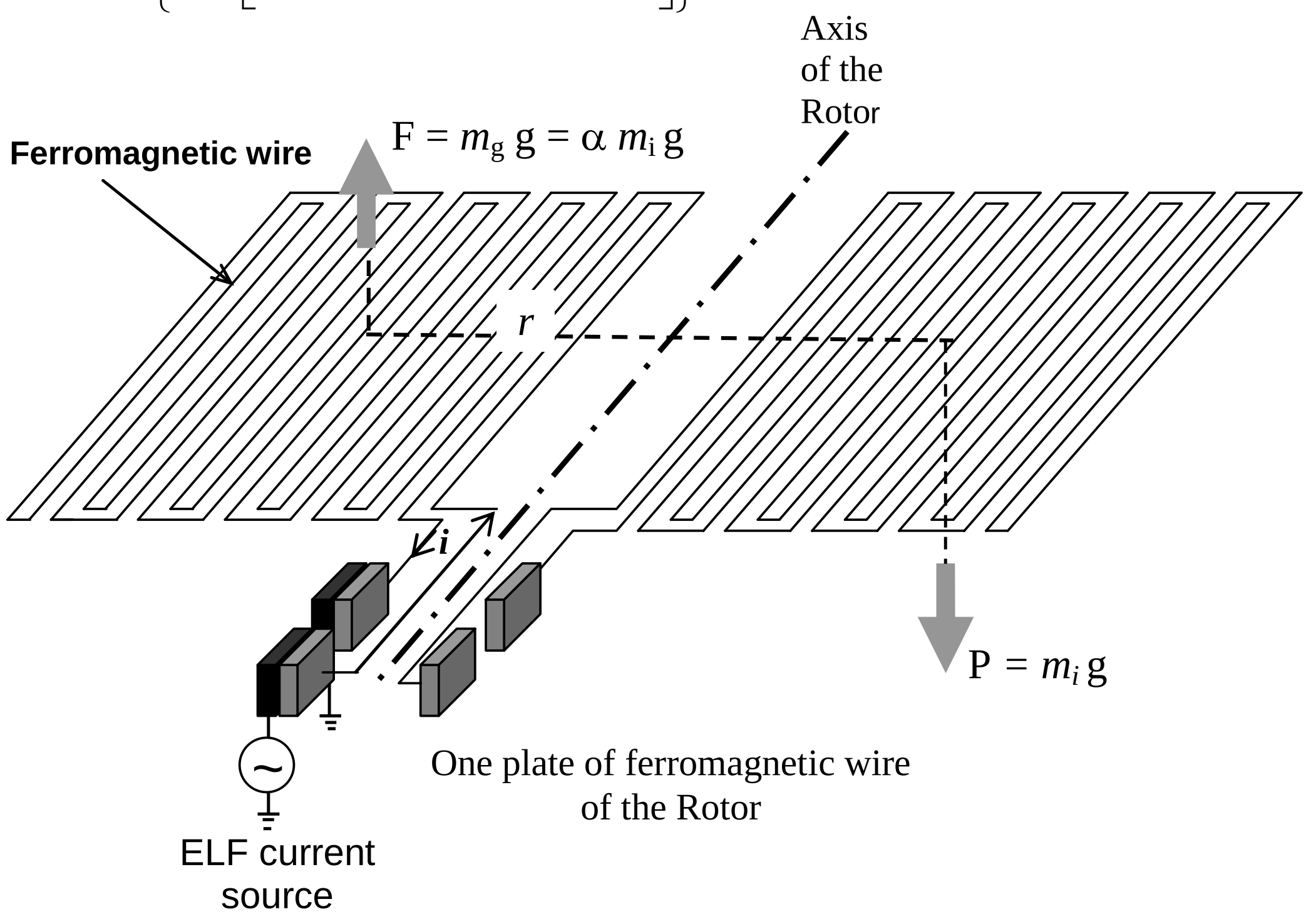

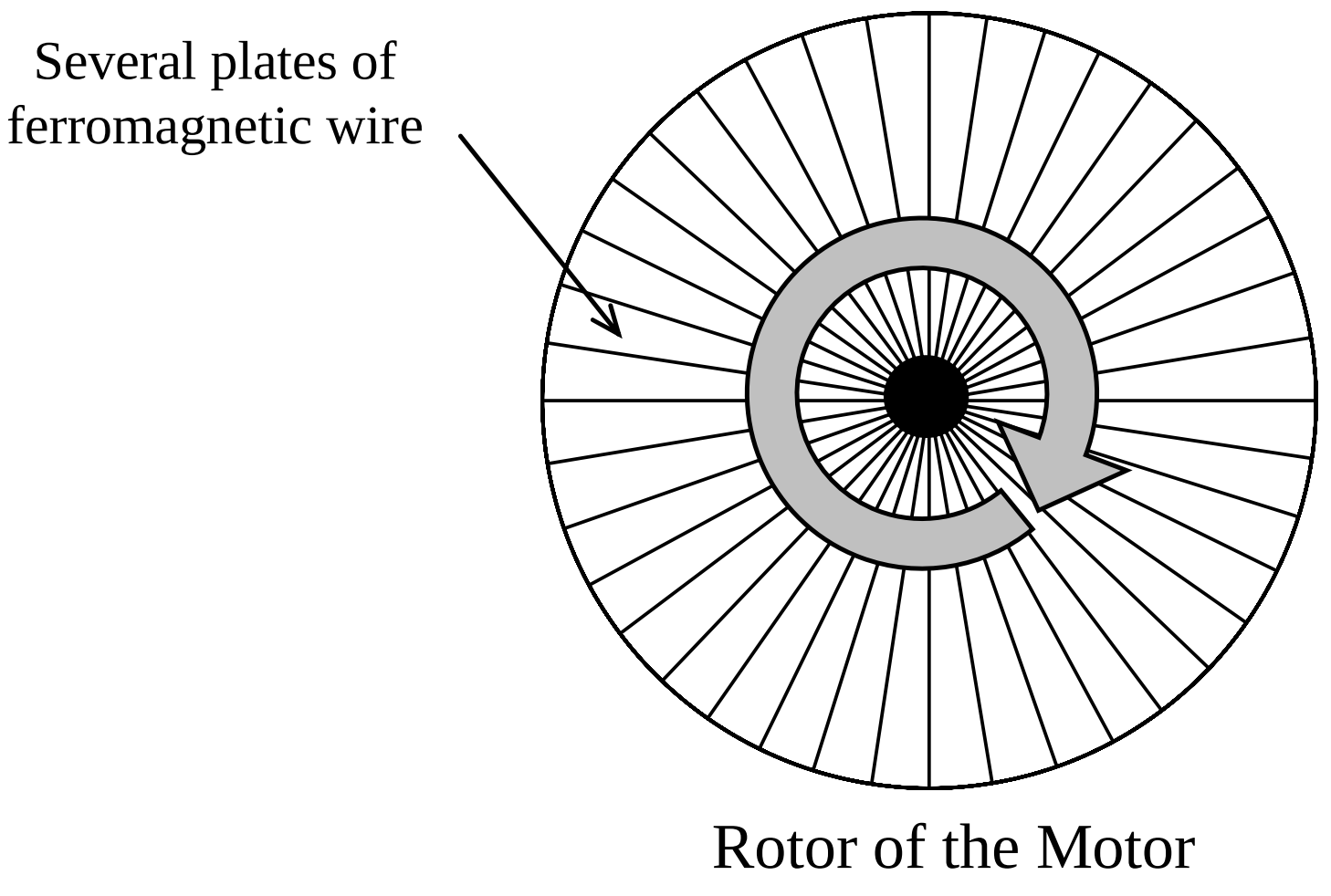

Fig. IV - Rotational Gravitational Motor

$$\Delta m_g = m_{g(initial)} - m_g =$$

$$= m_{i0} - \left\{1 - 2\left[\sqrt{1+\left(\frac{P^2}{2\rho^2 c v^3}\right)^2} - 1\right]\right\} m_{i0} =$$

$$= 2\left[\sqrt{1+\left(\frac{P^2}{2\rho^2 c v^3}\right)^2} - 1\right]\rho V \approx 10^{11} kg$$

The *transitory* loss of this great amount of gravitational mass may evidently produce a strong pressure variation and consequently a strong Earthquake.

Finally, we can evaluate the energy necessary to generate those sound waves. From (48) we can write $D_{max} = P_{max} v \approx 10^{16} W/m^2$. Thus, the released power is $P_0 = D_{max}\left(4\pi r_0^2\right) \approx 10^{21} W$ and the energy $\Delta E$ released at the time interval $\Delta t$ must be $\Delta E = P_0 \Delta t$. Assuming $\Delta t \approx 10^{-3} s$ we readily obtain

$$\Delta E = P_0 \Delta t \approx 10^{18}\ joules \approx 10^4\ Megatons$$

This is the amount of energy released by an earthquake of magnitude 9 $(M_s = 9)$, i.e., $E = 1.74 \times 10^{(5+1.44 M_s)} \cong 10^{18}\ joules$. The maximum magnitude in the *Richter* scale is 12. Note that the sole releasing of this energy at 1000km depth (without the effect of gravitational mass decreasing) cannot produce an Earthquake, since the sound waves reach 1km depth with pressures less than 10N/cm$^2$.

Let us now return to the Theory.

The equivalence between frames of non-inertial reference and gravitational fields assumed $m_g \equiv m_i$ because the inertial forces were given by $\vec{F}_i = m_i \vec{a}$, while the equivalent gravitational forces, by $\vec{F}_g = m_g \vec{g}$. Thus, to satisfy the equivalence ($\vec{a} \equiv \vec{g}$ and $\vec{F}_i \equiv \vec{F}_g$) it was *necessary that* $m_g \equiv m_i$. Now, the inertial force, $\vec{F}_i$, is given by Eq.(6), and from Eq.(13) we can obtain the gravitational force, $\vec{F}_g$. Thus, $\vec{F}_i \equiv \vec{F}_g$ leads to

$$\frac{m_g}{\left(1 - V^2/c^2\right)^{3/2}}\vec{a} \equiv G\frac{m_g'}{\left(r'\sqrt{1 - V^2/c^2}\right)^2}\frac{m_g}{\sqrt{1 - V^2/c^2}} \equiv$$

$$\equiv \left(G\frac{m_g'}{r'^2}\right)\frac{m_g}{\left(1 - V^2/c^2\right)^{3/2}} \equiv \vec{g}\frac{m_g}{\left(1 - V^2/c^2\right)^{3/2}} \qquad (61)$$

whence results

$$\vec{a} \equiv \vec{g} \qquad (62)$$

Consequently, the equivalence is evident, and therefore Einstein's equations from the General Relativity continue obviously valid.

The new expression for $F_i$ (Eqs. (5) and (6)) shows that the inertial forces are proportional to the *gravitational mass*, $m_g$. This means that these forces result from the gravitational interaction between the particle and the other gravitational masses of the Universe, just as *Mach's principle* predicts. Therefore the new expression for the inertial forces incorporates the Mach's principle into Gravitation Theory, and furthermore reveals that the inertial effects upon a particle can be reduced because, as we have seen, the gravitational mass may be reduced. When $m_g = m_{i0}$ the *nonrelativistic* equation for inertial forces, $\vec{F}_i = m_g \vec{a}$, reduces to $\vec{F}_i = m_{i0}\vec{a}$. This is the well-known *Newton's second law* for motion.

In Einstein's Special Relativity Theory the motion of a free-particle is described by means of $\delta S = 0$ [29]. Now based on Eq. (1), $\delta S = 0$ will be given by the following expression

$$\delta S = -m_g c \delta \int ds = 0. \qquad (63)$$

which also describes the motion of the particle inside the gravitational field. Thus, Einstein's equations from the General Relativity can be derived starting from $\delta\left(S_m + S_g\right) = 0$, where $S_g$ and $S_m$ refer to the *action* of the gravitational field and the action of the matter, respectively [30].

The variations $\delta S_g$ and $\delta S_m$ can be written as follows [31]:

$$\delta S_g = \frac{c^3}{16\pi G}\int\left(R_{ik} - \tfrac{1}{2}g_{ik}R\right)\delta g^{ik}\sqrt{-g}\,d\Omega \qquad (64)$$

$$\delta S_m = -\frac{1}{2c}\int T_{ik}\,\delta g^{ik}\sqrt{-g}\,d\Omega \qquad (65)$$

where $R_{ik}$ is the Ricci's tensor; $g_{ik}$ the metric tensor and $T_{ik}$ the matter's energy-momentum tensor:

$$T_{ik} = \left(P + \varepsilon_g\right)\mu_i\mu_k + Pg_{ik} \qquad (66)$$

where $P$ is the pressure and $\varepsilon_g = \rho_g c^2$ is now, the density of gravitational energy, $E_g$, of the particle; $\rho_g$ is then the density of gravitational mass of the particle, i.e., $M_g$ at the volume unit.

Substitution of (64) and (65) into $\delta S_m + \delta S_g = 0$ yields

$$\frac{c^3}{16\pi G}\int\left(R_{ik} - \tfrac{1}{2}g_{ik}R - \tfrac{8\pi G}{c^4}T_{ik}\right)\delta g^{ik}\sqrt{-g}\,d\Omega = 0$$

whence,

$$\left(R_{ik} - \tfrac{1}{2}g_{ik}R - \tfrac{8\pi G}{c^4}T_{ik}\right) = 0 \qquad (67)$$

because the $\delta g_{ik}$ are arbitrary.

Equations (67) in the following form

$$R_{ik} - \tfrac{1}{2}g_{ik}R = \tfrac{8\pi G}{c^4}T_{ik} \qquad (68)$$

or

$$R_i^k - \tfrac{1}{2}g\delta_i^k R = \tfrac{8\pi G}{c^4}T_i^k. \qquad (69)$$

are the Einstein's equations from the General Relativity.

It is known that these equations are *only valid* if the spacetime is *continuous*. We have shown at the beginning of this work that the spacetime *is not continuous* it is *quantized*. However, the spacetime can be considered approximately "*continuous*" when the *quantum number* $n$ is very large (Classical limit). Therefore, just under these circumstances the Einstein's equations from the General Relativity can be used in order to "*classicalize*" the quantum theory by means of approximated description of the spacetime.

Later on we will show that the length $d_{\min}$ of Eq. (29) is given by

$$d_{\min} = \tilde{k}\,l_{planck} = \tilde{k}\left(G\hbar/c^3\right)^{\frac{1}{2}} \approx 10^{-34}m \qquad (70)$$

(See Eq. (100)). On the other hand, we will find in the Eq. (129) the length scale of the *initial Universe*, i.e., $d_{initial} \approx 10^{-14}m$. Thus, from the Eq. (29) we get: $n = d_{initial}/d_{\min} = 10^{-14}/10^{-34} \approx 10^{20}$ this is the quantum number of the spacetime at *initial instant*. That quantum number is sufficiently large for the spacetime to be considered approximately "*continuous*" starting from the beginning of the Universe. Therefore Einstein's equations can be used even at the Initial Universe.

Now, it is easy to conclude why the attempt to quantize gravity starting from the General Relativity was a bad theoretical strategy.

Since the gravitational interaction can be repulsive, besides attractive, such as the electromagnetic interaction, then the *graviton* must have spin 1 (called *graviphoton*) and not 2. Consequently, the gravitational forces are also *gauge* forces because they are yielded by the exchange of the so-called "virtual" *quanta* of spin 1, such as the electromagnetic forces and the weak and strong nuclear forces.

Let us now deduce the *Entropy Differential Equation* starting from Eq. (55). Comparison of Eqs. (55) and (41) shows that $Un_r = \Delta pc$. For small velocities, i.e., $(V << c)$, we have $Un_r << m_{i0}c^2$. Under these circumstances, the development of Eq. (55) in power of $\left(Un_r / m_{i0}c^2\right)$ gives

$$m_g = m_{i0} - \left(\frac{Un_r}{m_{i0}c^2}\right)^2 m_{i0} \qquad (71)$$

In the particular case of *thermal radiation,* it is usual to relate the energy of the photons to the temperature, through the relationship $\langle h\nu\rangle \approx kT$ where $k = 1.38\times10^{-23} J/K$ is the Boltzmann's constant. Thus, in that case, the energy absorbed by the particle will be $U = \eta\langle h\nu\rangle \approx \eta kT$, where $\eta$ is a particle-dependent absorption/emission coefficient. Therefore, Eq.(71) may be rewritten in the following form:

$$m_g = m_{i0} - \left[\left(\frac{n_r\eta k}{c^2}\right)^2 \frac{T^2}{m_{i0}^{\ 2}}\right] m_{i0} \qquad (72)$$

For electrons at T=300K, we have

$$\left(\frac{n_r\eta k}{c^2}\right)^2 \frac{T^2}{m_e^{\ 2}} \approx 10^{-17}$$

Comparing (72) with (18), we obtain

$$E_{Ki} = \frac{1}{2}\left(\frac{n_r\eta k}{c}\right)^2 \frac{T^2}{m_{i0}}. \qquad (73)$$

The derivative of $E_{Ki}$ with respect to temperature $T$ is

$$\frac{\partial E_{Ki}}{\partial T} = \left(n_r\eta k/c\right)^2 \left(T/m_{i0}\right) \qquad (74)$$

Thus,

$$T\frac{\partial E_{Ki}}{\partial T} = \frac{\left(n_r\eta kT\right)^2}{m_{i0}c^2} \qquad (75)$$

Substitution of $E_{Ki} = E_i - E_{i0}$ into (75) gives

$$T\left(\frac{\partial E_i}{\partial T} + \frac{\partial E_{i0}}{\partial T}\right) = \frac{\left(n_r\eta kT\right)^2}{m_{i0}c^2} \qquad (76)$$

By comparing the Eqs.(76) and (73) and considering that $\partial E_{i0}/\partial T = 0$ because $E_{i0}$ does not depend on $T$, the Eq.(76) reduces to

$$T\left(\partial E_i/\partial T\right) = 2E_{Ki} \qquad (77)$$

However, Eq.(18) shows that $2E_{Ki} = E_i - E_g$. Therefore Eq.(77) becomes

$$E_g = E_i - T\left(\partial E_i/\partial T\right) \qquad (78)$$

Here, we can identify the energy $E_i$ with the *free-energy* of the system-F and $E_g$ with the *internal energy* of the system-U. Thus we can write the Eq.(78) in the following form:

$$U = F - T\left(\partial F/\partial T\right) \qquad (79)$$

This is the well-known equation of Thermodynamics. On the other hand, remembering that $\partial Q = \partial\tau + \partial U$ (1st principle of Thermodynamics) and

$$F = U - TS \qquad (80)$$

(Helmholtz's function), we can easily obtain from (79), the following equation

$$\partial Q = \partial\tau + T\partial S. \qquad (81)$$

For *isolated systems*, $\partial\tau = 0$, we have

$$\partial Q = T\partial S \qquad (82)$$

which is the well-known *Entropy Differential Equation*.

Let us now consider the Eq.(55) in the *ultra-relativistic case* where the inertial energy of the particle $E_i = M_i c^2$ is much larger than its inertial energy at rest $m_{i0}c^2$. Comparison of (4) and (10) leads to $\Delta p = E_i V/c^2$ which, in the ultra-relativistic case, gives $\Delta p = E_i V/c^2 \cong E_i/c \cong M_i c$. On the other hand, comparison of (55) and (41) shows that $Un_r = \Delta pc$. Thus $Un_r = \Delta pc \cong M_i c^2 >> m_{i0}c^2$. Consequently, Eq.(55) reduces to

$$m_g = m_{i0} - 2Un_r/c^2 \qquad (83)$$

Therefore, the *action* for such particle, in agreement with the Eq.(2), is

$$S = -\int_{t_1}^{t_2} m_g c^2 \sqrt{1 - V^2/c^2}\, dt =$$
$$= \int_{t_1}^{t_2} \left(-m_i + 2Un_r/c^2\right) c^2 \sqrt{1 - V^2/c^2}\, dt =$$
$$= \int_{t_1}^{t_2} \left[-m_i c^2 \sqrt{1 - V^2/c^2} + 2Un_r \sqrt{1 - V^2/c^2}\right] dt. \quad (84)$$

The integrant function is the *Lagrangean*, i.e.,

$$L = -m_{i0} c^2 \sqrt{1 - V^2/c^2} + 2Un_r \sqrt{1 - V^2/c^2} \qquad (85)$$

Starting from the Lagrangean we can find the Hamiltonian of the particle, by means of the well-known general formula:

$$H = V(\partial L/\partial V) - L.$$

The result is

$$H = \frac{m_{i0} c^2}{\sqrt{1 - V^2/c^2}} + Un_r \left[\frac{\left(4V^2/c^2 - 2\right)}{\sqrt{1 - V^2/c^2}}\right]. \qquad (86)$$

The second term on the right hand side of Eq.(86) results from the particle's interaction with the *electromagnetic field*. Note the similarity between the obtained Hamiltonian and the well-known Hamiltonian for the particle in an electromagnetic field [32]:

$$H = m_{i0} c^2 \Big/ \sqrt{1 - V^2/c^2} + Q\varphi. \qquad (87)$$

in which $Q$ is the electric charge and $\varphi$, the field's *scalar* potential. The quantity $Q\varphi$ expresses, as we know, the particle's interaction with the electromagnetic field in the same way as the second term on the right hand side of the Eq. (86).

It is therefore evident that it is the same quantity, expressed by different variables.

Thus, we can conclude that, in ultra-high energy conditions $\left(Un_r \cong M_i c^2 > m_{i0} c^2\right)$, the gravitational and electromagnetic fields can be described by the *same* Hamiltonian, i.e., in these circumstances they are *unified* !

It is known that starting from that Hamiltonian we may obtain a complete description of the electromagnetic field. This means that from the present theory for gravity we can also derive *the equations of the electromagnetic field*.

Due to $Un_r = \Delta pc \cong M_i c^2$ the second term on the right hand side of Eq.(86) can be written as follows

$$\Delta pc \left[\frac{\left(4V^2/c^2 - 2\right)}{\sqrt{1 - V^2/c^2}}\right] =$$
$$= \left[\frac{\left(4V^2/c^2 - 2\right)}{\sqrt{1 - V^2/c^2}}\right] M_i c^2 =$$
$$= Q\varphi = \frac{QQ'}{4\pi\varepsilon_0 R} = \frac{QQ'}{4\pi\varepsilon_0 r \sqrt{1 - V^2/c^2}}$$

whence

$$\left(4V^2/c^2 - 2\right) M_i c^2 = \frac{QQ'}{4\pi\varepsilon_0 r}$$

The factor $\left(4V^2/c^2 - 2\right)$ becomes equal to 2 in the ultra-relativistic case, then it follows that

$$2M_i c^2 = \frac{QQ'}{4\pi\varepsilon_0 r} \qquad (88)$$

From (44), we know that there is a minimum value for $M_i$ given by $M_{i(min)} = m_{i(min)}$. Eq.(43) shows that $m_{g(min)} = m_{i0(min)}$ and Eq.(23) gives $m_{g(min)} = \pm h/cL_{max}\sqrt{8} = \pm h\sqrt{3/8}/cd_{max}$.

Thus we can write

$$M_{i(min)} = m_{i0(min)} = \pm h\sqrt{3/8}/cd_{max} \qquad (89)$$

According to (88) the value $2M_{i(min)}c^2$ is correlated to $\left(QQ'/4\pi\varepsilon_0 r\right)_{min} = Q_{min}^2/4\pi\varepsilon_0 r_{max}$, i.e.,

$$\frac{Q_{min}^2}{4\pi\varepsilon_0 r_{max}} = 2M_{i(min)}c^2 \qquad (90)$$

where $Q_{min}$ is the *minimum electric charge* in the Universe ( therefore equal to minimum electric charge of the quarks, i.e., $\frac{1}{3}e$); $r_{max}$ is the *maximum distance* between $Q$ and $Q'$, which should be equal to the so-called "diameter", $d_c$, of the *visible* Universe ($d_c = 2l_c$ where $l_c$ is obtained from the Hubble's law for $V = c$, i.e., $l_c = c\tilde{H}^{-1}$). Thus, from (90) we readily obtain

$$Q_{min} = \sqrt{\pi\varepsilon_0 hc\sqrt{24}\left(d_c/d_{max}\right)} = \\ = \sqrt{\left(\pi\varepsilon_0 hc^2\sqrt{96}\tilde{H}^{-1}/d_{max}\right)} = \\ = \tfrac{1}{3}e \qquad (91)$$

whence we find

$$d_{max} = 3.4\times10^{30}\,m$$

This will be the maximum "diameter" that the Universe will reach. Consequently, Eq.(89) tells us that the *elementary quantum* of matter is

$$m_{i0(min)} = \pm h\sqrt{3/8}/cd_{max} = \pm 3.9\times10^{-73}kg$$

This is, therefore, the *smallest indivisible particle of matter*.

Considering that, the inertial mass of the *Observable Universe* is $M_U = c^3/2H_0G \cong 10^{53}kg$ and that its volume is $V_U = \frac{4}{3}\pi R_U^3 = \frac{4}{3}\pi\left(c/H_0\right)^3 \cong 10^{79}\,m^3$, where $H_0 = 1.75\times10^{-18}s^{-1}$ is the *Hubble constant*, we can conclude that the *number of these particles in the Observable Universe* is

$$n_U = \frac{M_U}{m_{i0(\min)}} \cong 10^{125}\ particles$$

By dividing this number by $V_U$, we get

$$\frac{n_U}{V_U} \cong 10^{46}\ particles/m^3$$

Obviously, the dimensions of the smallest indivisible particle of matter depend on its state of compression. In free space, for example, its volume is $V_U/n_U$. Consequently, its "radius" is $R_U/\sqrt[3]{n_U} \cong 10^{-15}m$.

If $N$ particles with diameter $\phi$ fill all space of $1m^3$ then $N\phi^3 = 1$. Thus, if $\phi \cong 10^{-15}m$ then the number of particles, with this diameter, necessary to fill all $1m^3$ is $N \cong 10^{45}\ particles$. Since the number of *smallest indivisible particles of matter* in the Universe is $n_U/V_U \cong 10^{46}particles/m^3$ we can conclude that these particles *fill all space* in the Universe, by forming a *Continuous*[4] *Universal Medium* or *Continuous Universal Fluid* (CUF), the density of which is

$$\rho_{CUF} = \frac{n_U\, m_{i0(\min)}}{V_U} \cong 10^{-27}kg/m^3$$

Note that this density is much smaller than the density of the *Intergalactic Medium* $\left(\rho_{IGM} \cong 10^{-26}kg/m^3\right)$.

The extremely-low density of the *Continuous Universal Fluid* shows that its *local gravitational mass* can be strongly affected by electromagnetic fields (including gravitoelectromagnetic fields), pressure, etc. (See Eqs. 57, 58, 59a, 59b, 55a, 55c and 60). The density of this fluid is clearly *not uniform* along the Universe, since it can be strongly compressed in several regions (galaxies, stars, blackholes, planets, etc). At the normal state (free space), the mentioned fluid is *invisible*. However, at *super compressed* state it can become *visible by giving origin* to the *known matter* since matter, as we have seen, is *quantized* and consequently, formed by an *integer number* of elementary quantum of matter with mass $m_{i0(\min)}$. Inside the proton, for example, there are $n_p = m_p/m_{i0(\min)} \cong 10^{45}$ *elementary quanta of matter* at supercompressed state, with volume $V_{proton}/n_p$ and "radius" $R_p/\sqrt[3]{n_p} \cong 10^{-30}m$.

Therefore, the solidification of the matter is just a *transitory state* of this Universal Fluid, which can back to the primitive state when the cohesion conditions disappear.

Let us now study another aspect of the present theory. By combination of gravity and the *uncertainty principle* we will derive the expression for the *Casimir force*.

An uncertainty $\Delta m_i$ in $m_i$ produces an uncertainty $\Delta p$ in $p$ and

---

[4] At *very small scale*.

therefore an uncertainty $\Delta m_g$ in $m_g$, which according to Eq.(41) , is given by

$$\Delta m_g = \Delta m_i - 2\left[\sqrt{1+\left(\frac{\Delta p}{\Delta m_i c}\right)^2}-1\right]\Delta m_i \quad (92)$$

From the uncertainty principle for position and momentum, we know that the product of the uncertainties of the simultaneously measurable values of the corresponding position and momentum components is at least of the magnitude order of $\hbar$, i.e.,

$$\Delta p \Delta r \sim \hbar$$

Substitution of $\Delta p \sim \hbar/\Delta r$ into (92) yields

$$\Delta m_g = \Delta m_i - 2\left[\sqrt{1+\left(\frac{\hbar/\Delta m_i c}{\Delta r}\right)^2}-1\right]\Delta m_i \quad (93)$$

Therefore if

$$\Delta r << \frac{\hbar}{\Delta m_i c} \quad (94)$$

then the expression (93) reduces to:

$$\Delta m_g \cong -\frac{2\hbar}{\Delta r c} \quad (95)$$

Note that $\Delta m_g$ does not depend on $m_g$.

Consequently, the uncertainty $\Delta F$ in the gravitational force $F = -G m_g m'_g / r^2$, will be given by

$$\Delta F = -G\frac{\Delta m_g \Delta m'_g}{(\Delta r)^2} =$$

$$= -\left[\frac{2}{\pi(\Delta r)^2}\right]\frac{hc}{(\Delta r)^2}\left(\frac{G\hbar}{c^3}\right) \quad (96)$$

The amount $\left(G\hbar/c^3\right)^{1/2} = 1.61\times 10^{-35} m$ is called the *Planck length*, $l_{planck}$ ,( the length scale on which quantum fluctuations of the metric of the space time are expected to be of order unity). Thus, we can write the expression of $\Delta F$ as follows

$$\Delta F = -\left(\frac{2}{\pi}\right)\frac{hc}{(\Delta r)^4} l^2_{planck} =$$

$$= -\left(\frac{\pi}{480}\right)\frac{hc}{(\Delta r)^4}\left[\left(\frac{960}{\pi^2}\right) l^2_{planck}\right] =$$

$$= -\left(\frac{\pi A_0}{480}\right)\frac{hc}{(\Delta r)^4} \quad (97)$$

or

$$F_0 = -\left(\frac{\pi A_0}{480}\right)\frac{hc}{r^4} \quad (98)$$

which is the expression of the *Casimir force* for $A = A_0 = \left(960/\pi^2\right) l^2_{planck}$ .

This suggests that $A_0$ is an *elementary area* related to the existence of a *minimum length* $d_{min} = \tilde{k} l_{planck}$ what is in accordance with the *quantization of space* (29) and which points out to the existence of $d_{min}$ .

It can be easily shown that the *minimum area* related to $d_{min}$ is the area of an *equilateral triangle* of side length $d_{min}$ ,i.e.,

$$A_{min} = \left(\tfrac{\sqrt{3}}{4}\right) d^2_{min} = \left(\tfrac{\sqrt{3}}{4}\right) \tilde{k}^2 l^2_{planck}$$

On the other hand, the *maximum area* related to $d_{min}$ is the area of a *sphere* of radius $d_{min}$ ,i.e.,

$$A_{max} = \pi d^2_{min} = \pi \tilde{k}^2 l^2_{planck}$$

Thus, the elementary area

$$A_0 = \delta_A d^2_{min} = \delta_A \tilde{k}^2 l^2_{planck} \quad (99)$$

must have a value between $A_{min}$ and $A_{max}$, i.e.,

$$\tfrac{\sqrt{3}}{4} < \delta_A < \pi$$

The previous assumption that $A_0 = \left(960/\pi^2\right) l^2_{planck}$ shows that $\delta_A \tilde{k}^2 = 960/\pi^2$ what means that

$$5.6 < \tilde{k} < 14.9$$

Therefore we conclude that

$$d_{min} = \tilde{k} l_{planck} \approx 10^{-34} m. \quad (100)$$

The $n-esimal$ area after $A_0$ is

$$A = \delta_A (n d_{min})^2 = n^2 A_0 \qquad (101)$$

It can also be easily shown that the *minimum volume* related to $d_{min}$ is the volume of a *regular tetrahedron* of edge length $d_{min}$, i.e.,

$$\Omega_{min} = \left(\tfrac{\sqrt{2}}{12}\right) d_{min}^3 = \left(\tfrac{\sqrt{2}}{12}\right) \tilde{k}^3 l_{planck}^3$$

The *maximum volume* is the volume of a *sphere* of radius $d_{min}$, i.e.,

$$\Omega_{max} = \left(\tfrac{4\pi}{3}\right) d_{min}^3 = \left(\tfrac{4\pi}{3}\right) \tilde{k}^3 l_{planck}^3$$

Thus, the elementary volume $\Omega_0 = \delta_V d_{min}^3 = \delta_V \tilde{k}^3 l_{planck}^3$ must have a value between $\Omega_{min}$ and $\Omega_{max}$, i.e.,

$$\left(\tfrac{\sqrt{2}}{12}\right) < \delta_V < \tfrac{4\pi}{3}$$

On the other hand, the $n-esimal$ volume after $\Omega_0$ is

$$\Omega = \delta_V (n d_{min})^3 = n^3 \Omega_0 \qquad n = 1,2,3,\ldots,n_{max}.$$

The existence of $n_{max}$ given by (26), i.e.,

$$n_{max} = L_{max}/L_{min} = d_{max}/d_{min} = \\ = \left(3.4\times10^{30}\right)/\tilde{k} l_{planck} \approx 10^{64}$$

shows that the Universe must have a *finite volume* whose value at the present stage is

$$\Omega_{Up} = n_{Up}^3 \Omega_0 = \left(d_p/d_{min}\right)^3 \delta_V d_{min}^3 = \delta_V d_p^3$$

where $d_p$ is the present length scale of the Universe. In addition as $\left(\tfrac{\sqrt{2}}{12}\right) < \delta_V < \tfrac{4\pi}{3}$ we conclude that the Universe must have a *polyhedral* space topology with volume between the volume of a *regular tetrahedron* of edge length $d_p$ and the volume of the *sphere of* diameter $d_p$.

A recent analysis of astronomical data suggests not only that the Universe is *finite*, but also that it has a *dodecahedral* space topology [33,34], what is in strong accordance with the previous theoretical predictions.

From (22) and (26) we have that $L_{max} = d_{max}/\sqrt{3} = n_{max} d_{min}/\sqrt{3}$. Since (100) gives $d_{min} \cong 10^{-34} m$ and $n_{max} \cong 10^{64}$ we conclude that $L_{max} \cong 10^{30} m$. From the *Hubble's law* and (22) we have that

$$V_{max} = \tilde{H} l_{max} = \tilde{H}(d_{max}/2) = \left(\sqrt{3}/2\right)\tilde{H} L_{max}$$

where $\tilde{H} = 1.7\times10^{-18} s^{-1}$. Therefore we obtain

$$V_{max} \cong 10^{12} m/s \qquad .$$

This is the speed upper limit imposed by the *quantization of velocity* (Eq. 36). It is known that the speed upper limit for *real* particles is equal to $c$. However, also it is known that *imaginary* particles can have velocities greater than $c$ (*Tachyons*). Thus, we conclude that $V_{max}$ is the speed upper limit for *imaginary particles in our ordinary space-time*. Later on, we will see that also exists a speed upper limit to the *imaginary* particles in the *imaginary* space-time.

Now, multiplying Eq. (98) (the expression of $F_0$) by $n^2$ we obtain

$$F = n^2 F_0 = -\left(\frac{\pi n^2 A_0}{480}\right)\frac{hc}{r^4} = -\left(\frac{\pi A}{480}\right)\frac{hc}{r^4} \qquad (102)$$

This is the general expression of the *Casimir force*.

Thus, we conclude that *the Casimir effect* is just a gravitational effect related to the *uncertainty principle*.

Note that Eq. (102) arises only when $\Delta m_i$ and $\Delta m_i'$ satisfy Eq.(94). If only $\Delta m_i$ satisfies Eq.(94), i.e., $\Delta m_i << \hbar/\Delta rc$ but $\Delta m_i' >> \hbar/\Delta rc$ then $\Delta m_g$ and $\Delta m_g'$ will be respectively given by

$$\Delta m_g \cong -2\hbar/\Delta rc \quad and \quad \Delta m_g' \cong \Delta m_i$$

Consequently, the expression (96) becomes

$$\Delta F=\frac{hc}{(\Delta r)^3}\left(\frac{G\Delta m_i'}{\pi c^2}\right)=\frac{hc}{(\Delta r)^3}\left(\frac{G\Delta m_i' c^2}{\pi c^4}\right)=$$
$$=\frac{hc}{(\Delta r)^3}\left(\frac{G\Delta E'}{\pi c^4}\right) \qquad (103)$$

However, from the uncertainty principle for *energy* and *time* we know that

$$\Delta E\sim \hbar/\Delta t \qquad (104)$$

Therefore, we can write the expression (103) in the following form:

$$\Delta F=\frac{hc}{(\Delta r)^3}\left(\frac{G\hbar}{c^3}\right)\left(\frac{1}{\pi\Delta t' c}\right)=$$
$$=\frac{hc}{(\Delta r)^3}l_{planck}^2\left(\frac{1}{\pi\Delta t' c}\right) \qquad (105)$$

From the General Relativity Theory we know that $dr = cdt/\sqrt{-g_{00}}$ . If the field is *weak* then $g_{00}=-1-2\phi/c^2$ and $dr = cdt/\left(1+\phi/c^2\right)=cdt/\left(1-Gm/r^2c^2\right)$. For $Gm/r^2c^2 \ll 1$ we obtain $dr \cong cdt$. Thus, if $dr = dr'$ then $dt = dt'$ . This means that we can change $(\Delta t' c)$ by $(\Delta r)$ into (105). The result is

$$\Delta F=\frac{hc}{(\Delta r)^4}\left(\frac{1}{\pi}l_{planck}^2\right)=$$
$$=\left(\frac{\pi}{480}\right)\frac{hc}{(\Delta r)^4}\underbrace{\left(\frac{480}{\pi^2}l_{planck}^2\right)}_{\frac{1}{2}A_0}=$$
$$=\left(\frac{\pi A_0}{960}\right)\frac{hc}{(\Delta r)^4}$$

or

$$F_0=\left(\frac{\pi A_0}{960}\right)\frac{hc}{r^4}$$

whence

$$F=\left(\frac{\pi A}{960}\right)\frac{hc}{r^4} \qquad (106)$$

Now, the Casimir force is *repulsive*, and its intensity is half of the intensity previously obtained (102).

Consider the case when both $\Delta m_i$ and $\Delta m_i'$ do not satisfy Eq.(94), and

$$\Delta m_i \gg \hbar/\Delta rc$$
$$\Delta m_i' \gg \hbar/\Delta rc$$

In this case, $\Delta m_g \cong \Delta m_i$ and $\Delta m_g' \cong \Delta m_i'$. Thus,

$$\Delta F=-G\frac{\Delta m_i\Delta m_i'}{(\Delta r)^2}=-G\frac{\left(\Delta E/c^2\right)\left(\Delta E'/c^2\right)}{(\Delta r)^2}=$$
$$=-\left(\frac{G}{c^4}\right)\frac{\left(\hbar/\Delta t\right)^2}{(\Delta r)^2}=-\left(\frac{G\hbar}{c^3}\right)\frac{hc}{(\Delta r)^2}\left(\frac{1}{c^2\Delta t^2}\right)=$$
$$=-\left(\frac{1}{2\pi}\right)\frac{hc}{(\Delta r)^4}l_{planck}^2=$$
$$=-\left(\frac{\pi}{1920}\right)\frac{hc}{(\Delta r)^4}\left(\frac{960}{\pi^2}l_{planck}^2\right)=-\left(\frac{\pi A_0}{1920}\right)\frac{hc}{(\Delta r)^4}$$

whence

$$F=-\left(\frac{\pi A}{1920}\right)\frac{hc}{r^4} \qquad (107)$$

The force will be *attractive* and its intensity will be the *fourth part* of the intensity given by the first expression (102) for the Casimir force.

We can also use this theory to explain some relevant cosmological phenomena. For example, the recent discovery that the cosmic expansion of the Universe may be *accelerating*, and not decelerating as many cosmologists had anticipated [35].

We start from Eq. (6) which shows that the *inertial forces*, $\vec{F}_i$, whose action on a particle, in the case of force and speed with *same direction*, is given by

$$\vec{F}_i=\frac{m_g}{\left(1-V^2/c^2\right)^{3/2}}\vec{a}$$

Substitution of $m_g$ given by (43) into the expression above gives

$$\vec{F}_i=\left(\frac{3}{\left(1-V^2/c^2\right)^{3/2}}-\frac{2}{\left(1-V^2/c^2\right)^2}\right)m_{i0}\vec{a}$$

whence we conclude that a particle with rest inertial mass, $m_{i0}$, subjected to a force, $\vec{F}_i$, acquires an acceleration $\vec{a}$ given by

$$\vec{a} = \frac{\vec{F}_i}{\left(\frac{3}{\left(1 - V^2/c^2\right)^{3/2}} - \frac{2}{\left(1 - V^2/c^2\right)^2}\right) m_{i0}}$$

By substituting the well-known expression of Hubble's law for velocity, $V = \tilde{H} l$, ($\tilde{H} = 1.7 \times 10^{-18} s^{-1}$ is the Hubble constant) into the expression of $\vec{a}$, we get *the acceleration for any particle in the expanding Universe*, i.e.,

$$\vec{a} = \frac{\vec{F}_i}{\left(\frac{3}{\left(1 - \tilde{H}^2 l^2/c^2\right)^{3/2}} - \frac{2}{\left(1 - \tilde{H}^2 l^2/c^2\right)^2}\right) m_{i0}}$$

Obviously, the distance $l$ increases with the expansion of the Universe. Under these circumstances, it is easy to see that the term

$$\left(\frac{3}{\left(1 - \tilde{H}^2 l^2/c^2\right)^{3/2}} - \frac{2}{\left(1 - \tilde{H}^2 l^2/c^2\right)^2}\right)$$

decreases, *increasing the acceleration* of the expanding Universe.

Let us now consider the phenomenon of gravitational deflection of light.

A distant star's light ray, under the Sun's gravitational force field describes the usual central force hyperbolic orbit. The deflection of the light ray is illustrated in Fig. V, with the bending greatly exaggerated for a better view of the angle of deflection.

The distance CS is the distance $d$ of closest approach. The angle of deflection of the light ray, $\delta$, is shown in the Figure V and is

$$\delta = \pi - 2\beta.$$

where $\beta$ is the angle of the asymptote to the hyperbole. Then, it follows that

$$\tan\delta = \tan(\pi - 2\beta) = -\tan 2\beta$$

From the Figure V we obtain

$$\tan\beta = \frac{V_y}{c}.$$

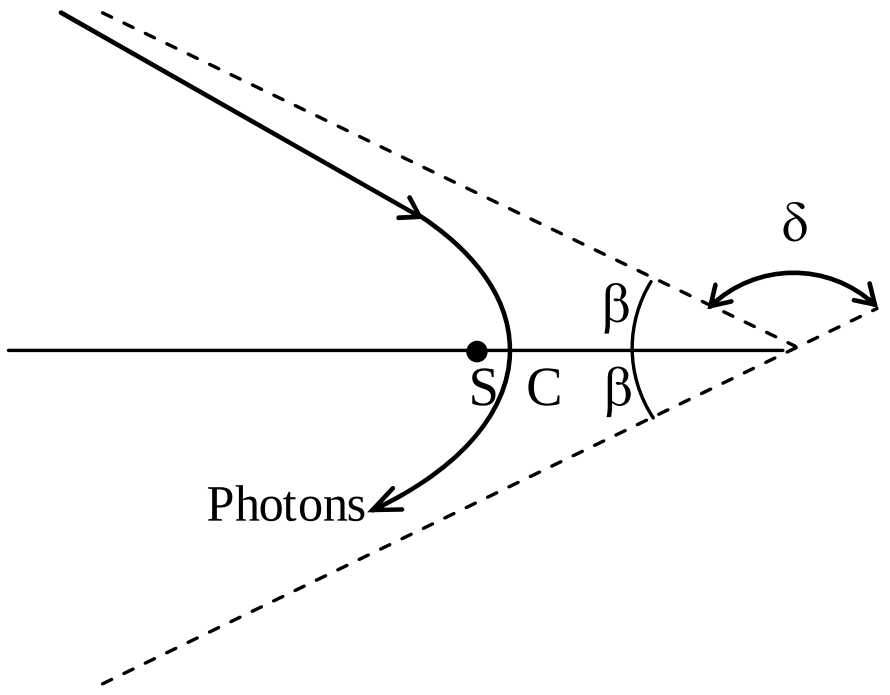

Fig. V – Gravitational deflection of light about the Sun.

Since $\delta$ and $\beta$ are very small we can write that

$$\delta = 2\beta \quad \text{and} \quad \beta = \frac{V_y}{c}$$

Then

$$\delta = \frac{2V_y}{c}$$

Consider the motion of the photons at some time $t$ after it has passed the point of closest approach. We impose Cartesian Co-ordinates with the origin at the point of closest approach, the x axis pointing along its path and the y axis towards the Sun. The gravitational pull of the Sun is

$$P = -G\frac{M_{gS} M_{gp}}{r^2}$$

where $M_{gp}$ is the relativistic *gravitational* mass of the photon and $M_{gS}$ the relativistic gravitational mass of the Sun. Thus, the component in a perpendicular direction is

$$F_y = -G\frac{M_{gS} M_{gp}}{r^2}\sin\beta =$$

$$= -G\frac{M_{gS} M_{gp}}{d^2 + c^2 t^2}\frac{d}{\sqrt{d^2 + c^2 t^2}}$$

According to Eq. (6) the expression of the force $F_y$ is

$$F_y = \left(\frac{m_{gp}}{\left(1 - V_y^2/c^2\right)^{\frac{3}{2}}}\right)\frac{dV_y}{dt}$$

By substituting Eq. (43) into this expression, we get

$$F_y = \left( \frac{3}{\left(1 - V_y^2/c^2\right)} - \frac{2}{\left(1 - V_y^2/c^2\right)^{\frac{3}{2}}} \right) M_{ip} \frac{dV_y}{dt}$$

For $V_y << c$, we can write this expression in the following form $F_y = M_{ip}\left(dV_y/dt\right)$. This force acts on the photons for a time $t$ causing an increase in the transverse velocity

$$dV_y = \frac{F_y}{M_{ip}} dt$$

Thus the component of transverse velocity acquired after passing the point of closest approach is

$$V_y = \frac{M_{gp}}{M_{ip}} \int \frac{d\left(-GM_{gS}\right)}{\left(d^2 + c^2t^2\right)^{\frac{3}{2}}} dt =$$

$$= \frac{-GM_{gS}}{dc}\left(\frac{M_{gp}}{M_{ip}}\right) = \frac{-GM_{gS}}{dc}\left(\frac{m_{gp}}{m_{ip}}\right)$$

Since the angle of deflection $\delta$ is given by

$$\delta = 2\beta = \frac{2V_y}{c}$$

we readily obtain

$$\delta = \frac{2V_y}{c} = \frac{-2GM_{gS}}{c^2 d}\left(\frac{m_{gp}}{m_{ip}}\right)$$

If $m_{gp}/m_{ip} = 2$, the expression above gives

$$\delta = -\frac{4GM_{gS}}{c^2 d}$$

As we know, this is the correct formula indicated by the experimental results.

Equation (4) says that

$$m_{gp} = \left\{1 - 2\left[\sqrt{1 + \left(\frac{\Delta p}{m_{ip}c}\right)^2} - 1\right]\right\} m_{ip}$$

Since $m_{gp}/m_{ip} = 2$ then, by making $\Delta p = h/\lambda$ into the equation above we get

$$m_{ip} = +\frac{2}{\sqrt{3}}\left(\frac{hf}{c^2}\right)i$$

Due to $m_{gp}/m_{ip} = 2$ we get

$$m_{gp} = +\frac{4}{\sqrt{3}}\left(\frac{hf}{c^2}\right)i$$

This means that the gravitational and inertial masses of the *photon* are *imaginaries*, and *invariants* with respect to speed of photon, i.e. $M_{ip} = m_{ip}$ and $M_{gp} = m_{gp}$.On the other hand, we can write that

$$m_{ip} = m_{ip(real)} + m_{ip(imaginary)} = \frac{2}{\sqrt{3}}\left(\frac{hf}{c^2}\right)i$$

and

$$m_{gp} = m_{gp(real)} + m_{gp(imaginary)} = \frac{4}{\sqrt{3}}\left(\frac{hf}{c^2}\right)i$$

This means that we must have

$$m_{ip(real)} = m_{gp(real)} = 0$$

The phenomenon of *gravitational deflection of light about the Sun* shows that *the gravitational interaction* between the Sun and the photons is *attractive*. Thus, due to the gravitational force between the Sun and a photon can be expressed by $F = -GM_{g(Sun)}m_{gp(imaginary)}/r^2$, where $m_{gp(imaginary)}$ is a quantity *positive* and *imaginary*, we conclude that the force $F$ will only be *attractive* if *the matter* $\left(M_{g(Sun)}\right)$ *has negative imaginary gravitational mass*.

The Eq. (41) shows that if the *inertial mass* of a particle is *null* then its *gravitational mass* is given by

$$m_g = \pm 2\Delta p/c$$

where $\Delta p$ is the *momentum* variation due to the energy absorbed by the particle. If the energy of the particle is *invariant*, then $\Delta p = 0$ and, consequently, its *gravitational mass* will also be null. This is the case of the photons, i.e., they have an invariant energy $hf$ and a *momentum* $h/\lambda$. As they cannot absorb additional energy, the variation in the *momentum* will be null $\left(\Delta p = 0\right)$ and, therefore, their *gravitational masses* will also be null.

However, if the energy of the particle is not invariant (it is able to absorb energy) then the absorbed energy will transfer the *amount of motion* (*momentum*) to the particle, and consequently its *gravitational mass* will be increased. This means that the *motion* generates gravitational mass*.*

On the other hand, if the *gravitational mass* of a particle is null then its *inertial mass*, according to Eq. (41), will be given by

$$m_i = \pm \frac{2}{\sqrt{5}} \frac{\Delta p}{c}$$

From Eqs. (4) and (7) we get

$$\Delta p = \left( \frac{E_g}{c^2} \right) \Delta V = \left( \frac{p_0}{c} \right) \Delta V$$

Thus we have

$$m_g = \pm \left( \frac{2p_0}{c^2} \right) \Delta V \text{ and } m_i = \pm \frac{2}{\sqrt{5}} \left( \frac{p_0}{c^2} \right) \Delta V$$

Note that, like the gravitational mass, the inertial mass is also directly related to the motion, i.e., it is also generated by the motion.

Thus, we can conclude that is the motion, or rather, the *velocity* is what makes the two types of mass.

In this picture, the fundamental particles can be considered as *immaterial vortex of velocity*; it is the velocity of these vortexes that causes the fundamental particles to have masses. That is, there exists not matter in the usual sense; but just *motion*. Thus, the difference between matter and energy just consists of the diversity of the motion direction; *rotating*, closed in itself, in the matter; *ondulatory*, with open cycle, in the energy (See Fig. VI).

Under this context, the *Higgs mechanism*† appears as a process, by which the velocity of an immaterial vortex can be increased or decreased by making the vortex (particle) *gain* or *lose mass*. If *real motion* is what makes *real mass* then, by analogy, we can say that *imaginary mass* is made by *imaginary motion*. This is not only a simple generalization of the process based on the theory of the *imaginary functions*, but also a fundamental conclusion related to the concept of *imaginary mass* that, as it will be shown, provides a coherent explanation for the *materialization* of the fundamental particles, in the beginning of the Universe.

It is known that the simultaneous disappearance of a pair (electron/positron) liberates an amount of energy, $2m_{i0e(real)}c^2$, under the form of two photons with frequency $f$, in such a way that

$$2m_{i0e(real)}c^2 = 2hf$$

Since the photon has *imaginary* masses associated to it, the phenomenon of transformation of the energy $2m_{i0e(real)}c^2$ into $2hf$ suggests that the imaginary energy of the photon, $m_{ip(imaginary)}c^2$, comes from the transformation of imaginary energy of the electron, $m_{i0e(imaginary)}c^2$, just as the real energy of the photon, $hf$, results from the transformation of real energy of the electron, i.e.,

$$2m_{i0e(imaginary)}c^2 + 2m_{i0e(real)}c^2 = \\ = 2m_{i0p(imaginary)}c^2 + 2hf$$

Then, it follows that

$$m_{i0e(imaginary)} = -m_{ip(imaginary)}$$

The sign (-) in the equation above, is due to the imaginary mass of the *photon* to be *positive*, on the contrary of the imaginary gravitational mass of the *matter*, which is *negative*, as we have already seen.

---

† The Standard Model is the name given to the current theory of fundamental particles and how they interact. This theory includes: *Strong interaction* and a combined theory of weak and electromagnetic interaction, known as *electroweak* theory. One part of the Standard Model is not yet well established. *What causes the fundamental particles to have masses*? The simplest idea is called the *Higgs mechanism*. This mechanism involves one additional particle, called the Higgs boson, and one additional force type, mediated by exchanges of this boson.

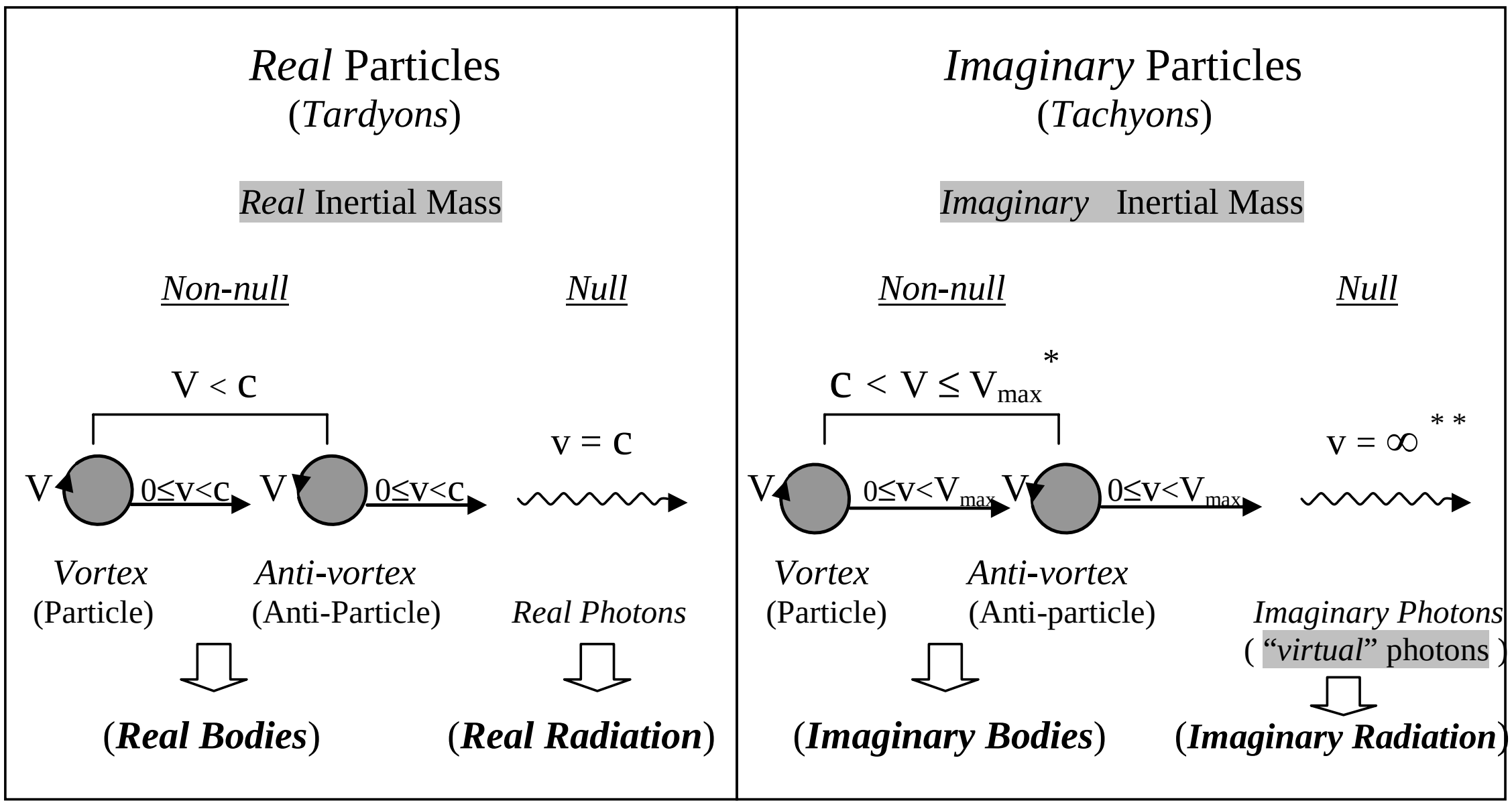

* $V_{max}$ is the *speed upper limit* for Tachyons with *non-null* imaginary inertial mass. It has been previously obtained starting from the *Hubble's law* and Eq.(22). The result is: $V_{max} = \left(\sqrt{3}/2\right)\tilde{H}L_{max} \cong 10^{12}\, m.s^{-1}$.

** In order to communicate instantaneously the *interactions* at infinite distance the velocity of the *quanta* ("virtual" photons) must be *infinity* and consequently their imaginary masses must be *null* .

Fig. VI - Real and Imaginary Particles.

Thus, we then conclude that

$$m_{i0e(imaginary)} = -m_{ip(imaginary)} =$$

$$= -\tfrac{2}{\sqrt{3}}\left(hf_e/c^2\right)\, i =$$

$$= -\tfrac{2}{\sqrt{3}}\left(h/\lambda_e c\right)\, i = -\tfrac{2}{\sqrt{3}}\, m_{i0e(real)} i$$

where $\lambda_e = h/m_{i0e(real)}\, c$ is the *Broglies' wavelength* for the electron.

By analogy, we can write for the *neutron* and the *proton* the following masses:

$$m_{i0neutron(imaginary)} = -\tfrac{2}{\sqrt{3}}\, m_{i0neutron(real)}\ i$$

$$m_{i0\,proton(imaginary)} = +\tfrac{2}{\sqrt{3}}\, m_{i0\,proton(real)}\ i$$

The sign (+) in the expression of $m_{i0\,proton(imaginary)}$ is due to the fact that $m_{i0neutron(imaginary)}$ and $m_{i0\,proton(imaginary)}$ must have contrary signs, as will be shown later on.

Thus, the *electron*, the *neutron* and the *proton* have respectively, the following masses:

*Electron*

$$m_{i0e(real)} = 9.11\times10^{-31}kg$$

$$m_{i0e(im)} = -\tfrac{2}{\sqrt{3}}\, m_{i0e(real)} i$$

$$m_{ge(real)} = \left\{1 - 2\left[\sqrt{1 + \left(\frac{U_{e(real)}}{m_{i0e(real)}c^2}\right)^2} - 1\right]\right\} m_{i0e(real)} =$$

$$= \chi_e m_{i0e(real)}$$

$$m_{ge(im)} = \left\{1 - 2\left[\sqrt{1 + \left(\frac{U_{e(im)}}{m_{i0e(im)}c^2}\right)^2} - 1\right]\right\} m_{i0e(im)} =$$

$$= \chi_e m_{i0e(im)}$$

### Neutron

$m_{i0n(real)} = 1.6747 \times 10^{-27} kg$

$m_{i0n(im)} = -\frac{2}{\sqrt{3}} m_{i0n(real)}\ i$

$$m_{gn(real)} = \left\{1 - 2\left[\sqrt{1 + \left(\frac{U_{n(real)}}{m_{i0n(real)}c^2}\right)^2} - 1\right]\right\} m_{i0n(real)} = $$
$$= \chi_n m_{i0n(real)}$$

$$m_{gn(im)} = \left\{1 - 2\left[\sqrt{1 + \left(\frac{U_{n(im)}}{m_{i0n(im)}c^2}\right)^2} - 1\right]\right\} m_{i0n(im)} = $$
$$= \chi_n m_{i0n(im)}$$

### Proton

$m_{i0pr(real)} = 1.6723 \times 10^{-27} kg$

$m_{i0pr(im)} = +\frac{2}{\sqrt{3}} m_{i0pr(real)}\ i$

$$m_{gpr(real)} = \left\{1 - 2\left[\sqrt{1 + \left(\frac{U_{pr(real)}}{m_{i0pr(real)}c^2}\right)^2} - 1\right]\right\} m_{i0pr(real)} = $$
$$= \chi_{pr} m_{i0pr(real)}$$

$$m_{gpr(im)} = \left\{1 - 2\left[\sqrt{1 + \left(\frac{U_{pr(im)}}{m_{i0pr(im)}c^2}\right)^2} - 1\right]\right\} m_{i0pr(im)} = $$
$$= \chi_{pr} m_{i0pr(im)}$$

where $U_{(real)}$ and $U_{(im)}$ are respectively, the *real* and *imaginary* energies absorbed by the particles.

When *neutrons*, *protons* and *electrons* were created after the Big-bang, they absorbed quantities of electromagnetic energy, respectively given by

$U_{n(real)} = \eta_n kT_n \qquad U_{n(imaginary)} = \eta_n kT_n\ i$

$U_{pr(real)} = \eta_{pr} kT_{pr} \qquad U_{pr(imaginary)} = \eta_{pr} kT_{pr}\ i$

$U_{e(real)} = \eta_e kT_e \qquad U_{e(imaginary)} = \eta_e kT_e\ i$

where $\eta_n$, $\eta_{pr}$ and $\eta_e$ are the *absorption factors* respectively, for the neutrons, protons and electrons; $k = 1.38 \times 10^{-23} J/{}^\circ K$ is the *Boltzmann constant*; $T_n$, $T_{pr}$ and $T_e$ are the temperatures of the Universe, respectively when neutrons, protons and electrons were created.

In the case of the electrons, it was previously shown that $\eta_e \cong 0.1$. Thus, by considering that $T_e \cong 6.2 \times 10^{31} K$, we get

$$U_{e(im)} = \eta_e kT_e\ i = 8.5 \times 10^{7}\ i$$

It is known that the protons were created at the same epoch. Thus, we will assume that

$$U_{pr(im)} = \eta_{pr} kT_{pr}\ i = 8.5 \times 10^{7}\ i$$

Then, it follows that

$$\chi_e = -1.8 \times 10^{21}$$
$$\chi_{pr} = -9.7 \times 10^{17}$$

Now, consider the gravitational forces, due to the *imaginary masses* of *two electrons,* $F_{ee}$, *two protons,* $F_{prpr}$, and *one electron and one proton,* $F_{epr}$, all at rest.

$$F_{ee} = -G\frac{m^2_{ge(im)}}{r^2} = -G\chi_e^2\frac{\left(-\frac{2}{\sqrt{3}} m_{i0e(real)} i\right)^2}{r^2} = $$
$$= +\frac{4}{3}G\chi_e^2\frac{m^2_{i0e(real)}}{r^2} = \frac{+2.3 \times 10^{-28}}{r^2} \quad (repulsion)$$

$$F_{prpr} = -G\frac{m^2_{gpr(im)}}{r^2} = -G\chi_{pr}^2\frac{\left(+\frac{2}{\sqrt{3}} m_{i0pr(real)} i\right)^2}{r^2} = $$
$$= +\frac{4}{3}G\chi_{pr}^2\frac{m^2_{i0pr(real)}}{r^2} = \frac{+2.3 \times 10^{-28}}{r^2} \quad (repulsion)$$

$$F_{epr} = -G\frac{m_{ge(im)} m_{gpr(im)}}{r^2} = $$
$$= -G\chi_e\chi_{pr}\frac{\left(-\frac{2}{\sqrt{3}} m_{i0e(real)} i\right)\left(+\frac{2}{\sqrt{3}} m_{i0pr(real)} i\right)}{r^2} = $$
$$= -\frac{4}{3}G\chi_e\chi_{pr}\frac{m_{i0e(real)} m_{i0pr(real)}}{r^2} = \frac{-2.3 \times 10^{-28}}{r^2}$$
$$(atraction)$$

Note that

$$F_{electric} = \frac{e^2}{4\pi\varepsilon_0 r^2} = \frac{2.3\times10^{-28}}{r^2}$$

Therefore, we can conclude that

$$F_{ee} = F_{prpr} \equiv F_{electric} = +\frac{e^2}{4\pi\varepsilon_0 r^2} \quad (repulsion)$$

and

$$F_{ep} \equiv F_{electric} = -\frac{e^2}{4\pi\varepsilon_0 r^2} \quad (atraction)$$

These correlations permit to define the *electric charge* by means of the following relation:

$$q = \sqrt{4\pi\varepsilon_0 G}\; m_{g(imaginary)}\, i$$

For example, in the case of the *electron*, we have

$$q_e = \sqrt{4\pi\varepsilon_0 G}\; m_{ge(imaginary)}\, i =$$
$$= \sqrt{4\pi\varepsilon_0 G}\left(\chi_e m_{i0e(imaginary)} i\right) =$$
$$= \sqrt{4\pi\varepsilon_0 G}\left(-\chi_e \tfrac{2}{\sqrt{3}} m_{i0e(real)} i^2\right) =$$
$$= \sqrt{4\pi\varepsilon_0 G}\left(\chi_e \tfrac{2}{\sqrt{3}} m_{i0e(real)}\right) = -1.6\times10^{-19} C$$

In the case of the *proton*, we get

$$q_{pr} = \sqrt{4\pi\varepsilon_0 G}\; m_{gpr(imaginary)}\, i =$$
$$= \sqrt{4\pi\varepsilon_0 G}\left(\chi_{pr} m_{i0pr(imaginary)} i\right) =$$
$$= \sqrt{4\pi\varepsilon_0 G}\left(+\chi_{pr} \tfrac{2}{\sqrt{3}} m_{i0pr(real)} i^2\right) =$$
$$= \sqrt{4\pi\varepsilon_0 G}\left(-\chi_{pr} \tfrac{2}{\sqrt{3}} m_{i0pr(real)}\right) = +1.6\times10^{-19} C$$

For the *neutron*, it follows that

$$q_n = \sqrt{4\pi\varepsilon_0 G}\; m_{gn(imaginary)}\, i =$$
$$= \sqrt{4\pi\varepsilon_0 G}\left(\chi_n m_{i0n(imaginary)} i\right) =$$
$$= \sqrt{4\pi\varepsilon_0 G}\left(-\chi_n \tfrac{2}{\sqrt{3}} m_{i0n(real)} i^2\right) =$$
$$= \sqrt{4\pi\varepsilon_0 G}\left(\chi_n \tfrac{2}{\sqrt{3}} m_{i0n(real)}\right)$$

However, based on the *quantization of the mass* (Eq. 44), we can write that

$$\chi_n \tfrac{2}{\sqrt{3}} m_{i0n(real)} = n^2 m_{i0(\min)} \qquad n \neq 0$$

Since $n$ can have only discrete values *different of zero* (See Appendix B), we conclude that $\chi_n$ cannot be null. However, it is known that the electric charge of the neutron is *null*. Thus, it is necessary to assume that

$$q_n = q_n^+ + q_n^- = \sqrt{4\pi\varepsilon_0 G}\; m^+_{gn(imaginary)}\, i +$$
$$+ \sqrt{4\pi\varepsilon_0 G}\; m^-_{gn(imaginary)}\, i =$$
$$= \sqrt{4\pi\varepsilon_0 G}\left(\chi_n m^+_{i0n(imaginary)}\, i\right) +$$
$$+ \sqrt{4\pi\varepsilon_0 G}\left(\chi_n m^-_{i0n(imaginary)}\, i\right) =$$
$$= \sqrt{4\pi\varepsilon_0 G}\left[\chi_n\left(+\tfrac{2}{\sqrt{3}} m_{i0n} i^2\right) + \chi_n\left(-\tfrac{2}{\sqrt{3}} m_{i0n} i^2\right)\right] = 0$$

We then conclude that in the neutron, *half* of the total amount of *elementary quanta of electric charge*, $q_{\min}$, is *negative*, while the other *half* is *positive*.

In order to obtain the value of the *elementary quantum of electric charge*, $q_{\min}$, we start with the expression obtained here for the electric charge, where we change $m_{g(imaginary)}$ by its quantized expression $m_{g(imaginary)} = n^2 m_{i0(imaginary)(\min)}$, derived from Eq. (44a). Thus, we get

$$q = \sqrt{4\pi\varepsilon_0 G}\; m_{g(imaginary)}\, i =$$
$$= \sqrt{4\pi\varepsilon_0 G}\; n^2 m_{i0(imaginary)(\min)}\, i =$$
$$= \sqrt{4\pi\varepsilon_0 G}\; \left[n^2\left(\pm \tfrac{2}{\sqrt{3}} m_{i0(\min)} i\right)\right] i =$$
$$= \mp \tfrac{2}{\sqrt{3}} \sqrt{4\pi\varepsilon_0 G}\; n^2 m_{i0(\min)}$$

This is the *quantized expression of the electric charge*.

For $n = 1$ we obtain the value of the *elementary quantum of electric charge*, $q_{\min}$, i.e.,

$$q_{\min} = \mp \tfrac{2}{\sqrt{3}} \sqrt{4\pi\varepsilon_0 G}\; m_{i0(\min)} = \mp 3.8\times10^{-83} C$$

where $m_{i0(\min)}$ is the *elementary quantum* of matter, whose value previously calculated, is $m_{i0(\min)} = \pm 3.9\times10^{-73} kg$.

The existence of *imaginary* mass associated to a *real* particle suggests the possible existence of *imaginary*

*particles* with imaginary masses in Nature.

In this case, the concept of *wave associated* to a particle (De Broglie's waves) would also be applied to the imaginary particles. Then, by analogy, the imaginary wave associated to an imaginary particle with imaginary masses $m_{i\psi}$ and $m_{g\psi}$ would be described by the following expressions

$$\vec{p}_\psi = \hbar \vec{k}_\psi$$

$$E_\psi = \hbar \omega_\psi$$

Henceforth, for the sake of simplicity, we will use the Greek letter $\psi$ to stand for the word *imaginary*; $\vec{p}_\psi$ is the *momentum* carried by the $\psi$ *wave* and $E_\psi$ its energy; $\left|\vec{k}_\psi\right| = 2\pi/\lambda_\psi$ is the propagation number and $\lambda_\psi$ the wavelength of the $\psi$ *wave;* $\omega_\psi = 2\pi f_\psi$ is the cyclical frequency.

According to Eq. (4), the *momentum* $\vec{p}_\psi$ is

$$\vec{p}_\psi = M_{g\psi}\vec{V}$$

where $V$ is the velocity of the $\psi$ particle.

By comparing the expressions of $\vec{p}_\psi$ we get

$$\lambda_\psi = \frac{h}{M_{g\psi}V}$$

It is known that the variable quantity which characterizes the De Broglie's waves is called *wave function,* usually indicated by symbol $\Psi$. The wave function associated with a material particle describes the dynamic state of the particle: its value at a particular point x, y, z, t is related to the probability of finding the particle in that place and instant. Although $\Psi$ does not have a physical interpretation, its square $\Psi^2$ (or $\Psi\ \Psi^*$) calculated for a particular point x, y, z, t *is proportional to the probability of finding the particle in that place and instant*.

Since $\Psi^2$ is proportional to the probability $P$ of finding the particle described by $\Psi$, the integral of $\Psi^2$ on the *whole space* must be finite – inasmuch as the particle is somewhere.

On the other hand, if

$$\int_{-\infty}^{+\infty} \Psi^2 dV = 0$$

the interpretation is that the particle will not exist. However, if

$$\int_{-\infty}^{+\infty} \Psi^2 dV = \infty \qquad (108)$$

*The particle will be everywhere simultaneously*.

In Quantum Mechanics, the wave function $\Psi$ corresponds, as we know, to the displacement $y$ of the undulatory motion of a rope. However, $\Psi$, as opposed to $y$, is not a measurable quantity and can, hence, be a complex quantity. For this reason, it is assumed that $\Psi$ is described in the $x-direction$ by

$$\Psi = \Psi_0 e^{-(2\pi i/h)(Et-px)}$$

This is the expression of the wave function for a *free* particle, with total energy $E$ and *momentum* $\vec{p}$, moving in the direction $+x$.

As to the imaginary particle, the *imaginary particle wave function* will be denoted by $\Psi_\psi$ and, by analogy the expression of $\Psi$, will be expressed by:

$$\Psi_\psi = \Psi_{0\psi} e^{-(2\pi i/h)(E_\psi t - p_\psi x)}$$

Therefore, the *general expression* of the wave function for a *free* particle can be written in the following form

$$\Psi = \Psi_{0(real)} e^{-(2\pi i/h)(E_{(real)}t - p_{(real)}x)} + \\ + \Psi_{0\psi} e^{-(2\pi i/h)(E_\psi t - p_\psi x)}$$

It is known that the *uncertainty principle* can also be written as a function of $\Delta E$ (uncertainty in the energy) and $\Delta t$ (uncertainty in the time), i.e.,

$$\Delta E.\Delta t \geq \hbar$$

This expression shows that a variation of energy $\Delta E$, during a time interval $\Delta t$, can only be detected if $\Delta t \geq \hbar/\Delta E$. Consequently, a variation of energy $\Delta E$, during a time interval $\Delta t < \hbar/\Delta E$, cannot be experimentally detected. This is a limitation imposed by Nature and not by our equipments.

Thus, a *quantum* of energy $\Delta E = hf$ that varies during a time interval $\Delta t = 1/f = \lambda/c < \hbar/\Delta E$ (wave period) cannot be experimentally detected. This is an *imaginary* photon or a "*virtual*" photon.

Now, consider a particle with energy $M_g c^2$. The DeBroglie's gravitational and inertial wavelengths are respectively $\lambda_g = h/M_g\, c$ and $\lambda_i = h/M_i\, c$. In Quantum Mechanics, particles of matter and quanta of radiation are described by means of *wave packet* (DeBroglie's waves) with average wavelength $\lambda_i$. Therefore, we can say that during a time interval $\Delta t = \lambda_i/c$, a *quantum* of energy $\Delta E = M_g c^2$ varies. According to the uncertainty principle, the particle will be detected if $\Delta t \geq \hbar/\Delta E$, i.e., if $\lambda_i/c \geq \hbar/M_g c^2$ or $\lambda_i \geq \lambda_g/2\pi$. This condition is usually satisfied when $M_g = M_i$. In this case, $\lambda_g = \lambda_i$ and obviously, $\lambda_i > \lambda_i/2\pi$. However, when $M_g$ decreases $\lambda_g$ increases and $\lambda_g/2\pi$ can become bigger than $\lambda_i$, making the particle *non-detectable* or *imaginary*.

According to Eqs. (7) and (41) we can write $M_g$ in the following form:

$$M_g = \frac{m_g}{\sqrt{1-V^2/c^2}} = \frac{\chi\ m_i}{\sqrt{1-V^2/c^2}} = \chi M_i$$

where

$$\chi = \left\{1 - 2\left[\sqrt{1+\left(\Delta p/m_{i0}c\right)^2} - 1\right]\right\}$$

Since the condition to make the particle *imaginary* is

$$\lambda_i < \frac{\lambda_g}{2\pi}$$

and

$$\frac{\lambda_g}{2\pi} = \frac{\hbar}{M_g c} = \frac{\hbar}{\chi M_i c} = \frac{\lambda_i}{2\pi\chi}$$

Then we get

$$\chi < \frac{1}{2\pi} = 0.159$$

However, $\chi$ can be *positive* or *negative* ($\chi < +0.159$ or $\chi > -0.159$).This means that when

$$-0.159 < \chi < +0.159$$

the particle becomes *imaginary*. Under these circumstances, we can say that the particle made a transition to the *imaginary space-time*.

Note that, when a particle becomes imaginary, its gravitational and inertial masses also become imaginary. However, the factor $\chi = M_{g(imaginary)}/M_{i(imaginary)}$ remains *real* because

$$\chi = \frac{M_{g(imaginary)}}{M_{i(imaginary)}} = \frac{M_g i}{M_i i} = \frac{M_g}{M_i} = real$$

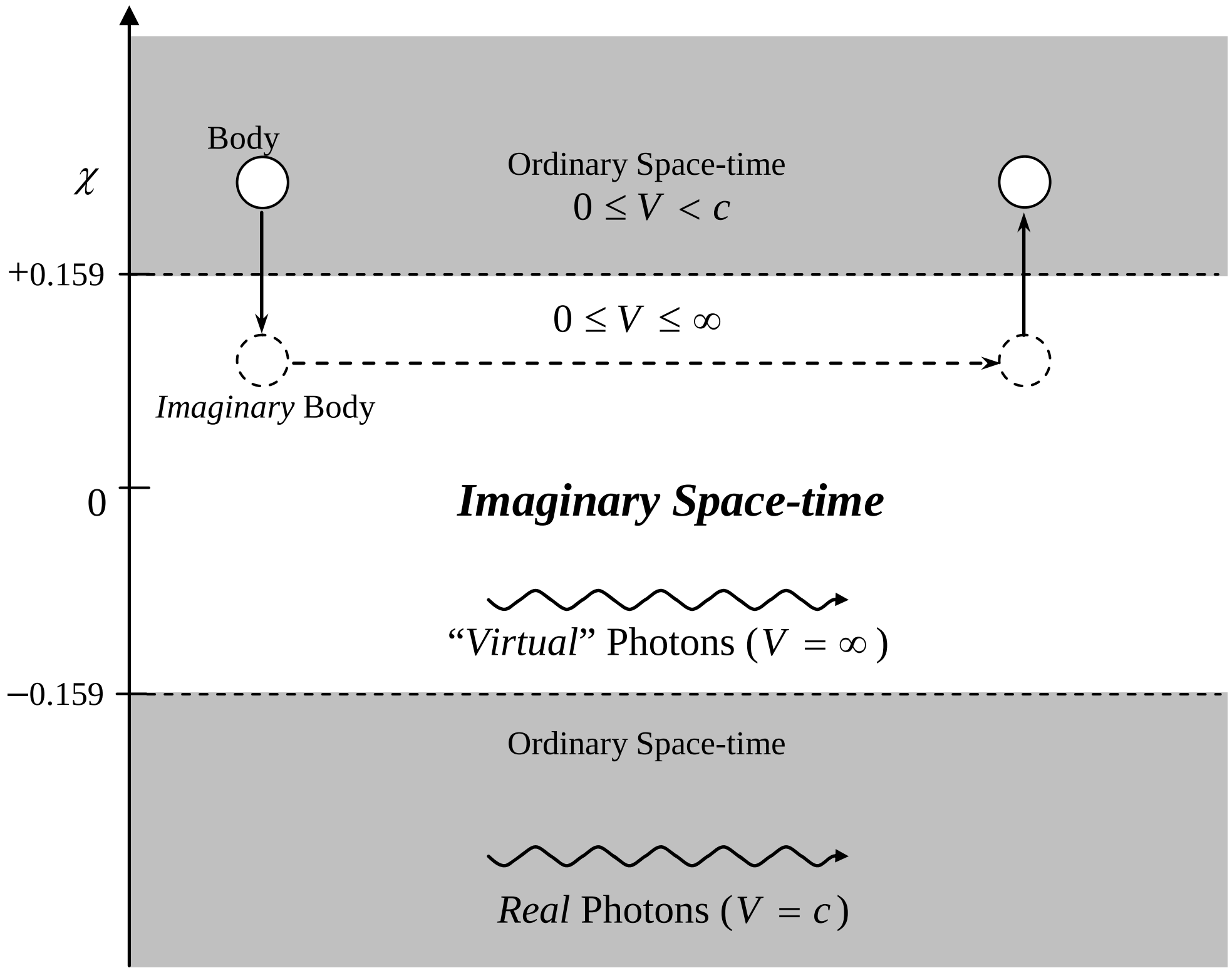

Fig. VII – *Travel in the imaginary space-time*. Similarly to the "*virtual*" photons, *imaginary* bodies can have *infinite speed* in the *imaginary space-time*.

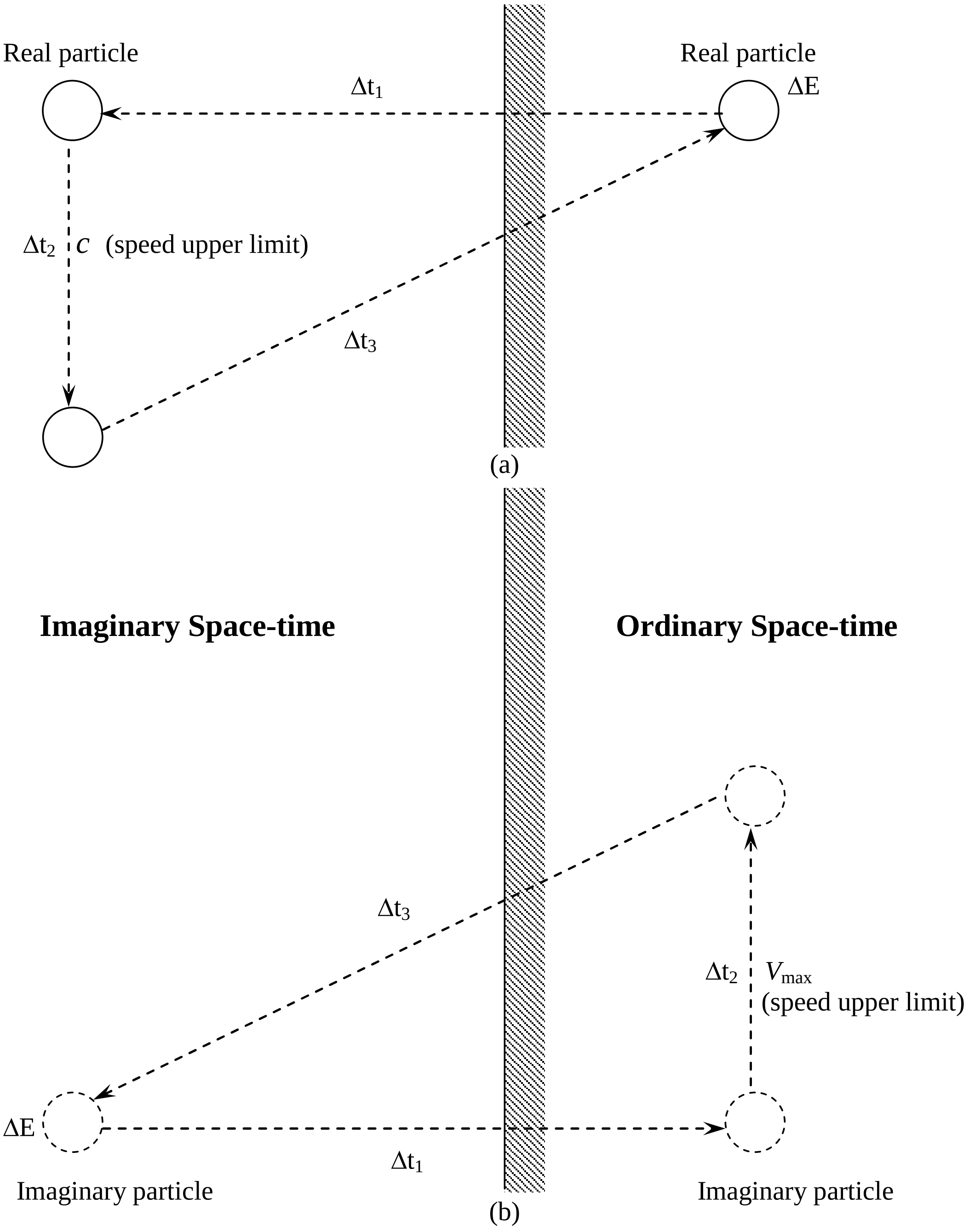

Fig. VIII – *"Virtual" Transitions* – (a) "Virtual" Transitions of a *real* particle to the *imaginary* space-time. The speed upper limit for *real* particle in the *imaginary* space-time is $c$.
(b) - "Virtual" Transitions of an *imaginary* particle to the *ordinary* space-time. The speed upper limit for *imaginary* particle in the *ordinary* space-time is $V_{max} \approx 10^{12} m.s^{-1}$

Note that to occur a "virtual" transition it is necessary that $\Delta t = \Delta t_1 + \Delta t_2 + \Delta t_3 < \hbar / \Delta E$ Thus, even at principle, it will be impossible to determine any variation of energy in the particle (*uncertainty principle*).

Thus, if the gravitational mass of the particle is reduced by means of the absorption of an amount of electromagnetic energy $U$, for example, we have

$$\chi = \frac{M_g}{M_i} = \left\{1 - 2\left[\sqrt{1 + \left(U/m_{i0}c^2\right)^2} - 1\right]\right\}$$

This shows that the energy $U$ of the electromagnetic field *remains acting on* the imaginary particle. In practice, this means that *electromagnetic fields act on imaginary particles*.

The gravity acceleration on a *imaginary* particle (due to the rest of the imaginary Universe) are given by

$$g'_j = \chi\ g_j \qquad j = 1,2,3,\ldots,n.$$

where $\chi = M_{g(imaginary)}/M_{i(imaginary)}$ and $g_j = -Gm_{gj(imaginary)}/r_j^2$. Thus, the gravitational forces acting on the particle are given by

$$F_{gj} = M_{g(imaginary)}g'_j = \\ = M_{g(imaginary)}\left(-\chi Gm_{gj(imaginary)}/r_j^2\right) = \\ = M_g i\left(-\chi Gm_{gj}i/r_j^2\right) = +\chi GM_g m_{gj}/r_j^2 .$$

Note that these forces are *real*. Remind that, the Mach's principle says that the *inertial effects* upon a particle are consequence of the gravitational interaction of the particle with the rest of the Universe. Then we can conclude that the *inertial forces* upon an *imaginary* particle are also real.

Equation (7) shows that , in the case of imaginary particles, the relativistic mass is

$$M_{g(imaginary)} = \frac{m_{g(imaginary)}}{\sqrt{1 - V^2/c^2}} = \\ = \frac{m_g i}{i\sqrt{V^2/c^2 - 1}} = \frac{m_g i}{\sqrt{V^2/c^2 - 1}}$$

This expression shows that *imaginary* particles can have velocities $V$ greater than $c$ in our ordinary space-time (Tachyons). The *quantization of velocity* (Eq. 36) shows that there is a speed upper limit $V_{max} > c$. As we have already calculated previously, $V_{max} \approx 10^{12} m.s^{-1}$, (Eq.102).

Note that this is the speed upper limit for *imaginary* particles *in our ordinary space-time* not in the *imaginary* space-time (Fig.7) because the *infinite* speed of the "virtual" *quanta* of the interactions shows that *imaginary* particles can have *infinite speed* in the *imaginary* space-time.

While the speed upper limit for imaginary particles in the ordinary space-time is $V_{max} \approx 10^{12} m.s^{-1}$, the speed upper limit for *real* particles in the *imaginary* space-time is $c$, because the relativistic expression of the mass shows that the velocity of *real* particles cannot be larger than $c$ in *any space-time*. The uncertainty principle permits that particles make "*virtual*" transitions, during a time interval $\Delta t$, if $\Delta t < \hbar/\Delta E$. The "*virtual*" transition of *mesons* emitted from nucleons that do not change of mass, during a time interval $\Delta t < \hbar/m_\pi c^2$, is a well-known example of "*virtual*" transition of particles. During a "virtual" transition of a *real* particle, the speed upper limit in the *imaginary* space-time is $c$, while the speed upper limit for an *imaginary* particle in the our ordinary space-time is $V_{max} \approx 10^{12} ms^{-1}$. (See Fig. 8).

There is a crucial cosmological problem to be solved: the problem of the *hidden mass*. Most theories predict that the amount of known matter, detectable and available in the universe, is only about 1/100 to 1/10 of the amount needed to close the universe. That is, to achieve the density sufficient to close-up the universe by maintaining the gravitational curvature (escape velocity equal to the speed of light) at the outer boundary.

Eq. (43) may solve this problem. We will start by substituting the expression of *Hubble's* law for velocity, $V = \tilde{H}l$, into Eq.(43). The expression obtained shows that particles which are at distances $l = l_0 = \left(\sqrt{5}/3\right)\left(c/\tilde{H}\right) = 1.3\times10^{26} m$ have *quasi null* gravitational mass $m_g = m_{g(min)}$; beyond this distance, the particles have *negative* gravitational mass. Therefore, there are two well-defined regions in the Universe; the region of the bodies with *positive* gravitational masses and the region of the bodies with *negative* gravitational mass. The total gravitational mass of the first region, in accordance with Eq.(45), will be given by

$$M_{g1} \cong M_{i1} = \frac{m_{i1}}{\sqrt{1-\bar{V}_1^2/c^2}} \cong m_{i1}$$

where $m_{i1}$ is the total *inertial mass* of the bodies of the mentioned region; $\bar{V}_1 << c$ is the average velocity of the bodies at region 1. The total gravitational mass of the second region is

$$M_{g2} = \left|1 - 2\left(\frac{1}{\sqrt{1-\bar{V}_2^2/c^2}} - 1\right)\right| M_{i2}$$

where $\bar{V}_2$ is the average velocity of the bodies ; $M_{i2} = m_{i2}\Big/\sqrt{1-\bar{V}_2^2/c^2}$ and $m_{i2}$ is the total *inertial mass* of the bodies of region 2.

Now consider that from Eq.(7), we can write

$$\xi = \frac{E_g}{V} = \frac{M_g c^2}{V} = \rho_g c^2$$

where $\xi$ is the *energy density* of matter.

Note that the expression of $\xi$ only reduces to the well-known expression $\rho c^2$, where $\rho$ is the sum of the inertial masses per volume unit, when $m_g = m_i$. Therefore, in the derivation of the well-known difference

$$\frac{8\pi G\rho_U}{3} - \tilde{H}^2$$

which gives the *sign of the curvature* of the Universe [36], we must use $\xi = \rho_{gU} c^2$ instead of $\xi = \rho_U c^2$.The result obviously is

$$\frac{8\pi G\rho_{gU}}{3} - \tilde{H}^2 \qquad (109)$$

where

$$\rho_{gU} = \frac{M_{gU}}{V_U} = \frac{M_{g1} + M_{g2}}{V_U} \qquad (110)$$

$M_{gU}$ and $V_U$ are respectively the total gravitational mass and the volume of the Universe.

Substitution of $M_{g1}$ and $M_{g2}$ into expression (110) gives

$$\rho_{gU} = \frac{m_{iU} + \left[\left|\frac{3}{\sqrt{1-\bar{V}_2^2/c^2}} - \frac{2}{1-\bar{V}_2^2/c^2}\right| m_{i2} - m_{i2}\right]}{V_U}$$

where $m_{iU} = m_{i1} + m_{i2}$ is the total inertial mass of the Universe.

The volume $V_1$ of the region 1 and the volume $V_2$ of the region 2, are respectively given by

$$V_1 = 2\pi^2 l_0^3 \qquad and \qquad V_2 = 2\pi^2 l_c^3 - V_1$$

where $l_c = c/\tilde{H} = 1.8\times10^{26}\,m$ is the so-called "radius" of the visible Universe. Moreover, $\rho_{i1} = m_{i1}/V_1$ and $\rho_{i2} = m_{i2}/V_2$. Due to the *hypothesis of the uniform distribution of matter in the space*, it follows that $\rho_{i1} = \rho_{i2}$.Thus, we can write

$$\frac{m_{i1}}{m_{i2}} = \frac{V_1}{V_2} = \left(\frac{l_0}{l_c}\right)^3 = 0.38$$

Similarly,

$$\frac{m_{iU}}{V_U} = \frac{m_{i2}}{V_2} = \frac{m_{i1}}{V_1}$$

Therefore,

$$m_{i2} = \frac{V_2}{V_U} m_{iU} = \left[1 - \left(\frac{l_0}{l_c}\right)^3\right] m_{iU} = 0.62 m_{iU}$$

and $m_{i1} = 0.38 m_{iU}$.

Substitution of $m_{i2}$ into the expression of $\rho_{gU}$ yields

$$\rho_{gU} = \frac{m_{iU} + \left|\frac{1.86}{\sqrt{1-\overline{V}_2^2/c^2}} - \frac{1.24}{1-\overline{V}_2^2/c^2} - 0.62\right| m_{iU}}{V_U}$$

Due to $\overline{V}_2 \cong c$, we conclude that the term between bracket *is much larger than* $10 m_{iU}$. The amount $m_{iU}$ is the mass of matter in the universe (1/10 to 1/100 of the amount needed to close the Universe).

Consequently, the total mass

$$m_{iU} + \left|\frac{1.86}{\sqrt{1-\overline{V}_2^2/c^2}} - \frac{1.24}{1-\overline{V}_2^2/c^2} - 0.62\right| m_{iU}$$

must be sufficient to close the Universe.

There is another cosmological problem to be solved: the problem of the *anomalies* in the spectral red-shift of certain galaxies and stars.

Several observers have noticed red-shift values that cannot be explained by the Doppler-Fizeau effect or by the Einstein effect (the gravitational spectrum shift, supplied by Einstein's theory).

This is the case of the so-called *Stefan's quintet* (a set of five galaxies which were discovered in 1877), whose galaxies are located at approximately the same distance from the Earth, according to very reliable and precise measuring methods. But, when the velocities of the galaxies are measured by its red-shifts, the velocity of one of them is much larger than the velocity of the others.

Similar observations have been made on the *Virgo constellation* and spiral galaxies. Also the Sun presents a red-shift greater than the predicted value by the Einstein effect.

It seems that some of these anomalies can be explained if we consider the Eq.(45) in the calculation of the *gravitational mass* of the point of emission.

The expression of the gravitational spectrum shift was previously obtained in this work. It is the same supplied by Einstein's theory [37], and is given by

$$\Delta\omega = \omega_1 - \omega_2 = \frac{\phi_2 - \phi_1}{c^2}\omega_0 = = \frac{-Gm_{g2}/r_2 + Gm_{g1}/r_1}{c^2}\omega_0 \qquad (111)$$

where $\omega_1$ is the frequency of the light at the point of emission ; $\omega_2$ is the frequency at the point of observation; $\phi_1$ and $\phi_2$ are respectively, the Newtonian gravitational potentials at the point of emission and at the point of observation.

In Einstein theory, this expression has been deduced from $T = t\sqrt{-g_{00}}$ [38] which correlates *own time* (real time), $t$, with the temporal coordinate $x^0$ of the space-time ( $t = x^0/c$ ).

When the gravitational field is *weak*, the temporal component $g_{00}$ of the metric tensor is given by $g_{oo} = -1 - 2\phi/c^2$[39].Thus, we readily obtain

$$T = t\sqrt{1 - 2Gm_g/rc^2} \qquad (112)$$

This is the same equation that we have obtained previously in this work.

Curiously, this equation tell us that we can have $T < t$ when $m_g > 0$ ; and $T > t$ for $m_g < 0$. In addition, if $m_g = c^2 r/2G$, i.e., if $r = 2Gm_g/c^2$ (*Schwarzschild radius*) we obtain $T$=0.

Let us now consider the well-known process of stars' *gravitational contraction*. It is known that the destination of the star is directly correlated to its mass. If the star's mass is less than $1.4M_{\odot}$ (Schemberg-Chandrasekhar's limit), it becomes a *white dwarf*. If its mass exceeds that limit, the pressure produced by the degenerate state of the matter no longer counterbalances the gravitational pressure, and the star's contraction continues. Afterwards there occurs the reactions between protons and electrons (capture of electrons), where *neutrons* and anti-neutrinos are produced.

The contraction continues until the system regains stability (when the pressure produced by the neutrons is sufficient to stop the gravitational collapse). Such systems are called *neutron stars*.

There is also a critical mass for the stable configuration of neutron stars. This limit has not been fully defined as yet, but it is known that it is located between $1.8M_{\odot}$ and $2.4M_{\odot}$. Thus, if the mass of the star exceeds $2.4M_{\odot}$ , the contraction will continue.

According to Hawking [40] collapsed objects cannot have mass less than $\sqrt{\hbar c/4G} = 1.1\times10^{-8}kg$. This means that, with the progressing of the compression, the neutrons cluster must become a cluster of *superparticles* where the *minimal inertial mass* of the superparticle is

$$m_{i(sp)} = 1.1\times10^{-8}kg. \qquad (113)$$

*Symmetry* is a fundamental attribute of the Universe that enables an investigator to study particular aspects of physical systems by themselves. For example, the assumption that space is homogeneous and isotropic is based on *Symmetry Principle*. Also here, by symmetry, we can assume that there are only *superparticles* with mass $m_{i(sp)} = 1.1\times10^{-8}kg$ in the cluster of *superparticles*.

Based on the mass-energy of the superparticles ( ~$10^{18}$ GeV ) we can say that they belong to a putative class of particles with mass-energy beyond the *supermassive* Higgs bosons ( the so-called X bosons). It is known that the GUT's theories predict an entirely new force mediated by a new type of boson, called simply X (or X boson ). The X bosons carry both electromagnetic and color charge, in order to ensure proper conservation of those charges in any interactions. The X bosons must be extremely massive, with mass-energy in the unification range of about $10^{16}$ GeV.

If we assume the superparticles *are not hyper*massive Higgs bosons then the possibility of the *neutrons cluster* become a *Higgs bosons cluster* before becoming a *superparticles cluster* must be considered. On the other hand, the fact that superparticles must be so massive also means that it is not possible to create them in any conceivable particle accelerator that could be built. They can exist as free particles only at a very early stage of the Big Bang from which the universe emerged.

Let us now imagine the Universe coming back to the past. There will be an instant in which it will be similar to a *neutrons cluster*, such as the stars at the final state of gravitational contraction. Thus, with the progressing of the compression, the *neutrons* cluster becomes a superparticles cluster. Obviously, this only can occur before $10^{-23}$s (after the Big-Bang).

The temperature T of the Universe at the $10^{-43}$s< t < $10^{-23}$s period can be calculated by means of the well-known expression[41]:

$$T \approx 10^{22}\left(t/10^{-23}\right)^{-1/2} \qquad (114)$$

Thus at $t \cong 10^{-43}s$ (at the *first* spontaneous breaking of symmetry) the temperature was $T \approx 10^{32}K$ (~$10^{19}$GeV).Therefore, we can assume that the absorbed electromagnetic energy by each *superparticle*, before $t \cong 10^{-43}s$, was $U = \eta kT > 1\times10^{9}J$ (see Eqs.(71) and (72)). By comparing with $m_{i(sp)}c^2 \cong 9\times10^{8}J$, we conclude that $U > m_{i(sp)}c$. Therefore, *the unification condition* $\left(U\eta_r \cong M_i c^2 > m_i c^2\right)$ is satisfied. This means that, *before* $t \cong 10^{-43}s$ *,the gravitational and electromagnetic interactions were unified*.

From the *unification condition* $\left(U\eta_r \cong M_i c^2\right)$, we may conclude that the superparticles' *relativistic inertial mass* $M_{i(sp)}$ is

$$M_{i(sp)} \cong \frac{U\eta_r}{c^2} = \frac{\eta n_r kT}{c^2} \approx 10^{-8}kg \qquad (115)$$

Comparing with the superparticles' *inertial mass at rest* (113), we conclude that

$$M_{i(sp)} \approx m_{i(sp)} = 1.1\times10^{-8}kg \qquad (116)$$

From Eqs.(83) and (115), we obtain the superparticle's *gravitational mass at rest*:

$$m_{g(sp)} = m_{i(sp)} - 2M_{i(sp)} \cong$$
$$\cong -M_{i(sp)} \cong -\frac{\eta n_r kT}{c^2} \qquad (117)$$

Consequently, the superparticle's *relativistic gravitational mass*, is

$$M_{g(sp)} = \frac{m_{g(sp)}}{\sqrt{1-V^2/c^2}} =$$
$$= \frac{\eta n_r kT}{c^2\sqrt{1-V^2/c^2}} \qquad (118)$$

Thus, the gravitational forces between two *superparticles* , according to (13), is given by:

$$\vec{F}_{12} = -\vec{F}_{21} = -G\frac{M_{g(sp)}M'_{g(sp)}}{r^2}\hat{\mu}_{21} =$$
$$= \left[\left(\frac{M_{i(sp)}}{m_{i(sp)}}\right)^2\left(\frac{G}{c^5\hbar}\right)\left(\eta n_r kT\right)^2\right]\frac{\hbar c}{r^2}\hat{\mu}_{21} \qquad (119)$$

Due to the *unification* of the gravitational and electromagnetic interactions at that period, we have

$$\vec{F}_{12} = -\vec{F}_{21} = G\frac{M_{g(sp)}M'_{g(sp)}}{r^2}\hat{\mu}_{21} =$$

$$= \left[\left(\frac{M_{i(sp)}}{m_{i(sp)}}\right)^2\left(\frac{G}{c^5\hbar}\right)(\eta\kappa T)^2\right]\frac{\hbar c}{r^2}\hat{\mu}_{21} =$$

$$= \frac{e^2}{4\pi\varepsilon_0 r^2} \qquad (120)$$

From the equation above we can write

$$\left[\left(\frac{M_{i(sp)}}{m_{i(sp)}}\right)^2\left(\frac{G}{c^5\hbar}\right)(\eta\kappa T)^2\hbar c\right] = \frac{e^2}{4\pi\varepsilon_0} \qquad (121)$$

Now assuming that

$$\left(\frac{M_{i(sp)}}{m_{i(sp)}}\right)^2\left(\frac{G}{c^5\hbar}\right)(\eta\kappa T)^2 = \psi \qquad (122)$$

the Eq. (121) can be rewritten in the following form:

$$\psi = \frac{e^2}{4\pi\varepsilon_0\hbar c} = \frac{1}{137} \qquad (123)$$

which is the well-known *reciprocal fine structure constant.*

For $T = 10^{32}K$ the Eq.(122) gives

$$\psi = \left(\frac{M_{i(sp)}}{m_{i(sp)}}\right)^2\left(\frac{G}{c^5\hbar}\right)(\eta n_r \kappa T)^2 \approx \frac{1}{100} \qquad (124)$$

This value has the same order of magnitude as the exact value(1/137) of the *reciprocal fine structure constant*.

From equation (120) we can write:

$$\left(G\frac{M_{g(sp)}M'_{g(sp)}}{\psi c\vec{r}}\right)\vec{r} = \hbar \qquad (125)$$

The term between parentheses has the same dimensions as the *linear momentum* $\vec{p}$. Thus, (125) tells us that

$$\vec{p}\cdot\vec{r} = \hbar. \qquad (126)$$

A component of the momentum of a particle cannot be precisely specified without loss of all knowledge of the corresponding component of its position at that time ,i.e., a particle cannot be precisely located in a particular direction without loss of all knowledge of its momentum component in that direction . This means that in intermediate cases the product of the uncertainties of the simultaneously measurable values of corresponding position and momentum components is *at least of the magnitude order* of $\hbar$,

$$\Delta p.\Delta r \geq \hbar \qquad (127)$$

This relation, *directly obtained here from the Unified Theory*, is the well-known relation of the *Uncertainty Principle* for position and momentum.

According to Eq.(83), the gravitational mass of the superparticles at the *center* of the cluster becomes *negative* when $2\eta n_r kT/c^2 > m_{i(sp)}$, i.e., when

$$T > T_{critical} = \frac{m_{i(sp)}c^2}{2\eta n_r k} \approx 10^{32}K.$$

According to Eq. (114) this temperature corresponds to $t_c \approx 10^{43}s$. With the progressing of the compression, more superparticles into the center will have *negative* gravitational mass. Consequently, there will be a critical point in which the *repulsive* gravitational forces between the superparticles with negative gravitational masses and the superparticles with positive gravitational masses will be so strong that an explosion will occur. This is the event that we call the Big Bang.

Now, starting from the Big Bang to the present time. Immediately after the Big Bang, the superparticles' *decompression*

begins. The gravitational mass of the most central superparticle will only be positive when the temperature becomes smaller than the critical temperature, $T_{critical} \approx 10^{32} K$. At the maximum state of compression (exactly at the Big Bang) the volumes of the superparticles were equal to the elementary volume $\Omega_0 = \delta_V d^3_{min}$ and the volume of the Universe was $\Omega = \delta_V (n d_{min})^3 = \delta_V d^3_{initial}$ where $d_{initial}$ was the *initial* length scale of the Universe. At this very moment the *average* density of the Universe was equal to the *average* density of the superparticles, thus we can write

$$\left(\frac{d_{initial}}{d_{min}}\right)^3 = \frac{M_{i(U)}}{m_{i(sp)}} \qquad (128)$$

where $M_{i(U)} \approx 10^{53} kg$ is the inertial mass of the Universe. It has already been shown that $d_{min} = \tilde{k} l_{planck} \approx 10^{-34} m.$ Then, from Eq.(128), we obtain:

$$d_{initial} \approx 10^{-14} m \qquad (129)$$

After the Big Bang the Universe expands itself from $d_{initial}$ up to $d_{cr}$ (when the temperature decrease reaches the critical temperature $T_{critical} \approx 10^{32} K$, and the gravity becomes *attractive*). Thus, it expands by $d_{cr} - d_{initial}$, under effect of the *repulsive* gravity

$$\bar{g} = \sqrt{g_{max} g_{min}} =$$

$$= \sqrt{\left[G\left(\tfrac{1}{2} M_{g(U)}\right) / \left(\tfrac{1}{2} d_{initial}\right)^2\right] \left[G \tfrac{1}{2} M_{i(U)} / \left(\tfrac{1}{2} d_{cr}\right)^2\right]} =$$

$$= \frac{2G\sqrt{M_{g(U)} M_{i(U)}}}{d_{cr} d_{initial}} = \frac{2G\sqrt{\sum m_{g(sp)} M_{i(U)}}}{d_{cr} d_{initial}} =$$

$$= \frac{2G\sqrt{\chi \sum m_{i(sp)} M_{i(U)}}}{d_{cr} d_{initial}} = \frac{2G M_{i(U)} \sqrt{\chi}}{d_{cr} d_{initial}}$$

during a period of time $t_c \approx 10^{-43} s$ .Thus,

$$d_{cr} - d_{initial} = \tfrac{1}{2} \bar{g} (t_c)^2 = \left(\sqrt{\chi}\right) \frac{G M_{i(U)}}{d_{cr} d_{initial}} (t_c)^2 \qquad (130)$$

The Eq.(83), gives

$$\chi = \frac{m_{g(sp)}}{m_{i(sp)}} = 1 - \frac{2 U n_r}{m_{i(sp)} c^2} = 1 - \frac{2 \eta n_r k T}{m_{i(sp)} c^2}$$

Calculations by Carr, B.J [41], indicate that it would seem reasonable to suppose that the fraction of initial *primordial black hole* mass ultimately converted into *photons* is about $0.11$. This means that we can take

$$\eta = 0.11$$

Thus, the amount $\eta M_{iU} c^2$, where $M_{iU}$ is the total inertial mass of the Universe, expresses the total amount of inertial energy converted into photons at the initial instant of the Universe(*Primordial Photons*).

It was previously shown that photons and also the matter have *imaginary* gravitational masses associated to them. The matter has *negative* imaginary gravitational mass, while the photons have *positive* imaginary gravitational mass, given by

$$M_{gp(imaginary)} = 2 M_{ip(imaginary)} = + \frac{4}{\sqrt{3}} \left(\frac{hf}{c^2}\right) i$$

where $M_{ip(imaginary)}$ is the *imaginary inertial* mass of the photons.

Then, from the above we can conclude that, at the initial instant of the Universe, an amount of *imaginary* gravitational mass, $M^{total}_{gm(imaginary)}$, which was associated to the fraction of the *matter* transformed into photons, has been converted into *imaginary* gravitational mass of the primordial photons, $M^{total}_{gp(imaginary)}$, while an amount of *real inertial* mass of the matter, $M^{total}_{im(real)} = \eta\ M_{iU} c^2$, has been converted into *real* energy of the primordial photons, $E_p = \sum_{j=1}^{N} hf_j$, i.e.,

$$M^{total}_{gm(imaginary)} + M^{total}_{im(real)} = $$
$$= M^{total}_{gp(imaginary)} + \underbrace{M^{total}_{ip(real)}}_{\frac{E_p}{c^2}}$$

where $M^{total}_{gm(imaginary)} = M^{total}_{gp(imaginary)}$ and

$$E_p / c^2 \equiv M^{total}_{ip(real)} = M^{total}_{im(real)} = \eta M_{iU} \cong 0.11 M_{iU}$$

It was previously shown that, for the *photons* equation: $M_{gp} = 2M_{ip}$, is valid. This means that

$$\underbrace{M_{gp(imagimary)} + M_{gp(real)}}_{M_{gp}} =$$
$$= 2\underbrace{\left(M_{ip(imagimary)} + M_{ip(real)}\right)}_{M_{ip}}$$

By substituting $M_{gp(imaginary)} = 2M_{ip(imaginary)}$ into the equation above, we get

$$M_{gp(real)} = 2M_{ip(real)}$$

Therefore we can write that

$$M^{total}_{gp(real)} = 2M^{total}_{ip(real)} = 0.22 M_{iU}$$

The phenomenon of *gravitational deflection of light about the Sun* shows that *the gravitational interaction* between the Sun and the photons is *attractive*. This is due to the gravitational force between the Sun and a photon, which is given by

$$F = -G\, M_{gSun(imaginary)} m_{gp(imaginary)} / r^2 ,$$

where $m_{gp(imaginary)}$ (the imaginary gravitational mass of the photon) is a quantity *positive* and *imaginary*, and $M_{gSun\,(imaginary\,)}$ (the imaginary gravitational mass associated to the matter of the Sun) is a quantity *negative* and imaginary.

The fact of the gravitational interaction between the imaginary gravitational masses of the primordial photons and the imaginary gravitational mass of the matter be *attractive* is highly relevant, because it shows that it is necessary to consider the effect of this gravitational interaction, which is equivalent to the gravitational effect produced by the amount of *real* gravitational mass, $M^{total}_{gp(real)} \cong 0.22 M_{iU}$, sprayed by all the Universe.

This means that this amount, which corresponds to 22% of the total inertial mass of the Universe, must be added to the overall computation of the *total mass of the matter* (stars, galaxies, etc., gas and dust of interstellar and intergalactic media). Therefore, this additional portion corresponds to what has been called *Dark Matter* (See Fig. IX).

On the other hand, the *total* amount of gravitational mass at the initial instant, $M^{total}_{g}$, according to Eq.(41), can be expressed by

$$M^{total}_{g} = \chi\ M_{iU}$$

This mass includes the total negative gravitational mass of the matter, $M^{total}_{gm(-)}$, plus the total gravitational mass, $M^{total}_{gp(real)}$, converted into primordial photons. This tells us that we can put

$$M^{total}_{g} = M^{total}_{gm(-)} + M^{total}_{gp(real)} = \chi\ M_{iU}$$

whence

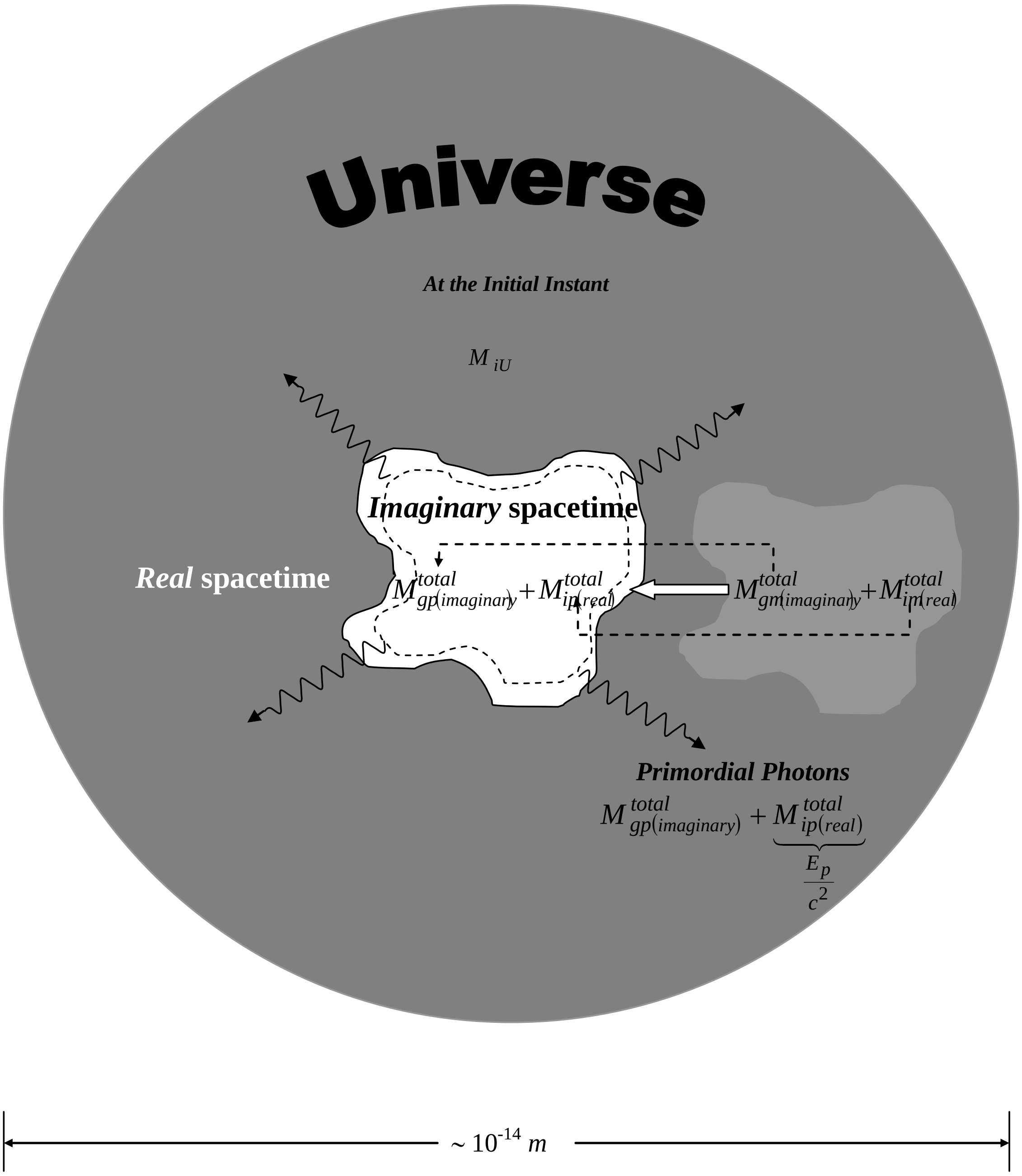

Fig. IX – Conversion of part of the *Real* Gravitational Mass of the Primordial Universe into *Primordial Photons*. The gravitational effect caused by the gravitational interaction of *imaginary* gravitational masses of the primordial photons with the *imaginary* gravitational mass associated to the matter is equivalent to the effect produced by the amount of *real* gravitational mass, $M^{total}_{gp(real)} \cong 0.22 M_{iU}$, sprayed by all Universe. This additional portion of mass corresponds to what has been called *Dark Matter*.

$$M_{gm(-)}^{total} = \chi \; M_{iU} - 0.22 M_{iU}$$

In order to calculate the value of $\chi$ we can start from the expression previously obtained for $\chi$, i.e.,

$$\chi = \frac{m_{g(sp)}}{m_{i(sp)}} = 1 - \frac{2\eta \; n_r kT}{m_{i(sp)}c^2} = 1 - \frac{T}{T_{critical}}$$

where

$$T_{critical} = \frac{m_{i(sp)}c^2}{2\eta n_r k} = 3.3 \times 10^{32} K$$

and

$$T = \frac{M_{i(sp)}c^2}{2\eta n_r k} = \frac{\left(\frac{m_{i(sp)}}{\sqrt{1-V^2/c^2}}\right)c^2}{2\eta n_r k} = \frac{T_{critical}}{\sqrt{1-V^2/c^2}}$$

We thus obtain

$$\chi = 1 - \frac{1}{\sqrt{1-V^2/c^2}}$$

By substitution of this expression into the equation of $M_{gm(-)}^{total}$, we get

$$M_{gm(-)}^{total} = \left(0.78 - \frac{1}{\sqrt{1-V^2/c^2}}\right) M_{iU}$$

On the other hand, the *Unification condition* ($U\eta_r \cong \Delta pc = M_{iU}c^2$) previously shown and Eq. (41) show that at the initial instant of the Universe, $M_{g(sp)}$ has the following value:

$$M_{g(sp)} = \left\{1 - 2\left[\sqrt{1 + \left(\frac{U\eta_r}{M_{i(sp)}}\right)} - 1\right]\right\} M_{i(sp)} \cong 0.1 M_{i(sp)}$$

Similarly, Eq.(45) tells us that

$$M_{g(sp)} = \left(1 - 2\left[\left(1 - V^2/c^2\right)^{-\frac{1}{2}}\right]\right) M_{i(sp)}$$

By comparing this expression with the equation above, we obtain

$$\frac{1}{\sqrt{1-V^2/c^2}} \cong 1.5$$

Substitution of this value into the expressions of $\chi$ and $M_{gm(-)}^{total}$ results in

$$\chi = -0.5$$

and

$$M_{gm(-)}^{total} \cong -0.72 M_{iU}$$

This means that 72% of the total energy of the Universe ($M_{iU}c^2$) is due to *negative gravitational mass* of the matter created at the initial instant.

Since the gravitational mass is correlated to the inertial mass (Eq. (41)), the energy related to the negative gravitational mass is where there is inertial energy (inertial mass). In this way, this negative gravitational energy permeates all space and tends to increase the rate of expansion of the Universe due to produce a strong gravitational repulsion between the material particles. Thus, this energy corresponds to what has been called *Dark Energy* (See Fig. X).

The value of $\chi = -0.5$ at the initial instant of the Universe shows that the gravitational interaction was *repulsive* at the Big-Bang. It remains repulsive until the temperature of the Universe is reduced down to the critical limit, $T_{critical}$. Below this temperature limit, the *attractive* component of the gravitational interaction became greater than the *repulsive* component, making *attractive* the resultant gravitational interaction. Therefore, at the beginning of the Universe – before the temperature decreased down to $T_{critical}$, there occurred an expansion of the Universe that was exponential in time rather than a normal power-law expansion. Thus, there was an evident *Inflation Period* during the beginning of the expansion of the Universe (See Fig. XI).

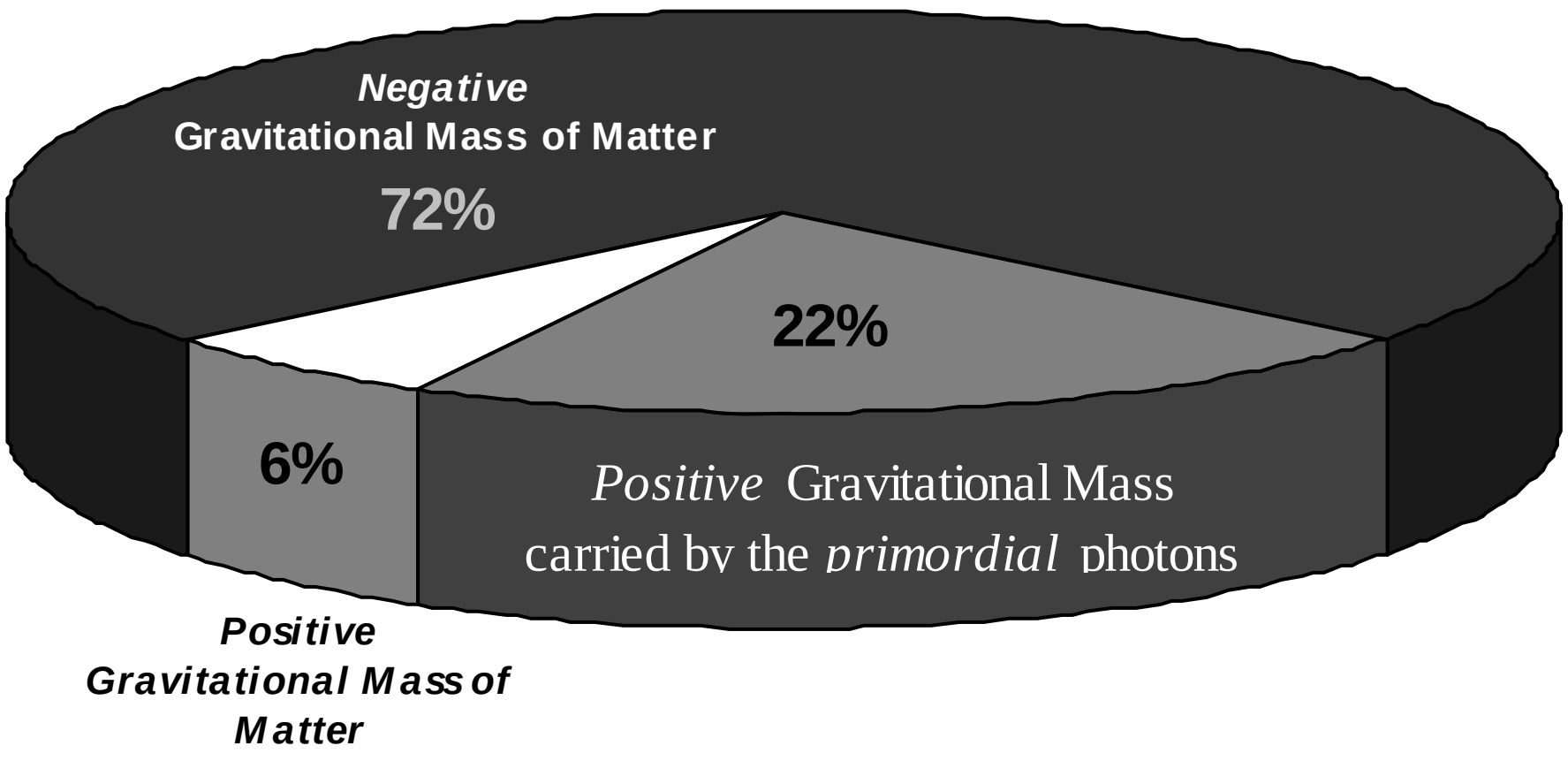

Fig. X - Distribution of Gravitational Masses in the Universe. The *total energy* related to *negative* gravitational mass of all the matter in the Universe corresponds to what has been called *Dark Energy*. While the *Dark Matter* corresponds to the *total* gravitational mass carried by the *primordial* photons, which is manifested in the *interaction of the imaginary* gravitational masses of the *primordial* photons with the imaginary mass of matter.

With the progressing of the *decompression* the *superparticles* cluster becomes a neutrons cluster. This means that the neutrons are created *without its antiparticle*, the antineutron. Thus, this solves the matter/antimatter dilemma that is unresolved in many cosmologies.

Now a question: How did the *primordial superparticles* appear at the beginning of the Universe?

It is a proven quantum fact that a wave function may collapse, and that, at this moment, all the possibilities that it describes are suddenly expressed in *reality*. This means that, through this process, particles can be suddenly *materialized*.

The materialization of the *primordial superparticles* into a critical volume denotes *knowledge* of what would happen with the Universe starting from that *initial condition*, a fact that points towards the *existence* of a Creator.

It was shown previously the possible existence of *imaginary particles* with imaginary masses in Nature. These particles can be associated with real particles, such as in case of the *photons* and *electrons*, as we have shown, or they can be associated with others imaginary particles by constituting the imaginary bodies. Just as the real particles constitute the real bodies.

The idea that we make about a *consciousness* is basically that of an *imaginary body* containing *psychic energy* and *intrinsic knowledge*. We can relate psychic energy with *psychic mass* (psychic mass= psychic energy/c$^2$). Thus, by analogy with the real bodies the psychic bodies would be constituted by psychic particles with psychic mass. Consequently, the psychic particles that constitute a consciousness would be equivalent to imaginary particles, and the *psychic mass* , $m_\Psi$ ,of the psychic particles would be equivalent to the *imaginary mass*, i.e.,

$$m_\Psi = m_{i(imaginary)} \qquad (131)$$

Thus, the imaginary masses associated to the *photons* and *electrons* would be *elementary psyche* actually, i.e.,

$$m_{\Psi photon} = m_{i(imaginary)photon} =$$
$$= \frac{2}{\sqrt{3}}\left(\frac{hf}{c^2}\right) i \qquad (132)$$

$$m_{\Psi electron} = m_{i(imaginary)electron} =$$
$$= -\frac{2}{\sqrt{3}}\left(\frac{hf_{electron}}{c^2}\right) i =$$
$$= -\frac{2}{\sqrt{3}} m_{i0(real)electron}\, i \qquad (133)$$

The idea that electrons have elementary psyche associated to themselves is not new. It comes from the pre-Socratic period.

By proposing the existence of psyche associated with matter, we are adopting what is called *panpsychic* posture. Panpsychism dates back to the pre-Socratic period;

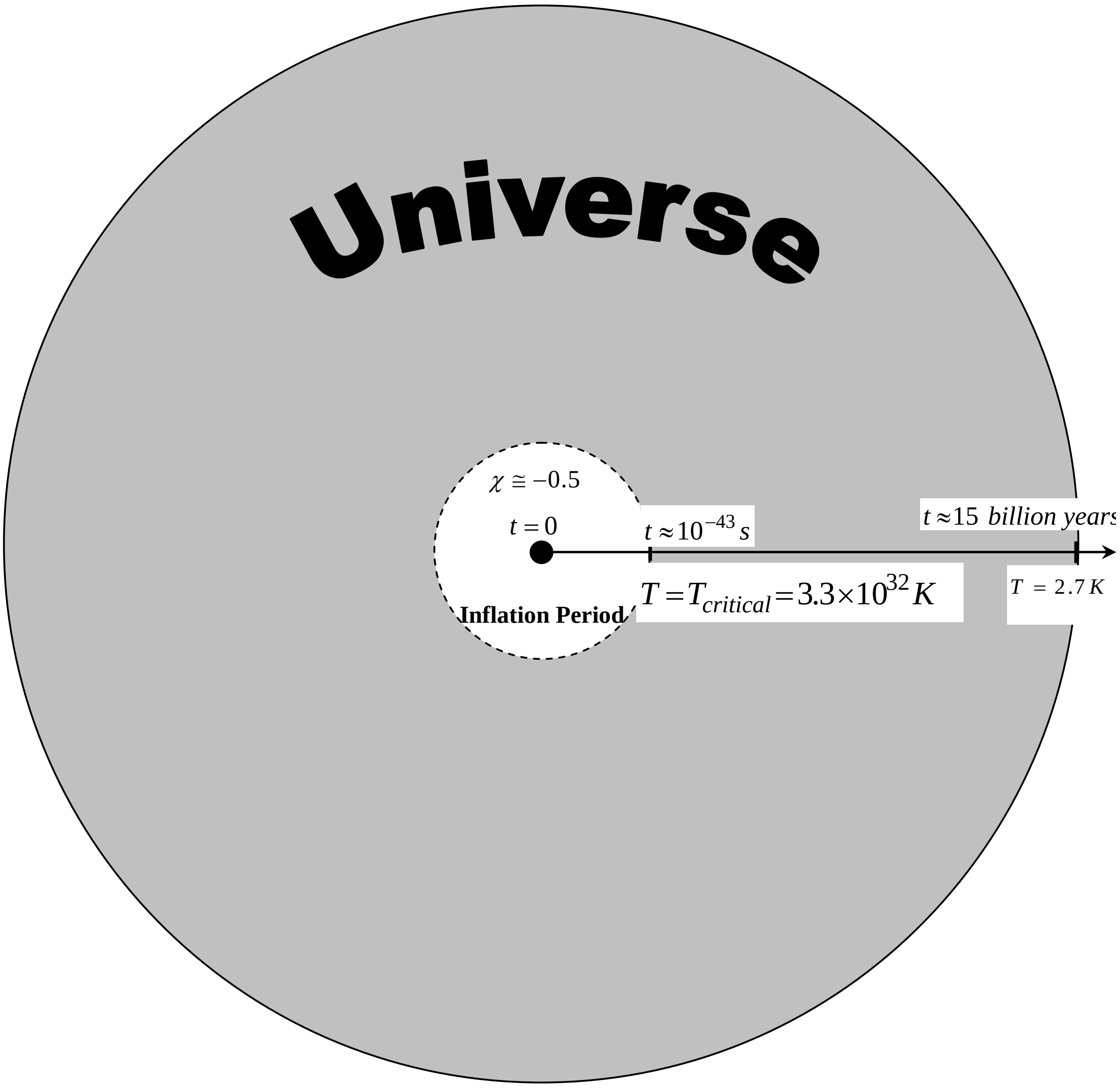

Fig. XI – *Inflation Period.* The value of $\chi \cong -0.5$ at the *Initial* Instant of the Universe shows that the gravitational interaction was repulsive at the Big-Bang. It remains repulsive until the temperature of the Universe is reduced down to the critical limit, $T_{critical}$. Below this temperature limit, the attractive component of the gravitational interaction became greater than the repulsive component, making attractive the resultant gravitational interaction. Therefore, at beginning of the Universe – before the temperature to be decreased down to $T_{critical}$, there occurred an expansion of the Universe that was exponential in time rather than a normal power-law expansion. Thus, there was an evident inflation period during the beginning of the expansion of the Universe.

remnants of organized panpsychism may be found in the Uno of Parmenides or in Heracleitus's Divine Flow. The scholars of Miletus's school were called *hylozoists*, that is, "those who believe that matter is alive". More recently, we will find the panpsychistic thought in Spinoza, Whitehead and Teilhard de Chardin, among others. The latter one admitted the existence of proto-conscious properties at the elementary particles' level.

We can find experimental evidences of the existence of psyche associated to electron in an experiment similar to that commonly used to show the wave duality of light. (Fig. XII). One merely substitutes an electron ray (fine electron beam) for the light ray. Just as in the experiment mentioned above, the ray which goes through the holes is detected as a wave if a *wave detector* is used (it is then observed that the interference pattern left on the detector screen is analogous with that produced by the light ray), and as a particle if a *particle detector* is used.

Since the electrons are detected on the other side of the metal sheet, it becomes obvious then that they passed through the holes. On the other hand, it is also evident that when they approached the holes, they had to decide which one of them to go through.

How can an electron "decide" which hole to go through? Where there is "choice", isn't there also *psyche*, by definition?

If the primordial superparticles that have been materialized at the beginning of the Universe came from the collapse of a primordial wave function, then the psychic form described by this wave function must have been generated in a consciousness with a psychic mass much greater than that needed to materialize the Universe (material and psychic).

This giant consciousness, in its turn, would not only be the greatest of all consciences in the Universe but also the *substratum* of everything that exists and, obviously, everything that exists would be entirely contained within it, including *all the spacetime*.

Thus, if the consciousness we refer to contains all the space, its volume is necessarily infinite, consequently having an *infinite psychic mass*.

This means that it contains all the existing psychic mass and, therefore, any other consciousness that exists will be contained in it. Hence, we may conclude that It is the *Supreme Consciousness* and that there is no other equal to It: It is *unique*.

Since the Supreme Consciousness also contains *all* time; past, present and future, then, for It the time does not flow as it flows for us.

Within this framework, when we talk about the Creation of the Universe, the use of the verb "to create" means that "something that was not" came into being, thus presupposing the concept of *time flow*. For the Supreme Consciousness, however, the instant of Creation is mixed up with all other times, consequently there being no "before" or "after" the Creation and, thus, the following question is not justifiable: "What did the Supreme Consciousness do before Creation?"

On the other hand, we may also infer, from the above that the existence of the Supreme Consciousness has no defined limit (beginning and end), what confers upon It the unique characteristic of *uncreated* and *eternal*.

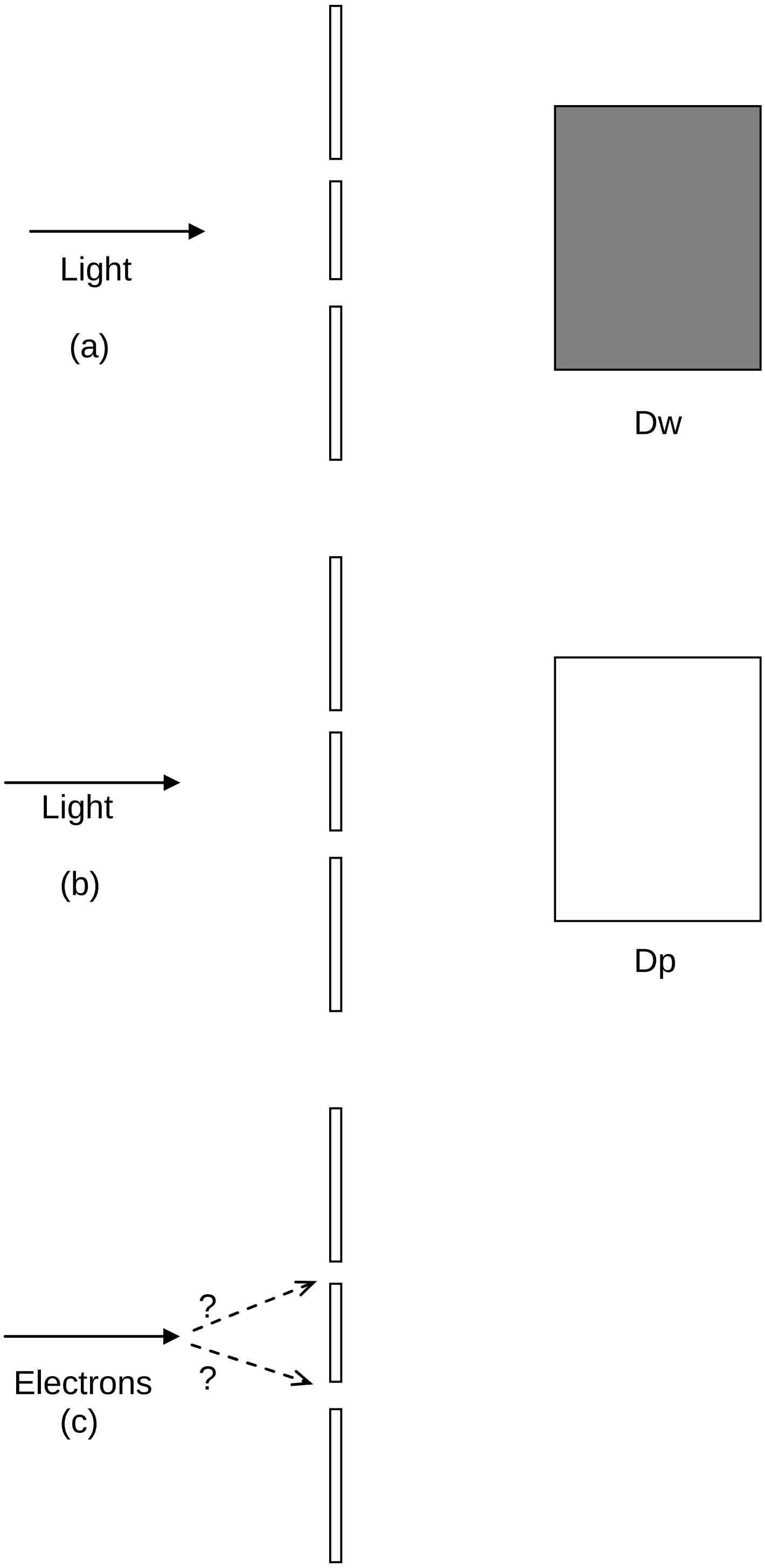

Fig. XII – A light ray, after going through the holes in the metal sheet, will be detected as a wave(a) by a wave detector Dw or as a particle if the wave detector is substituted for the wave detector Dp. Electron ray (c) has similar behavior as that of a light ray. However, before going through the holes, the electrons must “decide” which one to go through.

If the Supreme Consciousness is eternal, Its wave function $\Psi_{SC}$ shall never collapse (will never be null). Thus, for having an infinite psychic mass, the value of $\Psi_{SC}^2$ will be always infinite and, hence, we may write that

$$\int_{-\infty}^{+\infty} \Psi_{SC}^2 dV = \infty$$

By comparing this equation with Eq. (108) derived from Quantum Mechanics, we conclude that the Supreme Consciousness is simultaneously everywhere, i.e., It is *omnipresent*.

Since the Supreme Consciousness contains all consciences, it is expected that It also contain *all the knowledge.* Therefore, It is also *omniscient*. Consequently, It *knows* how to formulate well-defined mental images with psychic masses sufficient for its contents to *materialize*. In this way, It can materialize *everything* It wishes (*omnipotence*).

All these characteristics of the Supreme Consciousness (infinite, unique, uncreated, eternal, omnipresent, omniscient and omnipotent) coincide with those traditionally ascribed to *God* by most religions.

It was shown in this work that the "virtual" *quanta* of the *gravitational interaction* must have spin 1 and not 2, and that they are "virtual" photons (*graviphotons*) with *zero mass* outside the *coherent* matter. Inside the coherent matter the graviphoton mass is *non-zero*. Therefore, the gravitational forces are also *gauge* forces, because they are yielded by the exchange of "virtual" *quanta* of spin 1, such as the electromagnetic forces and the weak and strong nuclear forces.

Thus, the gravitational forces are produced by the exchange of "virtual" photons. Consequently, this is precisely the *origin of the gravity*.

Newton's theory of gravity does not explain *why* objects attract one another; it simply models this observation. Also Einstein's theory does not explain the origin of gravity. Einstein's theory of gravity only describes gravity with more precision than Newton's theory does.

Besides, there is nothing in both theories explaining the *origin of the energy* that produces the gravitational forces. Earth's gravity attracts all objects on the surface of our planet. This has been going on for well over 4.5 billions years, yet no known energy source is being converted to support this tremendous ongoing energy expenditure. Also is the enormous continuous energy expended by Earth's gravitational field for maintaining the Moon in its orbit - millennium after millennium. In spite of the ongoing energy expended by Earth's gravitational field to hold objects down on surface and the Moon in orbit, why the energy of the field never diminishes in strength or drains its energy source? Is this energy expenditure balanced by a conversion of energy from an unknown energy source?

The energy $W$ necessary to support the effort expended by the gravitational forces $F$ is well-known and given by

$$W = \int_{\infty}^{r} F dr = -G \frac{M_g m_g}{r}$$

According to the *principle of energy conservation*, this energy expenditure must be balanced by a conversion of energy from another energy type.

The Uncertainty Principle tells us that, due to the occurrence of exchange of *graviphotons* in a time interval $\Delta t < \hbar/\Delta E$ (where $\Delta E$ is the energy of the graviphoton), the energy variation $\Delta E$ cannot be detected in the system $M_g - m_g$. Since the total energy $W$ is the sum of the energy of the $n$ graviphotons, i.e., $W = \Delta E_1 + \Delta E_2 + ... + \Delta E_n$, then the energy $W$ *cannot be detected as well*. However, as we know it can be converted into another type of energy, for example, in rotational kinetic energy, as in the hydroelectric plants, or in the *Gravitational Motor*, as shown in this work.

It is known that a *quantum* of energy $\Delta E = hf$ which varies during a time interval $\Delta t = 1/f = \lambda/c < \hbar/\Delta E$ (wave period) cannot be experimentally detected. This is an *imaginary* photon or a "*virtual*" photon. Thus, the graviphotons are *imaginary* photons, i.e., the energies $\Delta E_i$ of the graviphotons are imaginaries energies and therefore the energy $W = \Delta E_1 + \Delta E_2 + ... + \Delta E_n$ is also an *imaginary* energy. Consequently, it belongs to the *imaginary space-time*.

According to Eq. (131), imaginary energy is equal to *psychic energy*. Consequently, the *imaginary space-time* is, in fact, the *psychic space-time*, which contains the Supreme Consciousness. Since the Supreme Consciousness has infinite psychic mass, then the *psychic space-time* has *infinite psychic energy*. This is highly relevant, because it confers to the *psychic space-time* the characteristic of *unlimited source of energy*.

This can be easily confirmed by the fact that, in spite of the enormous amount of energy expended by Earth's gravitational field to hold objects down on the surface of the planet and maintain the Moon in its orbit, the energy of Earth's gravitational field never diminishes in strength or drains its energy source.

If an experiment involves a large number of identical particles, all described by the same wave function $\Psi$, *real* density of mass $\rho$ of these particles in x, y, z, t is proportional to the corresponding value $\Psi^2$ ($\Psi^2$ is known as *density of probability*. If $\Psi$ is *complex* then $\Psi^2 = \Psi\Psi^*$. Thus, $\rho \propto \Psi^2 = \Psi.\Psi^*$). Similarly, in the case of psychic particles, the *density of psychic mass*, $\rho_\Psi$, in x, y, z, will be expressed by $\rho_\Psi \propto \Psi_\Psi^2 = \Psi_\Psi \Psi_\Psi^*$. It is known that $\Psi_\Psi^2$ is always *real* and *positive* while $\rho_\Psi = m_\Psi/V$ is an *imaginary* quantity. Thus, as the *modulus* of an imaginary number is always real and positive, we can transform the proportion $\rho_\Psi \propto \Psi_\Psi^2$, in equality in the following form:

$$\Psi_\Psi^2 = k|\rho_\Psi| \qquad (134)$$

where $k$ is a *proportionality constant* (real and positive) to be determined.

In Quantum Mechanics we have studied the *Superpositon Principle*, which affirms that, if a particle (or system of particles) is in a *dynamic state* represented by a wave function $\Psi_1$ and may also be in another dynamic state described by $\Psi_2$ then, the general dynamic state of the particle may be described by $\Psi$, where $\Psi$ is a linear combination(superposition)of $\Psi_1$ and $\Psi_2$, i.e.,

$$\Psi = c_1\Psi_1 + c_2\Psi_2 \qquad (135)$$

C*omplex constants* $c_1$ and $c_2$ respectively indicates the percentage of dynamic state, represented by $\Psi_1$ and $\Psi_2$ in the formation of the general dynamic state described by $\Psi$.

In the case of psychic particles (psychic bodies, consciousness, etc.),

by analogy, if $\Psi_{\Psi 1}$, $\Psi_{\Psi 2}$,..., $\Psi_{\Psi n}$ refer to the different dynamic states the psychic particle assume, then its general dynamic state may be described by the wave function $\Psi_{\Psi}$ , given by:

$$\Psi_{\Psi} = c_1\Psi_{\Psi 1} + c_2\Psi_{\Psi 2} + \ldots + c_n\Psi_{\Psi n} \quad (136)$$

The state of superposition of wave functions is, therefore, common for both psychic and material particles. In the case of material particles, it can be verified, for instance, when an electron changes from one orbit to another. Before effecting the transition to another energy level, the electron carries out "virtual transitions" [42]. A kind of *relationship* with other electrons before performing the real transition. During this relationship period, its wave function remains "*scattered*" by *a wide region of the space* [43] thus superposing the wave functions of the other electrons. In this relationship the electrons *mutually* influence one another, with the possibility of *intertwining* their wave functions‡‡. When this happens, there occurs the so-called *Phase Relationship* according to quantum-mechanics concept.

In the electrons "virtual" transition mentioned before, the "listing" of all the possibilities of the electrons is described, as we know, by *Schrödinger's wave equation.* Otherwise, it is general for material particles. By analogy, in the case of psychic particles, we may say that the "listing" of all the possibilities of the psyches involved in the relationship will be described by *Schrödinger's equation* – for psychic case, i.e.,

$$\nabla^2\Psi_{\Psi} + \frac{p_{\Psi}^2}{\hbar^2}\Psi_{\Psi} = 0 \quad (137)$$

Because the wave functions are capable of intertwining themselves, the quantum systems may "penetrate" each other, thus establishing an internal relationship where all of them are affected by the relationship, no longer being isolated systems but becoming an integrated part of a larger system. This type of internal relationship, which exists only in quantum systems, was called *Relational Holism* [44].

The equation of *quantization of mass* (33), in the generalized form, leads us to the following expression:

$$m_{i(imaginary)} = n^2 m_{i0(imagynary)(\min)}$$

Thus, we can also conclude that the *psychic mass is also quantized*, due to $m_{\Psi} = m_{i(imaginary)}$ (Eq. 131), i.e.,

$$m_{\Psi} = n^2 m_{\Psi(min)} \quad (138)$$

where

$$m_{\Psi(\min)} = -\tfrac{2}{\sqrt{3}}\left(hf_{\min}/c^2\right)i =$$
$$= -\tfrac{2}{\sqrt{3}}m_{i0(real)\min}\ i \quad (139)$$

It was shown that the *minimum quantum* of real inertial mass in the Universe, $m_{i0(real)\min}$, is given by:

$$m_{i0(real)\min} = \pm h\sqrt{3/8}/cd_{\max} =$$
$$= \pm 3.9\times 10^{-73}\,kg \quad (140)$$

By analogy to Eqs. (132) and (133), the expressions of the psychic masses associated to the *proton* and the *neutron* are respectively given by:

$$m_{\Psi proton} = m_{i(imaginary)proton} =$$
$$= +\tfrac{2}{\sqrt{3}}\left(hf_{proton}/c^2\right)i =$$
$$= +\tfrac{2}{\sqrt{3}}m_{i0(real)proton}\ i \quad (141)$$

‡‡ Since the electrons are simultaneously waves and particles, their wave aspects will interfere with each other; besides superposition, there is also the possibility of occurrence of *intertwining* of their wave functions.

$$m_{\Psi neutron} = m_{i(imaginary)neutron} =$$
$$= -\tfrac{2}{\sqrt{3}}\left(hf_{neutron}/c^2\right) i =$$
$$= -\tfrac{2}{\sqrt{3}} m_{i0(real)neutron}\; i \qquad (142)$$

The *imaginary* gravitational masses of the atoms must be *much smaller* than their *real* gravitational masses. On the contrary, the weight of the bodies would be very different of the observed values. This fact shows that $m_{i(imaginary)proton}$ and $m_{i(imaginary)neutron}$ must have *contrary* signs. In this way, the *imaginary* gravitational mass of an atom can be expressed by means of the following expression

$$m_{i(imaginary)atom} = N\left(m_e \pm \left(m_n - m_p\right) + \frac{\Delta E}{c^2}\right) i$$

where, $\Delta E$, is the interaction energy. By comparing this expression with the following expression

$$m_{i(real)atom} = N\left(m_e + m_p + m_n + \frac{\Delta E}{c^2}\right)$$

Thus,

$$\left|m_{i(imaginary)atom}\right| << m_{i(real)atom}$$

Now consider a monatomic body with *real* mass $M_{i(real)}$ and *imaginary* mass $M_{i(imaginary)}$. Then we have

$$\frac{M_{i(imaginary)}}{M_{i(real)}} = \frac{\Sigma\left(m_{i(imaginary)atom} + \frac{\Delta E_a i}{c^2}\right)}{\Sigma\left(m_{i(real)atom} + \frac{\Delta E_a}{c^2}\right)} =$$
$$= \frac{n\left(m_e \pm \left(m_n - m_p\right) + \frac{\Delta E}{c^2} + \frac{\Delta E_a}{c^2}\right) i}{n\left(m_e + m_p + m_n + \frac{\Delta E}{c^2} + \frac{\Delta E_a}{c^2}\right)} \cong$$
$$\cong \left(\frac{m_e \pm \left(m_n - m_p\right) + \frac{\Delta E}{c^2}}{m_e + m_p + m_n + \frac{\Delta E}{c^2}}\right) i$$

Since $\Delta E_a << \Delta E$.

The intensity of the gravitational forces between $M_{g(imaginary)}$ and an imaginary particle with mass $m_{g(imaginary)}$, *both at rest*, is given by

$$F = G M_{i(imaginary)} m_{i(imaginary)} / r^2 =$$
$$\cong \left(\frac{m_e \pm \left(m_n - m_p\right) + \frac{\Delta E}{c^2}}{m_e + m_p + m_n + \frac{\Delta E}{c^2}}\right) G \frac{M_{i(real)} i \;\; m_{i(real)} i}{r^2}$$

Therefore, the *total* gravity is

$$g_{real} + \Delta g_{(imaginary)} = -G\frac{M_{i(real)}}{r^2} -$$
$$-\left(\frac{m_e \pm \left(m_n - m_p\right) + \frac{\Delta E}{c^2}}{m_e + m_p + m_n + \frac{\Delta E}{c^2}}\right) G \frac{M_{i(real)}}{r^2}$$

Thus, the *imaginary* gravitational mass of a body produces an *excess* of gravity acceleration, $\Delta g$, given by

$$\Delta g \cong \left(\frac{m_e \pm \left(m_n - m_p\right) + \frac{\Delta E}{c^2}}{m_e + m_p + m_n + \frac{\Delta E}{c^2}}\right) G \frac{M_{i(real)}}{r^2}$$

In the case of *soft* atoms we can consider $\Delta E \cong 2\times10^{-13}\, joules$. Thus, in this case we obtain

$$\Delta g \cong 6\times10^{-4} G \frac{M_i}{r^2} \qquad (143)$$

In the case of the *Sun*, for example, there is an *excess* of gravity acceleration, due to its *imaginary* gravitational mass, given by

$$\Delta g \cong \left(6\times10^{-4}\right) G \frac{M_{iS}}{r^2}$$

At a distance from the Sun of $r = 1.0\times10^{13}\, m$ the value of $\Delta g$ is

$$\Delta g \cong 8\times10^{-10}\, m.s^{-2}$$

Experiments in the pioneer 10 spacecraft, at a distance from the Sun of about 67 AU or $r = 1.0\times10^{13}\, m$ [45], measured an excess acceleration towards the Sun of

$$\Delta g = 8.74 \pm 1.33\times10^{-10}\, m.s^{-2}$$

Note that the general expression for the gravity acceleration of the Sun is

$$g = \left(1 + \approx 6 \times 10^{-4}\right) G \frac{M_{iS}}{r^2}$$

Therefore, in the case of the *gravitational deflection of light about the Sun*, the new expression for the deflection of the light is

$$\delta = \left(1 + \approx 6 \times 10^{-4}\right) \frac{4GM_{iS}}{c^2 d} \qquad (144)$$

Thus, the increase in $\delta$ due to the excess acceleration towards the Sun can be considered *negligible*.

Similarly to the collapse of the real wave function, the collapse of the psychic wave function must suddenly also express in reality all the possibilities described by it. This is, therefore, *a point of decision* in which there occurs the compelling need of realization of the *psychic form*. Thus, this is moment in which the content of the psychic form realizes itself in the space-time. For an observer in space-time, something is *real* when it is in the form of matter or radiation. Therefore, the content of the psychic form may realize itself in space-time exclusively under the form of radiation, that is, it does not materialize. This must occur when the *Materialization Condition* is not satisfied, i.e., when the content of the psychic form is undefined (impossible to be defined by its own psychic) or it does not contain enough psychic mass to *materialize*[§§] the respective psychic contents.

Nevertheless, in both cases, there must always be a production of "virtual" photons to convey the psychic interaction to the other psychic particles, according to the quantum field theory, only through this type of quanta will interaction be conveyed, since it has an infinite reach and may be either attractive or repulsive, just as electromagnetic interaction which, as we know, is conveyed by the exchange of "virtual" photons.

If electrons, protons and neutrons have psychic mass, then we can infer that the psychic mass of the atoms are *Phase Condensates*[***]. In the case of the molecules the situation is similar. More molecular mass means more atoms and consequently, more psychic mass. In this case the phase condensate also becomes more structured because the great amount of elementary psyches inside the condensate requires, by stability reasons, a better distribution of them. Thus, in the case of molecules with very large molecular masses (*macromolecules*) it is possible that their psychic masses already constitute the most organized shape of a Phase Condensate, called Bose-Einstein Condensate[†††].

The fundamental characteristic of a Bose-Einstein condensate is, as we know, that the various parts making up the condensed system not only behave as a whole but also *become a whole*, i.e., in the psychic case, the various consciousnesses of the system become a *single consciousness* with psychic mass equal to the sum of the psychic masses of all the consciousness of the condensate. This obviously, increases the available knowledge in the system since it is proportional to the psychic mass of the consciousness. This unity confers an *individual* character to this type of consciousness. For this reason, from now on they will be called *Individual Material Consciousness.*

[§§] By this we mean not only materialization proper but also the movement of matter to realize its psychic content (including radiation).

[***] Ice and NaCl crystals are common examples of imprecisely-structured *phase condensates*. Lasers, super fluids, superconductors and magnets are examples of phase condensates more structured.

[†††] Several authors have suggested the possibility of the Bose-Einstein condensate occurring in the brain, and that it might be the physical base of memory, although they have not been able to find a suitable mechanism to underpin such a hypothesis. Evidences of the existence of Bose-Einstein condensates in living tissues abound (Popp, F.A Experientia, Vol. 44, p.576-585; Inaba, H., New Scientist, May89, p.41; Rattermeyer, M and Popp, F. A. Naturwissenschaften, Vol.68, N°5, p.577.)

We can derive from the above that most bodies do not possess individual material consciousness. In an iron rod, for instance, the cluster of elementary psyches in the iron molecules does not constitute Bose-Einstein condensate; therefore, the iron rod does not have an individual consciousness. Its consciousness is consequently, much more simple and constitutes just a phase condensate imprecisely structured made by the consciousness of the iron atoms.

The existence of consciousnesses in the atoms is revealed in the molecular formation, where atoms with strong mutual affinity (their consciousnesses) combine to form molecules. It is the case, for instance of the water molecules, in which two Hydrogen atoms join an Oxygen atom. Well, how come the combination between these atoms is always the same: the same grouping and the same invariable proportion? In the case of molecular combinations the phenomenon repeats itself. Thus, the chemical substances either mutually attract or repel themselves, carrying out specific motions for this reason. It is the so-called *Chemical Affinity*. This phenomenon certainly results from a specific interaction between the consciousnesses. From now on, it will be called *Psychic Interaction*.

*Mutual Affinity* is a dimensionless psychic quantity with which we are familiar and of which we have perfect understanding as to its meaning. The degree of *Mutual Affinity*, $A$, in the case of two consciousnesses, respectively described by $\Psi_{\Psi 1}$ and $\Psi_{\Psi 2}$, must be correlated to $\Psi^2_{\Psi 1}$ and $\Psi^2_{\Psi 2}$ [‡‡‡]. Only a simple algebraic form fills the requirements of interchange of the indices, the product

$$\Psi^2_{\Psi 1}.\Psi^2_{\Psi 2} = \Psi^2_{\Psi 2}.\Psi^2_{\Psi 1} = \\ = |A_{1,2}| = |A_{2,1}| = |A| \qquad (145)$$

In the above expression, $|A|$ is due to the product $\Psi^2_{\Psi 1}.\Psi^2_{\Psi 2}$ will be always positive. From equations (143) and (134) we get

$$|A| = \Psi^2_{\Psi 1}.\Psi^2_{\Psi 2} = k^2 |\rho_{\Psi 1}||\rho_{\Psi 2}| = \\ = k^2 \frac{|m_{\Psi 1}|}{V_1}\frac{|m_{\Psi 2}|}{V_2} \qquad (146)$$

The psychic interaction can be described starting from the psychic mass because the psychic mass is the source of the psychic field. Basically, *the psychic mass is gravitational mass*, $m_\Psi \equiv m_{g(imaginary)}$. In this way, the equations of the gravitational interaction are also applied to the Psychic Interaction. However, due to the psychic mass, $m_\Psi$, to be an *imaginary* quantity, it is necessary to put $|m_\Psi|$ into the mentioned equations in order to homogenize them, because as we know, the module of an imaginary number is always real and positive.

Thus, based on gravity theory, we can write the equation of the *psychic field* in *nonrelativistic* Mechanics.

$$\Delta\Phi = 4\pi G|\rho_\Psi| \qquad (147)$$

---

‡‡‡ Quantum Mechanics tells us that $\Psi$ do not have a physical interpretation or a simple meaning and also it cannot be experimentally observed. However such restriction does not apply to $\Psi^2$ , which is known as *density of probability* and represents the probability of finding the body, described by the wave function $\Psi$, in the point x, y, z at the moment t. A large value of $\Psi^2$ means a strong possibility to find the body, while a small value of $\Psi^2$ means a weak possibility to find the body.

It is similar to the equation of the gravitational field, with the difference that now instead of the density of gravitational mass we have the density of *psychic mass*. Then, we can write the general solution of Eq. (147), in the following form:

$$\Phi = -G\int \frac{|\rho_\Psi| dV}{r^2} \qquad (148)$$

This equation expresses, with nonrelativistic approximation, the potential of the psychic field of any distribution of psychic mass.

Particularly, for the potential of the field of only one particle with psychic mass $m_{\Psi 1}$, we get:

$$\Phi = -\frac{G|m_{\Psi 1}|}{r} \qquad (149)$$

Then the force produced by this field upon another particle with psychic mass $m_{\Psi 2}$ is

$$\left|\vec{F}_{\Psi 12}\right| = \left|-\vec{F}_{\Psi 21}\right| = -|m_{\Psi 2}|\frac{\partial \Phi}{\partial r} =$$

$$= -G\frac{|m_{\Psi 1}||m_{\Psi 2}|}{r^2} \qquad (150)$$

By comparing equations (150) and (146) we obtain

$$\left|\vec{F}_{\Psi 12}\right| = \left|-\vec{F}_{\Psi 21}\right| = -G|A|\frac{V_1 V_2}{k^2 r^2} \qquad (151)$$

In the *vectorial* form the above equation is written as follows

$$\vec{F}_{\Psi 12} = -\vec{F}_{\Psi 21} = -GA\frac{V_1 V_2}{k^2 r^2}\hat{\mu} \qquad (152)$$

Versor $\hat{\mu}$ has the direction of the line connecting the mass centers (psychic mass) of both particles and oriented from $m_{\Psi 1}$ to $m_{\Psi 2}$.

In general, we may distinguish and quantify two types of mutual affinity: *positive* and *negative* (*aversion*). The occurrence of the first type is synonym of psychic *attraction*, (as in the case of the atoms in the water molecule) while the aversion is synonym of *repulsion*. In fact, Eq. (152) shows that the forces $\vec{F}_{\Psi 12}$ and $\vec{F}_{\Psi 21}$ are attractive, if $A$ is *positive* (expressing *positive* mutual affinity between the two *psychic bodies*), and repulsive if $A$ is *negative* (expressing *negative* mutual affinity between the two *psychic bodies*). Contrary to the interaction of the matter, where the opposites attract themselves here, the *opposites repel themselves*.

A method and device to obtain images of *psychic bodies* have been previously proposed [46]. By means of this device, whose operation is based on the gravitational interaction and the piezoelectric effect, it will be possible to observe psychic bodies.

Expression (146) can be rewritten in the following form:

$$A = k^2 \frac{m_{\Psi 1}}{V_1}\frac{m_{\Psi 2}}{V_2} \qquad (153)$$

The psychic masses $m_{\Psi 1}$ and $m_{\Psi 2}$ are *imaginary* quantities. However, the product $m_{\Psi 1}.m_{\Psi 2}$ is a *real* quantity. One can then conclude from the previous expression that the degree of mutual affinity between two consciousnesses depends basically on the densities of their psychic masses, and that:

1) If $m_{\Psi 1} > 0$ and $m_{\Psi 2} > 0$ then $A > 0$ (positive mutual affinity between them)
2) If $m_{\Psi 1} < 0$ and $m_{\Psi 2} < 0$ then $A > 0$ (positive mutual affinity between them)
3) If $m_{\Psi 1} > 0$ and $m_{\Psi 2} < 0$ then $A < 0$ (negative mutual affinity between them)
4) If $m_{\Psi 1} < 0$ and $m_{\Psi 2} > 0$ then $A < 0$ (negative mutual affinity between them)

In this relationship, as occurs in the case of material particles ("virtual" transition of the electrons previously mentioned), the consciousnesses interact mutually, *intertwining* or not their wave functions. When this happens, there occurs the so-called *Phase Relationship* according to quantum-mechanics concept.

Otherwise a *Trivial Relationship* takes place.

The psychic forces such as the gravitational forces, must be very weak when we consider the interaction between two particles. However, in spite of the subtleties, those forces stimulate the relationship of the consciousnesses with themselves and with the Universe (Eq.152).

From all the preceding, we perceive that Psychic Interaction – unified with matter interactions, constitutes a single *Law* which links things and beings together and, in a network of continuous relations and exchanges, governs the Universe both in its material and psychic aspects. We can also observe that in the interactions the same principle reappears always identical. This *unity of principle* is the most evident expression of *monism* in the Universe.

## APPENDIX A: Allais effect explained

A Foucault-type pendulum slightly increases its period of oscillation at sites experiencing a *solar eclipse*, as compared with any other time. This effect was first observed by Allais [47] over 40 years ago. Also Saxl and Allen [48], using a torsion pendulum, have observed the phenomenon. Recently, an anomalous eclipse effect on gravimeters has become well-established [49], while some of the pendulum experiments have not. Here, we will show that the Allais gravity and pendulum effects during solar eclipses result from a *shielding effect of the Sun's gravity when the Moon is between the Sun and the Earth*.

The *interplanetary medium* includes interplanetary dust, cosmic rays and hot plasma from the solar wind. Its *density* is inversely proportional to the squared distance from the Sun, decreasing as this distance increases. Near the Earth-Moon system, this *density* is very low, with values about $5\,protons/cm^3\ \left(8.3\times10^{-21}kg/m^3\right)$. However, this density is *highly variable*. It can be increased up to $\sim 100\,protons/cm^3\ \left(1.7\times10^{-19}kg/m^3\right)$ [50].

The *atmosphere of the Moon* is very tenuous and insignificant in comparison with that of the Earth. The *average* daytime abundances of the elements known to be present in the lunar atmosphere, in atoms per cubic centimeter, are as follows: H <17, He 2-40x10$^3$, Na 70,K 17, Air 4x10$^4$, yielding ~8x10$^4$ total atoms per cubic centimeter $\left(\cong 10^{-16}kg.m^{-3}\right)$[51]. According to Öpik [52], near the Moon surface, the density of the lunar atmosphere can reach values up to $10^{-12}\,kg.m^{-3}$.The *minimum* possible density of the lunar atmosphere is in the top of the atmosphere and is essentially very close to the value of the *interplanetary medium*.

Since the density of the interplanetary medium is very small it cannot work as gravitational shielding. However, there is a top layer in the lunar atmosphere with density $\cong 1.3\times10^{-18}kg.m^{-3}$ that can work as a gravitational shielding and explain the Allais and pendulum effects. Below this layer, the density of the lunar atmosphere increases, making the effect of gravitational shielding negligible.

During the solar eclipses, when the Moon is between the Sun and the Earth, two *gravitational shieldings* $Sh1$ and $Sh2$, are established in the top layer of the lunar atmosphere (See Fig. 1A). In order to understand how these gravitational shieldings work (the gravitational shielding *effect*) see Fig. II. Thus, right after $Sh1$ (inside the system Moon-Lunar atmosphere), the *Sun's gravity acceleration*, $\vec{g}_S$, becomes $\chi\,\vec{g}_S$ where, according to Eq. (57) $\chi$ is given by

$$\chi = \left\{1-2\left[\sqrt{1+\left(\frac{n_r^2 D}{\rho c^3}\right)^2}-1\right]\right\} \qquad (1A)$$

The total density of *solar* radiation $D$ arriving at the top layer of the lunar atmosphere is given by

$$D = \sigma T^4 = 6.32\times10^7\,W/m^2$$

Since the temperature of the surface of the Sun is $T = 5.778\times10^3\,K$ and $\sigma = 5.67\times10^{-8}\,W.m^{-2}.K^{-4}$. The density of the top layer is $\rho \cong 1.3\times10^{-18}kg.m^{-3}$ then Eq. (1A) gives[§§§]

$$\chi = -1.1$$

The *negative* sign of $\chi$ shows that $\chi\vec{g}_S$, has *opposite* direction to $\vec{g}_S$. As previously showed (see Fig. II), after the second gravitational shielding

[§§§] The text in red in wrong. But the value of $\chi = -1.1$ is correct. It is not the solar radiation that produces the phenomenon. The exact description of the phenomenon starting from the same equation $(1A)$ is presented in the end of my paper: "*Scattering of Sunlight in Lunar Exosphere Caused by Gravitational Microclusters of Lunar Dust*" (2013).

$(Sh2)$ the gravity acceleration $\chi\vec{g}_S$ becomes $\chi^2\vec{g}_S$. This means that $\chi^2\vec{g}_S$ has the *same direction* of $\vec{g}_S$. In addition, right after $(Sh2)$ the lunar gravity becomes $\chi\vec{g}_{moon}$. Therefore, the *total gravity acceleration in the Earth* will be given by

$$\vec{g}' = \vec{g}_\oplus - \chi^2\vec{g}_S - \chi\vec{g}_{moon} \qquad (2A)$$

Since $g_S \cong 5.9\times10^{-3}\, m/s^2$ and $g_{moon} \cong 3.3\times10^{-5}\, m/s^2$ Eq. (2A), gives

$$\begin{aligned} g' &= g_\oplus - (-1.1)^2 g_S - (-1.1) g_{Moon} = \\ &\cong g_\oplus - 7.1\times10^{-3}\, m.s^{-2} = \\ &= \left(1 - 7.3\times10^{-4}\right) g_\oplus \end{aligned} \qquad (3A)$$

This decrease in $g$ increases the period $T = 2\pi\sqrt{l/g}$ of a *paraconical pendulum* (Allais effect) in about

$$T' = T\sqrt{\frac{g_\oplus}{\left(1-7.3\times10^{-4}\right)g_\oplus}} = 1.00037\ T$$

This corresponds to $0.037\%$ increase in the period, and is roughly the value (0.0372%) obtained by Saxl and Allen during the total solar eclipse in March 1970 [48].

As we have seen, the density of the interplanetary medium near the Moon is *highly variable* and can reach values up to $\sim 100\ protons\ /\ cm^3$ $\left(1.7\times10^{-19}\, kg/m^3\right)$.

When the density of the interplanetary medium increases, the top layer of the lunar atmosphere can also increase its density, by absorbing particles from the interplanetary medium due to the lunar gravitational attraction. In the case of a density increase of roughly 30% $\left(1.7\times10^{-18}\, kg/m^3\right)$, the value for $\chi$ becomes

$$\chi = -0.4$$

Consequently, we get

$$\begin{aligned} g' &= g_\oplus - (-0.4)^2 g_S - (-0.4) g_{Moon} = \\ &\cong g_\oplus - 9.6\times10^{-4}\, m.s^{-2} = \\ &= \left(1 - 9.7\times10^{-5}\right) g_\oplus \end{aligned} \qquad (4A)$$

This decrease in $g$ increases the pendulum's period by about

$$T' = T\sqrt{\frac{g_\oplus}{\left(1-9.4\times10^{-5}\right)g_\oplus}} = 1.000048\ T$$

This corresponds to $0.0048\%$ increase in the pendulum's period. Jun's abstract [53] tells us of a relative change less than 0.005% in the pendulum's period associated with the 1990 solar eclipse.

For example, if the density of the top layer of the lunar atmosphere increase up to $2.0917\times10^{-18}\, kg/m^3$, the value for $\chi$ becomes

$$\chi = -1.5\times10^{-3}$$

Thus, we obtain

$$\begin{aligned} g' &= g_\oplus - \left(-1.5\times10^{-3}\right)^2 g_S - \left(-1.5\times10^{-3}\right) g_{Moon} = \\ &\cong g_\oplus - 6.3\times10^{-8}\, m.s^{-2} = \\ &= \left(1 - 6.4\times10^{-9}\right) g_\oplus \end{aligned} \qquad (5A)$$

So, the total gravity acceleration in the Earth will *decrease* during the solar eclipses by about

$$6.4\times10^{-9}\, g_\oplus$$

The size of the effect, as measured with a *gravimeter*, during the 1997 eclipse, was roughly $(5-7)\times10^{-9}\, g_\oplus$ [54, 55].

The decrease will be even smaller for $\rho \gtrsim 2.0917\times10^{-18}$ kg.m$^{-3}$. The lower limit now is set by Lageos satellites, which suffer an anomalous acceleration of only about $3\times10^{-13}\, g_\oplus$, during "seasons" where the satellite experiences eclipses of the Sun by the Earth [56].

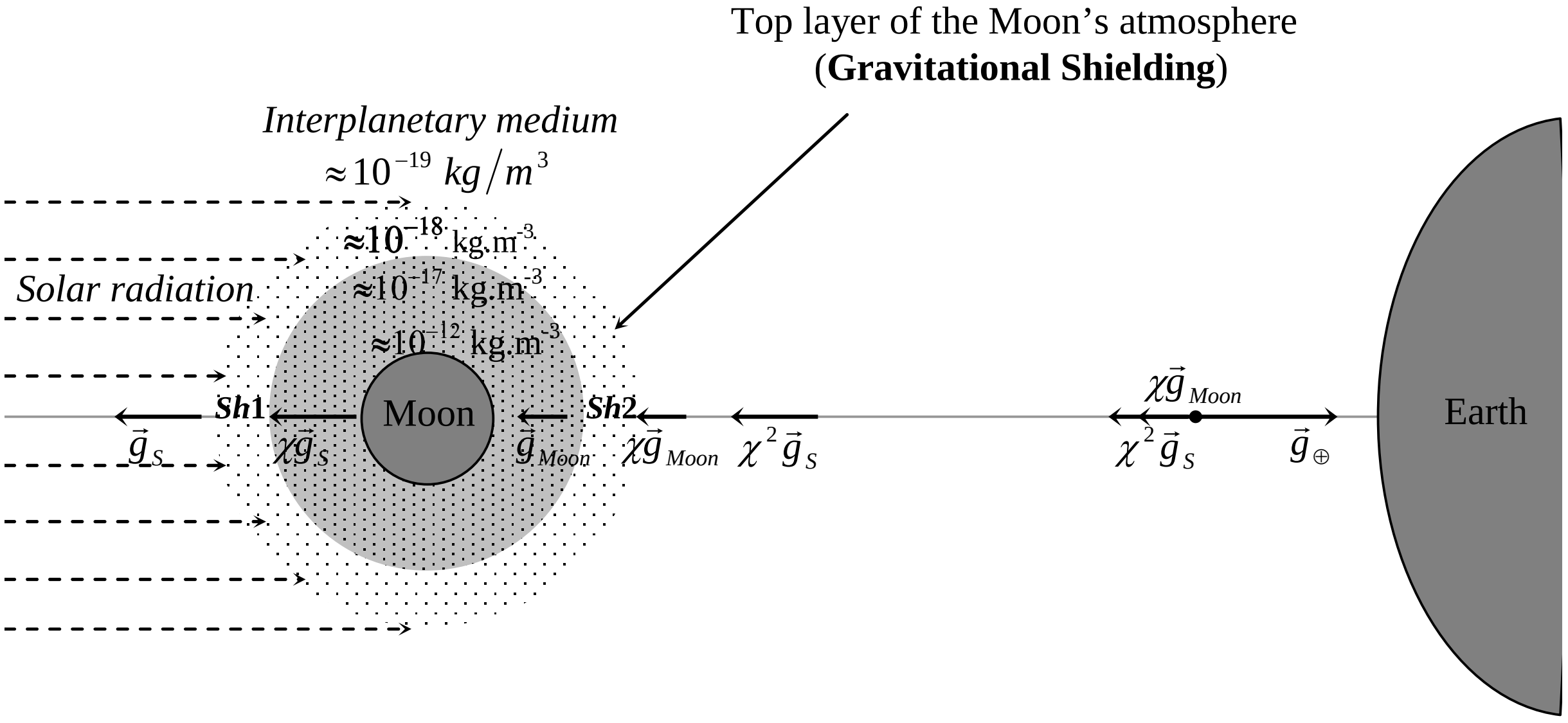

Fig. 1A – *Schematic diagram of the Gravitational Shielding around the Moon* – The top layer of the Moon's atmosphere with density of the order of $10^{-18}$ kg.m$^{-3}$ , produces a gravitational shielding when subjected to the radiation from the Sun. Thus, the solar gravity $\vec{g}_S$ becomes $\chi\ \vec{g}_S$ after the first shielding $Sh1$ and $\chi^2\vec{g}_S$ after the second shielding $Sh2$. The Moon gravity becomes $\chi\ \vec{g}_{Moon}$ after $Sh2$. Therefore the *total gravity acceleration in the Earth* will be given by $\vec{g}\,' = \vec{g}_\oplus - \chi^2\vec{g}_S - \chi\vec{g}_{moon}$.

## APPENDIX B

In this appendix we will show why, in the *quantized gravity equation* (Eq.34), $n = 0$ is *excluded* from the sequence of possible values of $n$. Obviously, the exclusion of $n = 0$, means that the gravity can have only discrete values *different of zero*.

Equation (33) shows that the gravitational mass is *quantized* and given by

$$M_g = n^2 m_{g(\min)}$$

Since Eq. (43) leads to

$$m_{g(\min)} = m_{io(\min)}$$

where

$$m_{i0(\min)} = \pm h\sqrt{3/8}/cd_{\max} = \pm 3.9\times 10^{-73} kg$$

is the *elementary quantum of inertial mass*. Then the equation for $M_g$ becomes

$$M_g = n^2 m_{g(\min)} = n^2 m_{i(\min)}$$

On the other hand, Eq. (44) shows that

$$M_i = n_i^2 m_{i0(\min)}$$

Thus, we can write that

$$\frac{M_g}{M_i} = \left(\frac{n}{n_i}\right)^2 \quad or \quad M_g = \eta^2 M_i \qquad (1B)$$

where $\eta = n/n_i$ is a *quantum number* different of $n$.

By multiplying both members of Eq. (1B) by $\sqrt{1 - V^2/c^2}$ we get

$$m_g = \eta^2 m_i \qquad (2B)$$

By substituting (2B) into Eq. (21) we get

$$E_n = \frac{n^2 h^2}{8 m_g L^2} = \frac{n^2 h^2}{8\eta^2 m_i L^2} \qquad (3B)$$

From this equation we can easily conclude that $\eta$ cannot be *zero* $\left(E_n \to \infty \;\; or \;\; E_n \to \frac{0}{0}\right)$. On the other hand, the Eq. (2B) shows that the exclusion of $\eta = 0$ means the exclusion of $m_g = 0$ as a possible value for the gravitational mass. Obviously, this also means the exclusion of $M_g = 0$ (Relativistic mass). Equation (33) tells us that $M_g = n^2 m_{g(\min)}$, thus we can conclude that the exclusion of $M_g = 0$ implies in the exclusion of $n = 0$ since $m_{g(\min)} = m_{i0(\min)} = finite\ value$ (*elementary* quantum of mass). Therefore Eq. (3B) is only valid for values of $n$ and $\eta$ different of zero. Finally, from the *quantized gravity equation* (Eq. 34),

$$g = -\frac{GM_g}{r^2} = n^2\left(-\frac{Gm_{g(\min)}}{(r_{\max}/n)^2}\right) = $$
$$= n^4 g_{\min}$$

we conclude that the exclusion of $n = 0$ means that *the gravity* can have only discrete values *different of zero*.